\documentclass[reprint,superscriptaddress,nofootinbib,amsmath,amssymb,aps,prx]{revtex4-2}

\usepackage{amsmath}
\usepackage{amssymb}
\usepackage{amsthm}
\usepackage{graphicx}
\graphicspath{{Images/}}
\usepackage[british]{babel}
\usepackage[breaklinks=true,hidelinks]{hyperref}
\hypersetup{colorlinks = true,linkcolor = blue,anchorcolor = blue,citecolor = blue,filecolor = blue,urlcolor = blue}
\usepackage{hypernat}
\usepackage{physics}
\usepackage[shortlabels]{enumitem}
\usepackage{commath}
\usepackage{cancel}
\allowdisplaybreaks
\usepackage[skip=1pt]{caption}
\usepackage{subcaption}
\usepackage{etoolbox}
\usepackage{array}
\usepackage{xcolor}
\usepackage{hhline}
\DeclareMathOperator{\sign}{sign}
\DeclareMathOperator{\erfc}{erfc}
\usepackage{bm}
\begin{document}
\title{\large First Principles Excitons in Periodic Systems with Gaussian Density Fitting and Ewald Potential Functions}
%\title{\large Bethe-Salpeter equation in the real-space RI-Gaussian scheme from first principles}

\author{M. A. García-Blázquez}
\email{manuelantonio.garcia@estudiante.uam.es}
\affiliation{Department of Condensed Matter Physics,
Universidad Aut\'onoma de Madrid, Madrid 28049, Spain}

\author{J. J. Palacios}
\email{juanjose.palacios@uam.es}
\affiliation{Department of Condensed Matter Physics,
Universidad Aut\'onoma de Madrid, Madrid 28049, Spain}
\affiliation{Condensed Matter Physics Center (IFIMAC) and Instituto Nicol\'as Cabrera (INC), Universidad Aut\'onoma de Madrid, Madrid 28049, Spain}
%%%%%%%%%%%%%%%%%%%%%%%%%%%%%%%%%%%%%%%%%%%%%%%%%%%%%%%%%%%%%%%%%%%%%%%%%%%%%%%%%
\begin{abstract}
Excitons, namely neutral excitations in a system of electrons arising from the electron-hole interaction, are often essential to explain optical measurements in materials. They are governed by the Bethe-Salpeter equation, which can be cast into a matrix form that is formally analogous to the one for electrons at the mean-field level. However, constructing the corresponding excitonic Hamiltonian in practice is challenging, specially from a computational perspective if one wishes to surpass effective models. Methods that enable such calculations from the different density-functional theory frameworks currently available are, therefore, convenient. In this work we present an approach to solve the BSE employing Gaussian basis functions starting from a self-consistent, possibly hybrid calculation in any non-metallic solid. It is based on the Gaussian density fitting or resolution of the identity approximation to reduce the initial quartic scaling in the basis dimension, in combination with the use of Ewald-type potential functions to automatically sum the conditionally convergent lattice series. As an illustration of the computational implementation, we provide examples of exciton spectra and optical absorption in some paradigmatic 2D and 3D materials where the single-particle approximation fails qualitatively. 
\end{abstract}
\maketitle
%%%%%%%%%%%%%%%%%%%%%%%%%%%%%%%%%%%%%%%%%%%%%%%%%%%%%%%%%%%%%%%%%%%%%%%%%%%%%%%%%
\section{Introduction}\label{IntroSec}
Correlation effects in a system with many particles, such as a solid, may be important both qualitatively and quantitatively in order to explain a wide range of experimental results, notably spectroscopic measurements including inelastic X-ray scattering (IXS) \cite{IXS1,IXS2,IXS3}, electron energy loss (EELS) \cite{EELS1,EELS2,EELS3}, attosecond transient absorption (ATAS) \cite{ATAS1,ATAS2,ATAS3}, double charge transfer (DCT) \cite{DCT1} or Auger electron spectroscopy (AES) \cite{AES1,AES2}. Incorporating such correlation functions between two particles is challenging from a theoretical perspective, often requiring simplifying approximations to obtain practical expressions, moreover in the computational facet any realistic calculation will inevitably turn demanding to some extent. However, the framework allows to study otherwise inaccessible phenomena, namely those that are dominated by quasi-particles emerging from the original interactions. Specifically, correlated electron-hole pairs, also known as excitons, have long been known to have a remarkable impact in optical properties of insulators and semiconductors \cite{Sham1980}, particularly, but not exclusively \cite{ARexp1962,rohlfing1998electron}, in low-dimensional materials \cite{LOWDIM1,LOWDIM2,LOWDIM3,LOWDIM4,NONLINEAR1,NONLINEAR2,ExpHBN}. 

The correlation function for excitons is governed by a Dyson-like expression, the Bethe-Salpeter equation (BSE) \cite{strinati1988}, which can be cast into matrix form to yield essentially an eigenvalue problem for the two-particle state, reminiscent of the Casida equation in time-dependent density-functional theory (TDDFT) \cite{Casida1995}. Some of the first realistic BSE calculations in materials were also carried out with Gaussians basis functions \cite{rohlfing_louie_2000} for the electronic wavefunctions, and in later years there has been progress in this regard \cite{PattersonEXCITON}. In the present day, BSE calculations may be considered part of the standard in first principles studies in periodic systems as attested by the existence of multiple public codes \cite{BGW2012,YAMBO2019,GPAW2024,EXCITING2019} that provide such functionality, however, it is mostly limited to formulations with (possibly augmented) plane-waves bases.  

In this work we propose a practical methodology to solve the BSE equation employing Gaussian-type functions (GTF) based on the Gaussian density fitting approximation, also known as resolution of the identity (RI) \cite{Vahtras1993,RI2005,Scheffler2012}, thereby reducing the impractical quartic scaling in the basis size originally required by the direct and exchange terms in the BSE. In order to carry out the calculations in practice it is necessary to address the conditional convergence of the Coulomb lattice series in two and three dimensions. To this effect we employ the Ewald \cite{Ewald1921,Saunders1992} (3D) and Parry \cite{Parry1975,Heyes1977} potential (2D) functions, which are absolutely convergent and can be implemented very generally for any material without multipole corrections. Since the density here is purely electronic and charge neutrality cannot be achieved by adding the nuclear contributions, this is at the cost of the necessity to evaluate the Ewald-type integrals in a direct-space supercell with the same scaling parameters as those employed in the reciprocal space discretization for the BSE. However, the latter is not an essentially limiting factor since it does not contribute to the dominant scaling of the 3-center integrals, moreover the matrix elements eventually decay in very large supercells and can be filtered out. 

In analogy to the single-particle framework, the choice of a GTF basis need not be inherently better or worse than other alternatives in general, instead it presents some advantages and disadvantages. In contrast to plane-waves, it avoids artificial lattice replication in low-dimensional materials (thus also the shape-dependent truncations of the Coulomb potential \cite{Rubio2006,BGW2012,YAMBO2019}), the reduced basis dimension allows for relatively fast internal summations and in particular a rapid convergence of the virtual states series in the polarizability \cite{RIG0W0}, the option of hybrid functionals as either a starting point or (in some cases) an alternative to GW corrections \cite{louie1986} is cheap, and the evaluation of the velocity operator for optical responses is straightforward even with non-local Hamiltonian terms \cite{garcia2023shift}. The main drawback is the need to perform an individualized basis optimization for each system, in our case also in the fitting auxiliary basis, while limited by numerical errors induced by (quasi-)linear dependencies and diffuse exponents (important to properly describe the conduction states), which obviously hinders automation or high-throughput scanning of materials. Furthermore, the general scarcity of studies in many-body properties of solids, particularly regarding excitons, makes the choice of bases and comparison of results harder, thereby we hope to contribute with a small set of examples.

In order to illustrate the methodology and prove its applicability we analyze the excitonic spectra and optical absorption obtained for some well known materials with two and three-dimensional periodicity, specifically a single-layer of black phosphorus and hexagonal boron nitride (hBN), in addition to solid argon. These systems exhibit different crystalline symmetries as well as single-particle band gaps, but they all present signals of strong electron-hole correlations in spectroscopic measurements. Through explicit comparisons we find that the results (notably the exciton textures in reciprocal space) are in overall good agreement, both qualitatively and quantitavely, with previous computational studies mainly employing plane-waves, and by extension with experimental measurements when substrate effects are not important.

The outline of this paper is as follows. Section \ref{BSEsec} is devoted to a basis-agnostic, practical formulation of the BSE. In Section \ref{CorrLSec} we introduce the general theory of two-particle correlations and the BSE, particularizing for the simultaneous electron-hole pair with screened Coulomb interactions as given by the GW approximation in the static limit, and employing the random-phase approximation (RPA). In Section \ref{BSEMatSec} we formulate the BSE in the standard matrix representation to define the exciton spectrum in terms of an eigenvalue problem. Section \ref{GaussianRIsec} explains in detail the proposed method employing GTF to construct the BSE Hamiltonian. In Section \ref{BSEgauss} we particularize the general results in the basis of eigenfunctions by expanding in two sets of GTF, one for the single-particle calculation and another (auxiliary) to fit the electronic density; the resulting integrals collected in Section \ref{GaussianIntSec} and expressed conveniently. In Section \ref{EwaldTheorySec} we address the conditional convergence problem by explaining the use of Ewald potential functions in a general case, with the introduction of supercells to obtain chargeless lattice distributions for the suitable sectors of the BSE. In Section \ref{Ewald3Dsec} and \ref{Ewald2Dsec} we consider, respectively, the habitual Ewald potential function for the 3D case and the modified Parry version for the important 2D case, as well as their integrations with GTF. In Section \ref{MetricsSec} we discuss several possible metric options for the RI. In Section \ref{SecDivergence} we tackle the divergent terms in the BSE, for which the previous formalism can still be applied albeit with some careful modifications.
Section \ref{ApplicationsSec} is devoted to the actual implementation and results in specific materials. In Section \ref{ImplementationSec} we outline the central details of the ab-initio implementation, followed in \ref{ConductivitySec} by a discussion on the optical absorption in relation to the solutions of the BSE. In Sections \ref{PhoSec}, \ref{hBNSec} and \ref{ArSec} we present results for black phosphorus, hBN and solid argon, respectively. The conclusions and outlook of the present work can be found in Section \ref{ConclusionSec}. Finally, Appendices \ref{ScreeningApp}, \ref{GaussianApp}, \ref{TDAApp} discuss in more detail some technicalities and extensions. The Supplemental Material contains the band structures, input files for the self-consistent calculations and auxiliary basis sets for the case studies in Section \ref{ApplicationsSec}.

\section{The BSE in the GW-RPA approximation}\label{BSEsec}
\subsection{Correlation Functions}\label{CorrLSec}
The two-particle correlation function is generally defined for fermions as \cite{strinati1988}
\begin{equation}\label{Lcorrelator}
L(1,2,1',2')=-G_{2}(1,2,1',2')+G(1,1')G(2,2')
\end{equation}
where $(1)\equiv(\bm{x}_{1},t_{1})\equiv(\bm{r}_{1},\sigma_{1},t_{1})$ is a compact notation for the space, spin and time variables, $G(1,2)=-i\langle T[\hat{\psi}(1)\hat{\psi}^{\dagger}(2)]\rangle$ and $G_{2}(1,2,1',2')=-\langle T[\hat{\psi}(1)\hat{\psi}(2)\hat{\psi}^{\dagger}(2')\hat{\psi}^{\dagger}(1')]\rangle$ are the time-ordered one and two-particle Green's functions, respectively, and $\hat{\psi}(1),\hat{\psi}^{\dagger}(1)$ are the (in our case, electronic) field operators in the Heisenberg picture. We hereafter assume an interacting system of electrons with a time-independent Hamiltonian $\hat{H}$ and at $T=0$, so that the expectation values are taken in the many-body ground state. In the limit of non-interacting particles, $\hat{H}=\hat{H}^{0}$ with eigenfunctions $\ket{\psi_{n}}$ and energies $\varepsilon_{n}$, the ground state is the Fermi sea (devoid of excitons), which we assume to be gaped, and the one-body Green's function $G=G^{0}$ admits the following spectral representation in the frequency domain
\begin{equation}\label{G0}
G^{0}(\bm{x}_{1},\bm{x}_{2};\omega)=\sum_{n}\frac{\psi_{n}(\bm{x}_{1})\psi_{n}^{*}(\bm{x}_{2})}{\hbar\omega-\varepsilon_{n}+i\eta^{+}\sign(\varepsilon_{n}-\varepsilon_{F})}
\end{equation}
where $\varepsilon_{F}$ is the Fermi energy. Hence it can be written in matrix form in the (truncated) orthonormal basis of single-particle eigenstates as $G^{0}_{l;l'}(z)=[(zI-H^{0})^{-1}]_{l,l'}$, where $z\rightarrow\hbar\omega+i\eta^{+}\sign(\varepsilon_{l}-\varepsilon_{F})$, $I$ is the identity matrix and the following shorthand notation for matrix elements of two-point functions has been introduced
\begin{equation}\label{Int2}
F_{m;n}\equiv\iint\psi^{*}_{m}(\bm{r}_{1})F(\bm{r}_{1},\bm{r}_{2})\psi_{n}(\bm{r}_{2})d\bm{r}_{1}d\bm{r}_{2}
\end{equation}
%The domain of all spatial integrals is understood to be $\mathbb{R}^{3}$ throughout the text. 
A similar matrix expression can be formulated for the interacting $G$ involving the many-body Hamiltonian $\hat{H}$.

In analogy with $G(1,2)$, the correlation function \eqref{Lcorrelator} satisfies a Dyson-like equation which is now 4-point, namely the BSE \cite{strinati1988,martin2016}
\begin{equation*}\begin{aligned}
&L(1,2,1',2')=L_{0}(1,2,1',2')\\
&+\int L_{0}(1,4,1',3)\Xi(3,5,4,6)L(6,2,5,2')d(3456)
\end{aligned}\end{equation*}
where $L_{0}(1,2,1',2')=G(1,2')G(2,1')$ is the correlation function for non-interacting particles and $\Xi$ is the interaction kernel, which in the GW approximation is given by \cite{rohlfing_louie_2000}
\begin{equation}\begin{aligned}\label{kernel}
\Xi(3,5,4,6)=&-i\delta(3,4)\delta(5,6)v(3,5)\\
&+i\delta(3,6)\delta(4,5)W(3,5)
\end{aligned}\end{equation}
where $v(1,2)=\delta(t_{1}-t_{2})/\abs{\bm{r}_{1}-\bm{r}_{2}}$ (in atomic units) and $W$ are the bare and screened Coulomb potentials, respectively. In \eqref{kernel}, the variation of $W$ under perturbations has been neglected. For optical processes, which conserve the number of particles in the system, it is sufficient to consider simultaneous creation and simultaneous annihilation of an electron and a hole, which is given by the limit $t_{1'}=t_{1}^{+}$ (indicated by $1'=1^{+}$) and $t_{2'}=t_{2}^{+}$ in $L(1,2,1',2')$. For this section of the correlation function, the time structure of the BSE is simplified to a single variable in the frequency domain. Furthermore, by neglecting the dynamical effects of $W$, which is justified if the plasmon frequencies are much larger than the binding energies \cite{rohlfing_louie_2000}, one obtains the static electron-hole version of the BSE \cite{martin2016}
\begin{equation}\begin{aligned}\label{BSEstatic}
&L(\bm{x}_{1},\bm{x}_{2},\bm{x}_{1'},\bm{x}_{2'};\omega)=L_{0}(\bm{x}_{1},\bm{x}_{2},\bm{x}_{1'},\bm{x}_{2'};\omega)\\
&+\int L_{0}(\bm{x}_{1},\bm{x}_{4},\bm{x}_{1'},\bm{x}_{3};\omega)\Xi(\bm{x}_{3},\bm{x}_{5},\bm{x}_{4},\bm{x}_{6})\\
&\cdot L(\bm{x}_{6},\bm{x}_{2},\bm{x}_{5},\bm{x}_{2'};\omega)d(\bm{x}_{3}\bm{x}_{4}\bm{x}_{5}\bm{x}_{6})
\end{aligned}\end{equation}
In this case the screening is determined by the corresponding Hedin's equation \cite{Hedin1965} in the static limit
\begin{equation}\begin{aligned}\label{WDyson}
&W(\bm{x}_{1},\bm{x}_{2})=v(\bm{x}_{1},\bm{x}_{2})\\
&+\iint v(\bm{x}_{1},\bm{x}_{3})P(\bm{x}_{3},\bm{x}_{4})W(\bm{x}_{4},\bm{x}_{2})d\bm{x}_{3}d\bm{x}_{4}
\end{aligned}\end{equation} 
where 
\begin{equation}\label{P1}
P(1,2)=-iG(1,2^{+})G(2,1^{+})=-iL_{0}(1,2,1^{+},2^{+})
\end{equation}
is the irreducible polarizability in the GW approximation. Both $P$ and $L_{0}$ are often simplified by introducing the RPA \cite{louie1986}, namely performing the substitution $G(1,2)\rightarrow G^{0}(1,2)$ in \eqref{P1} for both interacting Green's functions. Employing \eqref{G0}, the corresponding spectral representations are, relabelling $P$ as $P_{0}$ in the RPA,
\begin{equation}\begin{aligned}\label{P0}
&P_{0}(\bm{x}_{1},\bm{x}_{2};z)=\\
&\sum_{m,n}(f_{m}-f_{n})\frac{\psi_{m}^{*}(\bm{x}_{1})\psi_{n}(\bm{x}_{1})\psi_{m}(\bm{x}_{2})\psi_{n}^{*}(\bm{x}_{2})}{z-(\varepsilon_{n}-\varepsilon_{m})},
\end{aligned}\end{equation}
\begin{equation}\begin{aligned}
&L_{0}(\bm{x}_{1},\bm{x}_{2},\bm{x}_{3},\bm{x}_{4};z)=\\
&i\sum_{m,n}(f_{n}-f_{m})\frac{\psi_{m}(\bm{x}_{1})\psi_{n}(\bm{x}_{2})\psi_{n}^{*}(\bm{x}_{3})\psi_{m}^{*}(\bm{x}_{4})}{z-(\varepsilon_{m}-\varepsilon_{n})}
\end{aligned}\end{equation}
where $f_{n}$ is the Fermi distribution and $z\rightarrow\hbar\omega+i\eta^{+}\sign(f_{m}-f_{n})$ at $T=0$. Due to the neglect of dynamical screening effects, here we consider only $P_{0}(\bm{x}_{1},\bm{x}_{2};0)\equiv P_{0}(\bm{x}_{1},\bm{x}_{2})$, which as can be seen from \eqref{P0} is a real-valued quantity.

The static electron-hole BSE \eqref{BSEstatic} can now be written in tensor form by projecting in the orthonormal set of single-particle eigenfunctions $\set{\ket{\psi_{n}}}$, which yields \cite{martin2016,rubio2002}
\begin{equation}\label{BSEL}
L_{m,m';n,n'}(z)=i(f_{m'}-f_{n'})[(zI-H^{\text{e-h}})^{-1}]_{m,m';n,n'}
\end{equation}
where
\begin{equation}\label{He-h}
H^{\text{e-h}}_{m,m';n,n'}=(\varepsilon_{n}-\varepsilon_{m})\delta_{m,n'}\delta_{m',n}+i(f_{m}-f_{n})\Xi_{m,m';n,n'}
\end{equation}
In analogy with \eqref{Int2}, we have introduced the following shorthand convention for matrix elements of four-point functions,
\begin{equation}\begin{aligned}\label{Int4}
&F_{m,m';n,n'}\equiv \int\psi^{*}_{m}(\bm{r}_{1})\psi_{m'}^{*}(\bm{r}_{2})\cdot\\
&F(\bm{r}_{1},\bm{r}_{2},\bm{r}_{3},\bm{r}_{4})\psi_{n}(\bm{r}_{3})\psi_{n'}(\bm{r}_{4})d(\bm{r}_{1}\bm{r}_{2}\bm{r}_{3}\bm{r}_{4}),
\end{aligned}\end{equation}
which applied to the GW interaction kernel \eqref{kernel} yields
\begin{equation}\begin{aligned}\label{KernelMatrix}
&\Xi_{m,m';n,n'}=\\
&-i\left[\iint\psi^{*}_{m}(\bm{r})\psi_{n}(\bm{r})v(\bm{r},\bm{r}')\psi_{n'}(\bm{r}')\psi^{*}_{m'}(\bm{r}')d\bm{r}d\bm{r}'\right.\\
&\left.-\iint\psi^{*}_{m}(\bm{r})\psi_{n'}(\bm{r})W(\bm{r},\bm{r}')\psi_{n}(\bm{r}')\psi^{*}_{m'}(\bm{r}')d\bm{r}d\bm{r}'\right]
\end{aligned}\end{equation}

\subsection{The BSE in Matrix Form}\label{BSEMatSec}
The 4-point tensors in \eqref{BSEL}, \eqref{He-h} can be regarded as regular matrices upon grouping of the indices ($m,n$) and also of ($n',m'$), as suggested by the definition of the 4-point identity $I_{m,m';n,n'}=\delta_{m,n'}\delta_{m',n}$ (which stems from $F(\bm{r}_{1},\bm{r}_{2},\bm{r}_{3},\bm{r}_{4})=\delta(\bm{r}_{1}-\bm{r}_{4})\delta(\bm{r}_{2}-\bm{r}_{3})$ in \eqref{Int4}). Because of the Fermi factors in \eqref{BSEL} and \eqref{He-h}, $(m,n)$ consists of one occupied ($v$) and one unoccupied ($c$) state, and likewise for $(n',m')$. Therefore, \eqref{He-h} can be explicitly written in block matrix form, with the two indices $(m,n)$ and $(n',m')$ spanning $\set{(v,c)}\cup\set{(c,v)}$ (in that order), as
\begin{equation}\label{He-hMatrix}
H^{\text{e-h}}=\begin{pmatrix} H^{\text{res}} & H^{\text{coupl}} \\[0.2cm] -(H^{\text{coupl}})^{*} & -(H^{\text{res}})^{*} \end{pmatrix},
\end{equation}
\begin{equation}\label{Hres}
H^{\text{e-h}}_{v,c';c,v'}=H^{\text{res}}_{v,c';c,v'}=(\varepsilon_{c}-\varepsilon_{v})\delta_{c,c'}\delta_{v,v'}+i\Xi_{v,c';c,v'},
\end{equation}
\begin{equation}\label{Hcoupl}
H^{\text{e-h}}_{v,v';c,c'}=H^{\text{coupl}}_{v,c';c,v'}=i\Xi_{v,v';c,c'}
\end{equation}
where we have used that $i\Xi_{m,m';n,n'}=(i\Xi_{n,n';m,m'})^{*}$, noting that the static RPA screening is real-valued due to \eqref{WDyson} and \eqref{P0}. Also, noting that $i\Xi_{m,m';n,n'}=i\Xi_{m',m;n',n}$ due to the symmetry of the screening under $\bm{r}\leftrightarrow\bm{r}'$, one finds that $H^{\text{res}}=H^{\text{res}\dagger}$ and $H^{\text{coupl}}=H^{\text{coupl}\dagger*}$. From \eqref{He-hMatrix} it then follows that $H^{\text{e-h}}$ is pseudo-hermitian with $H^{\text{e-h}\dagger}=\begin{pmatrix}I&0\\0&-I\end{pmatrix}H^{\text{e-h}}\begin{pmatrix}I&0\\0&-I\end{pmatrix}$, hence its eigenvalues are still real. 

In analogy with the one-body Green's function, $H^{\text{e-h}}$ is identified with a two-particle (electron and hole) or BSE Hamiltonian, with spectrum 
\begin{equation}\label{ExcSpectrum}
\sum_{m',n'}H^{\text{e-h}}_{m,m';n,n'}A_{\chi}^{n',m'}=E_{\chi}A_{\chi}^{m,n}
\end{equation}
where $\chi$ labels the electron-hole bound states, with energies $E_{\chi}$ and wavefunctions $\Psi_{\chi}(\bm{r},\bm{r}')=\sum_{m,n}A_{\chi}^{m,n}\psi_{m}(\bm{r})\psi_{n}^{*}(\bm{r}')$. Finally, the causal $L$ can be built with the spectral representation  
\begin{equation*}
L^{R}_{m,m';n,n'}(\omega)=i\sum_{\chi,\chi'}(f_{m'}-f_{n'})\frac{A^{m,n}_{\chi}N_{\chi,\chi'}^{-1}A^{n',m'*}_{\chi'}}{\omega-E_{\chi}+i\eta^{+}}
\end{equation*}
where $N_{\chi,\chi'}=\sum_{m_{1},n_{1}}A^{m_{1},n_{1}*}_{\chi}A^{m_{1},n_{1}}_{\chi'}$ is the excitonic overlap matrix.

In a crystalline solid, the single-particle eigenstates are further labelled by the crystal momentum $\bm{k}$ ($m\rightarrow m\bm{k}$) in the Brillouin zone (BZ), which we assume to be discrete with $N_{\bm{k}}$ points by formally considering a supercell with periodic boundary conditions. Noting that $W(\bm{r},\bm{r}')=W(\bm{r}+\bm{R},\bm{r}'+\bm{R})$ for any lattice vector $\bm{R}$, it can be shown that $\Xi_{m\bm{k}_{1},m'\bm{k}_{2};n\bm{k}_{3},n'\bm{k}_{4}}\propto\delta_{\bm{k}_{1}+\bm{k}_{2},\bm{k}_{3}+\bm{k}_{4}}$ up to a reciprocal lattice vector. Then\footnote{Note that this parametrization of the crystal momenta spans exactly the non-vanishing $H^{\text{e-h}}$ entries according to the selection rule above, assuming that $\bm{k},\bm{k}',\bm{Q}$ share the same grid and that the latter is a periodic vector space, as e.g. in \eqref{MPgrid}.} $H^{\text{e-h}}_{m\bm{k},m'\bm{k}'+\bm{Q}';n\bm{k}+\bm{Q},n'\bm{k}'}\propto\delta_{\bm{Q},\bm{Q}'}$, whence $\bm{Q}\in\text{BZ}$ (up to a reciprocal lattice vector) is a good quantum number for the exciton states, in particular labelling the irreducible representation $\bm{R}\rightarrow\exp(-i\bm{QR})$ of the translation group acting on both space coordinates simultaneously. The single-particle basis can then be reordered to bring the full BSE Hamiltonian into block form as
\begin{equation}\label{fullHBSE}
H^{\text{e-h}}=\begin{pmatrix} \boxed{Q_{1}}& &0&\\ &\boxed{Q_{2}}& &\cdots \\0& &\boxed{Q_{3}}& \\ &\vdots & &\ddots \end{pmatrix},
\end{equation}
\begin{equation}\label{Qblock}
\boxed{Q_{i}}=\begin{pmatrix}
H^{\text{res}}(\bm{Q}_{i}) & 
H^{\text{coupl}}(\bm{Q}_{i}) \\[0.2cm] -H^{\text{coupl}}(\bm{Q}_{i})^{\dagger} & H^{\text{ares}}(\bm{Q}_{i})
\end{pmatrix}
\end{equation}
The entries of each block matrix in \eqref{Qblock} are defined by varying
$v,c,\bm{k},v',c',\bm{k}'$ in their respective domains, while $\bm{Q}_{i}$ is fixed. Therefore, employing \eqref{KernelMatrix} and \eqref{Hres} we can write 
\begin{equation}\begin{aligned}\label{HQres}
&H^{\text{res}}(\bm{Q})_{v\bm{k},c'\bm{k}';c\bm{k},v'\bm{k}'}\equiv
H^{\text{res}}_{v\bm{k},c'\bm{k}'+\bm{Q};c\bm{k}+\bm{Q},v'\bm{k}'}=\\[0.2cm]
&(\varepsilon_{c\bm{k}+\bm{Q}}-\varepsilon_{v\bm{k}})\delta_{c,c'}\delta_{v,v'}\delta_{\bm{k},\bm{k}'}\\[0.2cm]
&-D_{v\bm{k},c\bm{k}+\bm{Q}}^{v'\bm{k}',c'\bm{k}'+\bm{Q}}+X_{v\bm{k},v'\bm{k}'}^{c\bm{k}+\bm{Q},c'\bm{k}'+\bm{Q}}, 
\end{aligned}\end{equation}
with the direct and exchange terms (respectively)
\begin{equation}\begin{aligned}\label{ResonantD}
&D_{v\bm{k},c\bm{k}+\bm{Q}}^{v'\bm{k}',c'\bm{k}'+\bm{Q}}=\\
&\iint\psi^{*}_{v\bm{k}}(\bm{r})\psi_{v'\bm{k}'}(\bm{r})W(\bm{r},\bm{r}')\psi_{c\bm{k}+\bm{Q}}(\bm{r}')\psi^{*}_{c'\bm{k}'+\bm{Q}}(\bm{r}')d\bm{r}d\bm{r}', 
\end{aligned}\end{equation}
\begin{equation}\begin{aligned}\label{ResonantX}
&X_{v\bm{k},v'\bm{k}'}^{c\bm{k}+\bm{Q},c'\bm{k}'+\bm{Q}}=\\
&\iint\psi^{*}_{v\bm{k}}(\bm{r})\psi_{c\bm{k}+\bm{Q}}(\bm{r})v(\bm{r},\bm{r}')\psi_{v'\bm{k}'}(\bm{r}')\psi^{*}_{c'\bm{k}'+\bm{Q}}(\bm{r}')d\bm{r}d\bm{r}'
\end{aligned}\end{equation}
In insulators or semiconductors not of narrow-gap, the assumption that the interacting ground state remains the Fermi sea is often reasonable. Then the coupling part $H^{\text{coupl}}$ in \eqref{He-hMatrix} (or \eqref{Qblock}) can be neglected and the BSE Hamiltonian $H^{\text{e-h}}$ (or $\boxed{Q}$) can be identified with the hermitian resonant part $H^{\text{res}}$ (or $H^{\text{res}}(\bm{Q})$ \eqref{HQres}, respectively). This is the widely employed Tamm-Dancoff approximation (TDA), and in particular from \eqref{ExcSpectrum} it yields the exciton wavefunctions in the form
\begin{equation}\label{WavefunctionTDA}
\Psi_{\chi\bm{Q}}(\bm{r}_{h},\bm{r}_{e})=\sum_{v,c,\bm{k}}A_{\chi}^{v\bm{k},c\bm{k}+\bm{Q}}
\psi_{v\bm{k}}(\bm{r}_{h}) \psi^{*}_{c\bm{k}+\bm{Q}}(\bm{r}_{e})
\end{equation} 
In Appendix \ref{TDAApp} we complete the discussion on the BSE Hamiltonian without the TDA, which may yield significant corrections in the finite $\bm{Q}$ case \cite{Kresse2015}, although we adhere to the TDA for the remaining of this work.

For spinless electrons, namely in non-magnetic ground states with negligible spin-orbit coupling (SOC) and in the absence of external magnetic fields, it is convenient to rearrange the $\set{(v,c)}$ basis for the exciton states originally spanning all spin bands as $\set{(v\uparrow c\uparrow),(v\uparrow c\downarrow),(v\downarrow c\uparrow),(v\downarrow c\downarrow)}$ simultaneously for all $\bm{k}$ and for any $\bm{Q}$, such that $v,c$ now span only bands with the corresponding spin projection. From \eqref{kernel} it then follows that the BSE Hamiltonian can be block-diagonalized into two sections with $1/4$ of the original dimension: the triplet
\begin{equation}\begin{aligned}\label{Htriplet}
&H^{\text{e-h}}_{\text{triplet}}(\bm{Q})_{v\bm{k},c'\bm{k}';c\bm{k},v'\bm{k}'}=
(\varepsilon_{c\bm{k}+\bm{Q}}-\varepsilon_{v\bm{k}})\delta_{c,c'}\delta_{v,v'}\delta_{\bm{k},\bm{k}'}\\[0.2cm]
&-D_{v\bm{k},c\bm{k}+\bm{Q}}^{v'\bm{k}',c'\bm{k}'+\bm{Q}}
\end{aligned}\end{equation}
at the subspaces $\{(v\uparrow c\downarrow)\},\{(v\downarrow c\uparrow)\},\{[(v\uparrow c\uparrow)-(v\downarrow c\downarrow)]/\sqrt{2}\}$, and the singlet
\begin{equation}\begin{aligned}\label{Hsinglet}
&H^{\text{e-h}}_{\text{singlet}}(\bm{Q})_{v\bm{k},c'\bm{k}';c\bm{k},v'\bm{k}'}=
(\varepsilon_{c\bm{k}+\bm{Q}}-\varepsilon_{v\bm{k}})\delta_{c,c'}\delta_{v,v'}\delta_{\bm{k},\bm{k}'}\\[0.2cm]
&-D_{v\bm{k},c\bm{k}+\bm{Q}}^{v'\bm{k}',c'\bm{k}'+\bm{Q}}
+2X_{v\bm{k},v'\bm{k}'}^{c\bm{k}+\bm{Q},c'\bm{k}'+\bm{Q}}
\end{aligned}\end{equation} 
at the subspace $\{[(v\uparrow c\uparrow)+(v\downarrow c\downarrow)]/\sqrt{2}\}$. In this work we consider only the spinless scenario, hence solving the triplet and singlet configurations separately.

Note that \eqref{HQres} depends on the gauge choice of the single-particle eigenfunctions $\psi_{n\bm{k}}$, but only through a unitary transformation that does not alter the excitonic energies $E_{\chi}$. In order to fix a gauge for the exciton eigenfunctions, we impose the condition $\sum_{\mu}c^{\bm{k}}_{\mu n}\in\mathbb{R}$ $\forall n$ $\forall\bm{k}$ on the single-particle eigenvectors, see \eqref{EigenSP}.

In order to actually solve the BSE we adopt a homogeneous Monkhorst-Pack grid \cite{MonkPack} of the primitive cell in reciprocal space as
\begin{equation}\label{MPgrid}
\bm{k}\in\set{\sum_{i=1}^{\text{dim}}\frac{\alpha_{i}}{N_{i}}\bm{G}_{i}\;\vert\;\alpha_{i}=0,\dots,N_{i}-1}
\end{equation}
where $\bm{G}_{i}$ are the reciprocal basis vectors. This choice makes the addition or subtraction of any two points in the grid also be a point in the grid, up to a reciprocal lattice vector.

\section{Gaussian Basis \& Density Fitting}\label{GaussianRIsec}
\subsection{Formulation of the BSE with the RI Approximation}\label{BSEgauss}
We expand the single-particle states in a basis of local functions
\begin{equation}\begin{aligned}\label{EigenSP}
\psi_{n\bm{k}}(\bm{r})&=\frac{1}{\sqrt{N_{\bm{k}}}}\sum_{\mu}c_{\mu n}^{\bm{k}}\sum_{\bm{R}}e^{i\bm{kR}}\varphi_{\mu}^{\bm{R}}(\bm{r})\\
&\equiv
\frac{1}{\sqrt{N_{\bm{k}}}}\sum_{\mu}c_{\mu n}^{\bm{k}}\varphi_{\mu\bm{k}}(\bm{r})
\end{aligned}\end{equation}
where $\varphi_{\mu}^{\bm{R}}(\bm{r})\equiv\varphi_{\mu}(\bm{r}-\bm{R})$ is centered in the unit cell $\bm{R}$ \footnote{The supercell is constructed around the reference cell at $\bm{R}=0$.} and $N_{\bm{k}}=\prod_{i}N_{i}$, see \eqref{MPgrid}. In our case, the local functions $\varphi_{\mu}:\mathbb{R}^{3}\rightarrow\mathbb{R}$ are GTF \cite{dovesicrystal23} and the coefficients $c_{\mu n}^{\bm{k}}$ are determined self-consistently in a density-functional theory (DFT) framework, possibly hybrid (HF/DFT) with a fraction of Fock exchange. The eigenstates \eqref{EigenSP} are normalized as $\bra{\psi_{n\bm{k}}}\ket{\psi_{n'\bm{k}'}}=\delta_{n,n'}\delta_{\bm{k},\bm{k}'}$ in the supercell. %\footnote{The resulting Gaussian integrals are, as usual, performed over the whole space.}. 
Inserting \eqref{EigenSP} into \eqref{ResonantD} and employing the lattice periodicity it follows that 
\begin{equation*}\begin{aligned}
&D_{v\bm{k},c\bm{k}+\bm{Q}}^{v'\bm{k}',c'\bm{k}'+\bm{Q}}=\frac{1}{N_{\bm{k}}}\sum_{\bm{R},\bm{R}',\bm{R}''}e^{i\bm{kR''}}e^{i(\bm{k}'+\bm{Q})\bm{R}'}e^{-i\bm{k}'\bm{R}}\cdot\\[0.2cm]
&\sum_{\mu,\mu',\nu,\nu'}c^{\bm{k}+\bm{Q}}_{\mu c}\:c^{\bm{k}'+\bm{Q}*}_{\nu,c'}\:c^{\bm{k}'}_{\mu'v'}\:c^{\bm{k}*}_{\nu'v}\cdot\\[0.2cm]
&\iint\varphi^{\bm{0}}_{\mu}(\bm{r})\varphi^{\bm{R}'}_{\nu}(\bm{r})W(\bm{r},\bm{r}')\varphi^{\bm{R}}_{\mu'}(\bm{r}')\varphi^{\bm{R}''}_{\nu'}(\bm{r}')d\bm{r}d\bm{r}'
\end{aligned}\end{equation*}
and an analogous expression is found for the exchange \eqref{ResonantX}. In order to reduce the a priori $\mathcal{O}(N^{4}_{\mu}N^{3}_{\bm{R}}N^{2}_{\bm{k}}N^{2}_{v}N^{2}_{c})$ scaling of the whole direct and exchange contributios to the BSE Hamiltonian\footnote{For the direct term it is actually more severe due to the additional series introduced by the Dyson equation \eqref{WDyson}.} we employ the Gaussian density fitting or RI approximation, by which the charge density (or equivalently the product of crystalline orbitals) is expanded in a larger, auxiliary GTF basis $\set{\phi_{P}}$ as \cite{RI2005,Burow2017,Patterson2020}
\begin{equation}\label{RI}
\varphi^{*}_{\mu\bm{k}}(\bm{r})\varphi_{\mu'\bm{k}'}(\bm{r})\approx\sum_{P}B^{\mu\bm{k},\mu'\bm{k}'}_{P}\phi_{P\bm{k}'-\bm{k}}(\bm{r})
\end{equation}
%Note that an exact expansion with the Clebsch-Gordan coefficients could not be possible due to the multiple atomic centerings. 
with $\phi_{P\bm{k}}(\bm{r})$ defined as in \eqref{EigenSP}. The $B$ coefficients are determined by minimizing the difference between both sides of \eqref{RI} with respect to a certain norm defined by the (translationally invariant) metric $m:\mathbb{R}^{3}\rightarrow\mathbb{R}^{+}\cup\set{\infty
}$ ($m(\bm{r},\bm{r}')=m(\bm{r}-\bm{r}')$), yielding
\begin{equation}\label{RIcoeffs}
B^{\mu\bm{k},\mu'\bm{k}'}_{P}[m]=\sum_{P'}M_{P,P'\bm{k}'-\bm{k}}^{-1}[m]L^{\mu\bm{k},\mu'\bm{k}'}_{P'}[m],
\end{equation}
\begin{equation}\label{Mmatrix}
M_{P,P'\bm{k}}\equiv\frac{1}{N_{\bm{k}}}\iint\phi_{P-\bm{k}}(\bm{r})m(\bm{r}-\bm{r}')\phi_{P'\bm{k}}(\bm{r}')d\bm{r}d\bm{r}',
\end{equation}
\begin{equation}\begin{aligned}\label{Lmatrix}
&L^{\mu\bm{k},\mu'\bm{k}'}_{P}\equiv\\[0.2cm]
&\frac{1}{N_{\bm{k}}}\iint\phi_{P\bm{k}-\bm{k}'}(\bm{r})m(\bm{r}-\bm{r}')\varphi^{*}_{\mu\bm{k}}(\bm{r}')\varphi_{\mu'\bm{k}'}(\bm{r}')d\bm{r}d\bm{r}'
\end{aligned}\end{equation}
The functional dependence on the metric $m$ has been, and will omitted for simplicity. The dominant term in the scaling of the direct and exchange full matrices is now $\mathcal{O}(N^{2}_{\mu}N_{P}N^{2}_{\bm{R}}N^{2}_{\bm{k}}N^{2}_{v}N^{2}_{c})$ (disregarding still the effect of the Dyson series for $W$) as given by the 3-center integrals in \eqref{Lmatrix}, which is a substantial advantage with respect to the original formulation even for small unit cells. In order to perform the inversion of the matrix \eqref{Mmatrix} in the auxiliary basis, whose augmented size favours the appearance of numerical linear dependencies, a small regularization parameter $\alpha\in\mathbb{R}$ can be introduced as $M^{-1}_{P,P'\bm{k}}\to M^{-1,\alpha}_{P,P'\bm{k}}\equiv[(M_{\bm{k}}+\alpha I)^{-1}]_{P,P'}$ \cite{CP2KGW}. It is readily seen that these quantities obey the following properties: $M_{P,P'\bm{k}}=M_{P',P-\bm{k}}=M_{P,P'-\bm{k}}^{*}$, in particular \eqref{Mmatrix} is a hermitian matrix (and also positive definite, albeit in bases with quasi-linear-depdendencies a finite $\alpha$ is needed to realise this in practice), and $L^{\mu\bm{k},\mu'\bm{k}'}_{P}=(L^{\mu'\bm{k}',\mu\bm{k}}_{P})^{*}$.

The 4-center Coulomb integrals appearing in the projected kernel, first term of \eqref{KernelMatrix}, can then be expanded as
\begin{equation}\begin{aligned}\label{CoulombInt}
&(m\bm{k},n\bm{k}+\bm{Q}\vert n'\bm{k}',m'\bm{k}'+\bm{Q})\equiv \\[0.2cm]
&\iint\psi^{*}_{m\bm{k}}(\bm{r})\psi_{n\bm{k}+\bm{Q}}(\bm{r})v(\bm{r},\bm{r}')\psi_{n'\bm{k}'}(\bm{r}')\psi^{*}_{m'\bm{k}'+\bm{Q}}(\bm{r}')d\bm{r}d\bm{r}'\\[0.2cm]
%&=\frac{1}{N_{\bm{k}}^{2}}\sum_{\mu,\mu',\mu'',\mu'''}c^{\bm{k}*}_{\mu m}c^{\bm{k}+\bm{Q}}_{\mu' n}c^{\bm{k}'}_{\mu'' n'}c^{\bm{k}'+\bm{Q}*}_{\mu''' m'}\sum_{R}v^{\mu\bm{k},\mu'\bm{k}+\bm{Q}}_{R}v^{\mu'''\bm{k}'+\bm{Q},\mu''\bm{k}'}_{R}=\\
&=\sum_{R}v_{R}^{m\bm{k},n\bm{k}+\bm{Q}}v_{R}^{m'\bm{k}'+\bm{Q},n'\bm{k}'}
\end{aligned}\end{equation}
where 
\begin{equation}\label{vnmatrix}
v_{R}^{m\bm{k}_{1},m'\bm{k}_{2}}=\frac{1}{\sqrt{N_{\bm{k}}}}\sum_{\mu,\mu'}c^{\bm{k}_{1}*}_{\mu m}c^{\bm{k}_{2}}_{\mu'm'}v^{\mu\bm{k}_{1},\mu'\bm{k}_{2}}_{R},
\end{equation}
\begin{equation}\label{vmatrix}
v_{R}^{\mu\bm{k}_{1},\mu'\bm{k}_{2}}\equiv\sum_{P,P'}J^{1/2}_{R,P\bm{k}_{2}-\bm{k}_{1}}M^{-1}_{P,P'\bm{k}_{2}-\bm{k}_{1}}L^{\mu\bm{k}_{1},\mu'\bm{k}_{2}}_{P'},
\end{equation}
\begin{equation}\label{Jmatrix}
J_{R,P\bm{k}}\equiv\frac{1}{N_{\bm{k}}}\iint\phi_{R-\bm{k}}(\bm{r})v(\bm{r},\bm{r}')\phi_{P\bm{k}}(\bm{r}')d\bm{r}d\bm{r}'
\end{equation}
and $J_{R,P\bm{k}}=M_{R,P\bm{k}}[v]=\sum_{Q}J^{1/2}_{R,Q\bm{k}}J^{1/2}_{Q,P\bm{k}}$, which can also be realised as a Cholesky decomposition (possibly aided by the regularization parameter $\alpha$). From the previous properties it follows that 
\begin{equation}\label{vsymm}
v^{m\bm{k}_{1},m'\bm{k}_{2}}_{R}=\left(v^{m'\bm{k}_{2},m\bm{k}_{1}}_{R}\right)^{*}%=v_{R}^{m'-\bm{k}_{2},m-\bm{k}_{1}}
\end{equation}
In particular, the exchange terms in \eqref{ResonantX} read
\begin{equation}\label{ExchangeV}
X_{v\bm{k},v'\bm{k}'}^{c\bm{k}+\bm{Q},c'\bm{k}'+\bm{Q}}=\sum_{R}v_{R}^{v\bm{k},c\bm{k}+\bm{Q}}\left(v_{R}^{v'\bm{k}',c'\bm{k}'+\bm{Q}}\right)^{*}
\end{equation}

On the other hand, the integrals involving the screened interaction in the projected kernel, second term of \eqref{KernelMatrix}, can be expanded employing Hedin's equation \eqref{WDyson} with the RPA irreducible polarizability \eqref{P0} as 
\begin{equation}\begin{aligned}\label{ScreenedInt}
&\iint\psi^{*}_{m\bm{k}}(\bm{r})\psi_{n'\bm{k}'}(\bm{r})W(\bm{r},\bm{r}')\psi_{n\bm{k}+\bm{Q}}(\bm{r}')\psi^{*}_{m'\bm{k}'+\bm{Q}}(\bm{r}')d\bm{r}d\bm{r}'\\[0.2cm]
&=\sum_{R,R'}v_{R}^{m\bm{k},n'\bm{k}'}\left[(I-\Pi_{\bm{k}-\bm{k}'})^{-1}\right]_{R,R'}v_{R'}^{m'\bm{k}'+\bm{Q},n\bm{k}+\bm{Q}}
\end{aligned}\end{equation}
where we have introduced the irreducible polarizability matrix $\Pi_{R,R'\bm{k}}$ (in the auxiliary basis), which in the absence of SOC or in the presence of inversion and/or time-reversal symmetry takes the form (see Appendix \ref{ScreeningApp} for the derivation and the general case) 
\begin{equation}\begin{aligned}\label{Pimatrix}
\Pi_{R,R'\bm{k}}=\frac{2N_{\bm{k}}}{N_{\tilde{\bm{k}}}}\sum_{\tilde{\bm{k}}}\sum_{\tilde{v},\tilde{c}}\frac{v_{R}^{\tilde{v}\tilde{\bm{k}},\tilde{c}\tilde{\bm{k}}+\bm{k}}\left(v_{R'}^{\tilde{v}\tilde{\bm{k}},\tilde{c}\tilde{\bm{k}}+\bm{k}}\right)^{*}}{\varepsilon_{\tilde{v}\tilde{\bm{k}}}-\varepsilon_{\tilde{c}\tilde{\bm{k}}+\bm{k}}}
\end{aligned}\end{equation}
Note that the $N_{\bm{k}}$ dependence is cancelled out with the prefactors in \eqref{vnmatrix}, and that a global $2$ factor arises in the spinless case when reducing to the triplet \eqref{Htriplet} and singlet \eqref{Hsinglet} subspaces since $\tilde{v},\tilde{c}$ originally span both spin states in each Kramer's pairs. Furthermore, $I-\Pi_{\bm{k}}$ can be indentified with the static dielectric matrix $\epsilon(\bm{k})\equiv\epsilon(\bm{k},\omega=0)$ and analogously for the inverse, which in the present case are represented in the auxiliary GTF basis instead of the usual one formed by reciprocal lattice vectors $\bm{G}$. The direct terms in \eqref{ResonantD} can therefore be expressed as
\begin{equation}\begin{aligned}\label{DirectV}
D_{v\bm{k},c\bm{k}+\bm{Q}}^{v'\bm{k}',c'\bm{k}'+\bm{Q}}=\sum_{R,R'}v_{R}^{v\bm{k},v'\bm{k}'}\epsilon^{-1}_{R,R'}(\bm{k}-\bm{k}')\left(v_{R'}^{c\bm{k}+\bm{Q},c'\bm{k}'+\bm{Q}}\right)^{*}
\end{aligned}\end{equation}

\subsection{Gaussian Integrals}\label{GaussianIntSec}
As shown above, the electron-hole Hamiltonian \eqref{He-hMatrix} can be computed from a HF/DFT calculation employing the GTF basis $\set{\varphi_{\mu}}$ (for the single-particle eigenstates) by computing the auxiliary tensors $v^{m\bm{k}_{1},m'\bm{k}_{2}}_{R}$ \eqref{vmatrix} with the aid of an auxiliary GTF basis $\set{\phi_{P}}$ (for the density). This quantity is built from 3 different types of crystalline integrals, namely the 2-center \eqref{Mmatrix}, \eqref{Jmatrix} and the 3-center \eqref{Lmatrix}, which can in turn be expanded in Gaussian integrals as\footnote{Note that the $\bm{R}-$summations should technically be converged within the supercell if the latter is to properly represent the ideal crystal, but as usual we sum as many vectors as needed.}
\begin{equation}\begin{aligned}\label{JmatrixG}
&J_{P,P'\bm{k}}=\sum_{\bm{R}}e^{i\bm{kR}}J_{P,P'\bm{R}}\equiv\\
&\sum_{\bm{R}}e^{i\bm{kR}}\iint_{\mathbb{R}^{3}}\phi_{P}^{\bm{0}}(\bm{r})v(\bm{r},\bm{r}')\phi_{P'}^{\bm{R}}(\bm{r}')d\bm{r}d\bm{r}' ,
\end{aligned}\end{equation}
\begin{equation}\begin{aligned}\label{MmatrixG}
&M_{P,P'\bm{k}}=\sum_{\bm{R}}e^{i\bm{kR}}M_{P,P'\bm{R}}\equiv\\
&\sum_{\bm{R}}e^{i\bm{kR}}\iint_{\mathbb{R}^{3}}\phi_{P}^{\bm{0}}(\bm{r})m(\bm{r}-\bm{r}')\phi_{P'}^{\bm{R}}(\bm{r}')d\bm{r}d\bm{r}' ,
\end{aligned}\end{equation}
\begin{equation}\begin{aligned}\label{LmatrixG}
&L^{\mu\bm{k},\mu'\bm{k}'}_{P}=\sum_{\bm{R},\bm{R}'}e^{i(\bm{k}'\bm{R}'-\bm{kR})}L^{\mu\bm{R},\mu'\bm{R}'}_{P}\equiv\sum_{\bm{R},\bm{R}'}e^{i(\bm{k}'\bm{R}'-\bm{kR})}\\
&\cdot\iint_{\mathbb{R}^{3}}\phi_{P}^{\bm{0}}(\bm{r})m(\bm{r}-\bm{r}')\varphi_{\mu}^{\bm{R}}(\bm{r}')\varphi_{\mu'}^{\bm{R}'}(\bm{r}')d\bm{r}d\bm{r}'
\end{aligned}\end{equation}

In order to enable the analytical evaluation of these integrals, we introduce the Hermite Gaussians \cite{Helgaker} with exponent $p$ and center $\mathbf{P}$
\begin{equation}\label{HermiteGaussian}
\Lambda_{t,u,v}(\bm{r},p,\mathbf{P})=\frac{\partial^{t}}{\partial\text{P}_{x}^{t}}\frac{\partial^{u}}{\partial\text{P}_{y}^{u}}\frac{\partial^{v}}{\partial\text{P}_{z}^{v}}e^{-p\:\abs{\bm{r}-\mathbf{P}}^{2}}
\end{equation}
as well as the Hermite Coulomb integrals \cite{Helgaker}
\begin{equation}\label{HermiteCoulomb}
R_{t,u,v}(p,\mathbf{P})=\frac{\partial^{t}}{\partial\text{P}_{x}^{t}}\frac{\partial^{u}}{\partial\text{P}_{y}^{u}}\frac{\partial^{v}}{\partial\text{P}_{z}^{v}}F_{0}(p\abs{\mathbf{P}}^{2})
\end{equation}
where $F_{0}(x)=\frac{\sqrt{\pi}}{2}\frac{\erf(\sqrt{x})}{\sqrt{x}}$ is the zeroth order Boys function. In Appendix \ref{GaussianApp} we explicitly show the expansion of GTF integrals into Hermite Gaussian integrals for general 2 and 3-center cases. The main advantage of working with Hermite Gaussians over regular GTF is that the integrals involving arbitrarily high multipoles (angular momenta) can be obtained from the simplest, pure $s-$type case by differentiating with respect to a given center as in \eqref{HermiteGaussian}. A comment on the evaluation of \eqref{HermiteCoulomb} in practice is also given at the end of Appendix \ref{GaussianApp}.

\subsection{Conditional Convergence \& Ewald Potential}\label{EwaldTheorySec}
The 2-center Coulomb integrals \eqref{JmatrixG} are always present in the current RI-based formulation of the BSE irrespective of the metric, however, they cannot be evaluated directly in real space. Indeed, \eqref{JmatrixG} represents the electrostatic interaction energy between the charge distributions $\phi_{P}^{\bm{0}}$ (local) and $\sum_{\bm{R}}e^{i\bm{kR}}\phi_{P'}^{\bm{R}}$ (extended), which is guaranteed to be absolutely convergent in the lattice sum only if \cite{Piela1982} 
\begin{equation}\label{ConvCond}
k_{\text{min}}+l_{\text{min}}\geq \text{dim}
\end{equation}
where $k_{\text{min}}$, $l_{\text{min}}$ are the lowest non-vanishing multipole moments of $\phi_{P}$, $\phi_{P'}$ respectively (the unit cell centering is irrelevant), and $\text{dim}$ is the lattice dimensionality. In particular, the series in \eqref{JmatrixG} is apparently divergent for two $s-$type GTF (which are charged) in any periodic system, conditionally convergent for one $s$ and one $p-$type GTF in 2D, and also for two $p-$types or one $s$ and one $d-$type GTF in 3D. 

The former, in principle divergent case is automatically solved for $\bm{k}$ points that are fractions of reciprocal lattice vectors\footnote{Generalizations to arbitrary $\bm{k}$ also exist \cite{Varga2018}, but they are not considered here.} by performing a partition in the smallest commensurate supercell. In particular, for the whole Monkhorst-Pack grid \eqref{MPgrid} (excluding $\bm{k}=0$, see below) it suffices to take
\begin{equation}\label{seriesSUP}
\sum_{\bm{R}}f(\bm{R})=\sum_{\bm{R}^{N_{i}}}\:\sum_{\bm{R}'}^{\text{SUP}(\set{N_{i}})}\:f(\bm{R}^{N_{i}}+\bm{R}')
\end{equation}
\begin{equation}\label{SUP}
\begin{aligned}
&\text{SUP}(\set{N_{i}})=\left\{\sum_{i=1}^{\text{dim}}m_{i}\bm{R}_{i}\;\vert\; \bm{R}_{i}\bm{G}_{j}=2\pi\delta_{ij},\right. \\
&\left.m_{i}=\left\lceil-\frac{N_{i}}{2}\right\rceil,\dots,\left\lceil\frac{N_{i}}{2}\right\rceil-1 \right\},
\end{aligned}
\end{equation}
where the truncation is done only in the outer summation in $\bm{R}^{N_{i}}$, which spans the integer linear combinations of $N_{i}\bm{R}_{i}$ and thus fixes a supercell. The inner summation within each supercell is performed to completion, hence for $\bm{k}\neq\bm{0}$
\begin{equation*}
\sum_{\bm{R}}e^{i\bm{kR}}\int_{\mathbb{R}^{3}}\phi_{P'}^{\bm{R}}(\bm{r})d\bm{r}=C_{P'}\sum_{\bm{R}^{N_{i}}}e^{i\bm{kR}^{N_{i}}}\sum_{\bm{R}'}^{\text{SUP}(\set{N_{i}})}e^{i\bm{kR}}=0
\end{equation*}
where $C_{P'}=\int\phi_{P'}^{\bm{R}}(\bm{r})d\bm{r}$ is the charge of the GTF $\phi_{P'}$ (non-vanishing if and only if $\phi_{P'}$ is of $s-$type), and we have noted that the sum over the $N_{i}-$th (or any divisor of it) roots of unity is null. Employing this prescription, the extended charge distribution is effectively chargeless (it is when restricted to the orbitals inside each supercell \eqref{SUP}) and $k_{\text{min}}+l_{\text{min}}\geq1$ in \eqref{ConvCond}. Therefore, the 1D case can be readily computed with the standard Coulomb potential summing over a desired number of indivisible supercells \eqref{SUP}. For $\bm{k}=\bm{0}$ a supercell cannot be found such that the density given by the enclosed GTF is chargeless, as the lattice sum no longer translates into a sum over the roots of unity. This case is addressed separately in Section \ref{SecDivergence}. 

In the case of conditional convergence inevitably found in 2D and 3D, the result of the lattice summation depends crucially on the definition of the unit cell or a spreading transformation \cite{Harris1975,Piela1982} for the extended charge density that removes the relevant multipole moments\footnote{And in 3D, also the trace of the spherical second moment.}. Alternatively, we employ the Ewald potential function $A(\bm{r}-\bm{r}')$, defined in \eqref{Ewald3D} for 3D and in \eqref{Ewald2D} for 2D, to automatize the treatment of the conditionally convergent series in any periodic system. For a chargeless lattice density $\rho^{\text{latt}}(\bm{r})=\sum_{\bm{R}}\rho^{\bm{R}}(\bm{r})$, where $\rho^{\bm{R}}$ is an exponentially localized neutral density centered at cell $\bm{R}$, the electrostatic and Ewald potentials are coincident after a suitable spreading transformation $\rho^{\text{latt}}\to\tilde{\rho}^{\text{latt}}$ has been applied \cite{Saunders1992}, namely
\begin{equation}\label{EwaldSubs}
\int_{\mathbb{R}^{3}}\tilde{\rho}^{\text{latt}}(\bm{r}')v(\bm{r}-\bm{r}')d\bm{r}'=\int_{\mathbb{R}^{3}}\rho^{\bm{0}}(\bm{r}')A(\bm{r}-\bm{r}')d\bm{r}'
\end{equation}
The lattice summation is now implicitly performed inside $A(\bm{r}-\bm{r}')$, where it is absolutely convergent, see Section \ref{Ewald3Dsec} or \ref{Ewald2Dsec}. Note that the Ewald substitution \eqref{EwaldSubs} relies on the theoretical existence of a spreading transformation for the lattice density (which is guaranteed due to charge neutrality \cite{Harris1975,Saunders1992}), but in practice it does not require such transformation to be found. Therefore, \eqref{JmatrixG} can be written with the aid of \eqref{seriesSUP} as
\begin{equation*}\begin{aligned}
&J_{P,P'\bm{k}}=\sum_{\bm{R}^{N_{i}}}\:\sum_{\bm{R}'}^{\text{SUP}(\set{N_{i}})}e^{i\bm{kR}^{N_{i}}}\iint\phi^{\bm{0}}_{P}(\bm{r})v(\bm{r},\bm{r}')e^{i\bm{kR}'}\cdot\\[0.2cm]
&\cdot\tilde{\phi}^{\bm{R}^{N_{i}}+\bm{R}'}_{P'}(\bm{r}')d\bm{r}d\bm{r}'=
\iint\phi^{\bm{0}}_{P}v(\bm{r},\bm{r}')\tilde{\rho}^{\text{latt}}_{P'\bm{k}}(\bm{r}')d\bm{r}d\bm{r}'= \\[0.2cm]
&\iint\phi^{\bm{0}}_{P}(\bm{r})A(\bm{r}-\bm{r}')\rho^{\bm{0}}_{P'\bm{k}}(\bm{r}')d\bm{r}d\bm{r}'
\end{aligned}\end{equation*}
where for the second equality we have used that $e^{i\bm{qR}^{N_{i}}}=1$ by construction, and we have introduced the lattice density with supercell periodicity
\begin{equation}
\rho^{\text{latt}}_{P\bm{k}}(\bm{r})=\sum_{\bm{R}^{N_{i}}}\rho_{P\bm{k}}^{\bm{R}^{N_{i}}}(\bm{r}) \;,
\end{equation}
\begin{equation}
\rho^{\bm{R}^{N_{i}}}_{P\bm{k}}(\bm{r})=\sum_{\bm{R}'}^{\text{SUP}(\set{N_{i}})}e^{i\bm{kR}'}\phi_{P}^{\bm{R}'+\bm{R}^{N_{i}}}(\bm{r})
\end{equation}
As stated above, $\rho^{\bm{R}^{N_{i}}}_{P\bm{k}}$ is indeed a neutral and exponentially localized density centered at supercell $\bm{R}^{N_{i}}$. Finally, the 2-center Coulomb integrals \eqref{JmatrixG} can be computed at any $\bm{k}\neq\bm{0}$ as
\begin{equation}\label{JmatrixGEwald}
J_{P,P'\bm{k}}=\sum_{\bm{R}'}^{\text{SUP}(\set{N_{i}})}e^{i\bm{kR}'}\iint_{\mathbb{R}^{3}}\phi^{\bm{0}}_{P}(\bm{r})A(\bm{r}-\bm{r}')\phi^{\bm{R}'}_{P'}(\bm{r}')d\bm{r}d\bm{r}'
\end{equation}
That is, one pre-computes the Ewald potential function matrix elements in the auxiliary basis with one function restricted to the original unit cell and the other spanning exactly the supercell \eqref{SUP} defined by the BSE $\bm{k}-$grid \eqref{MPgrid}. 

\subsection{Ewald Potential Function in 3D}\label{Ewald3Dsec}
The Ewald method for lattice sums has been known for a long time \cite{Ewald1921,Ziman1964}. It is based on the separation of the Coulomb potential employing the error function
\begin{equation*}
\frac{1}{r}=\frac{\erfc(\sqrt{\gamma}r)}{r}+\frac{\erf(\sqrt{\gamma}r)}{r}
\end{equation*}
with arbitrary screening parameter $\gamma\geq0$. The first, short-range term is treated in real-space while the second, long-range term is treated in reciprocal space exploiting the periodicity of the lattice, ultimately yielding the Ewald potential function \cite{Saunders1992}
\begin{equation}\label{Ewald3D}
\begin{aligned}
&A(\bm{s})=-\frac{\pi}{\gamma V}+\\
&\sum_{\bm{R}}^{'}\frac{\erfc(\sqrt{\gamma}\abs{\bm{s}-\bm{R}})}{\abs{\bm{s}-\bm{R}}}+\frac{4\pi}{V}\sum_{\bm{G}\neq\bm{0}}\frac{1}{\abs{\bm{G}}^{2}}e^{-\frac{\abs{\bm{G}}^{2}}{4\gamma}+i\bm{Gs}} 
\end{aligned}
\end{equation}
where $V$ is the unit cell volume and the prime in the direct lattice sum indicates that any possible term with $\bm{R}=\bm{s}$ is excluded. 

The integral of \eqref{Ewald3D} with two Hermite Gaussians \eqref{HermiteGaussian} was found in \cite{Saunders1992}. Here we adapt it to our context in \eqref{JmatrixGEwald},  for which the unit of periodicity that defines both the $\bm{R}$ and $\bm{G}$ sums in \eqref{Ewald3D} is the supercell \eqref{SUP}, whence 
\begin{equation}\label{Ewald3DInt}
\begin{aligned}
&\iint\Lambda_{t,u,v}(\bm{r},\alpha,\bm{r}_{1})A(\bm{r}-\bm{r}')\Lambda_{t',u',v'}(\bm{r}',\beta,\bm{r}_{2})d\bm{r}d\bm{r}'=\\[0.2cm]
&\frac{2\pi^{5/2}(-1)^{t'+u'+v'}}{(\alpha\beta)^{3/2}}\left\{\sum_{\bm{R}^{N_{i}}}\left[\sqrt{\mu}R_{t+t',u+u',v+v'}(\mu,\bm{r}_{12}-\bm{R}^{N_{i}})  \right.  \right. \\[0.2cm]
&\left.\left.-\sqrt{\tilde{\gamma}}R_{t+t',u+u',v+v'}(\tilde{\gamma},\bm{r}_{12}-\bm{R}^{N_{i}}) \right] + 
\right. \\[0.2cm]
&\left.\frac{4\pi^{3/2}}{V^{(N_{i})}}\sum_{\bm{G}^{N_{i}}\neq0}^{''}\frac{(G_{x}^{N_{i}})^{t+t'}(G_{y}^{N_{i}})^{u+u'}(G_{z}^{N_{i}})^{v+v'}}{\abs{\bm{G}^{N_{i}}}^{2}}e^{-\frac{\abs{\bm{G}^{N_{i}}}^{2}}{4\tilde{\gamma}}}\cdot\right.\\[0.2cm]
&\left.\cos^{(t+t'+u+u'+v+v')}(\bm{G}^{N_{i}}\bm{r}_{12})
 \right\}
\end{aligned}
\end{equation}
where $V^{(N_{i})}=N_{\bm{k}}V$ is the supercell volume, the summation in $\bm{G}^{N_{i}}=(G^{N_{i}}_{x},G^{N_{i}}_{y},G^{N_{i}}_{z})$ spans the integer linear combinations of $\frac{1}{N_{i}}\bm{G}_{i}$ (see \eqref{MPgrid}), the double prime in the sum indicating that only one element from each $(\bm{G}^{N_{i}},-\bm{G}^{N_{i}})$ pair is included, and we have introduced
\begin{equation}\label{mu}
\mu\equiv\frac{\alpha\beta}{\alpha+\beta},
\end{equation}
\begin{equation}\label{tildegamma}
\tilde{\gamma}\equiv\left(\mu^{-1}+\gamma^{-1}\right)^{-1},
\end{equation}
\begin{equation}\label{r12}
\bm{r}_{12}\equiv\bm{r}_{1}-\bm{r}_{2}
\end{equation}
\begin{equation}\label{dercos}
\cos^{(n)}(r_{0})\equiv\left.\frac{\partial^{n}\cos(r)}{\partial r^{n}}\right\vert_{r=r_{0}}
\end{equation}
The constant term in \eqref{Ewald3D} does not contribute to \eqref{Ewald3DInt} due to the charge neutrality of the (supercell) lattice distribution. 

Both the direct and reciprocal sums in \eqref{Ewald3DInt} are exponentially convergent and the result is independent of $\gamma$, however, the choice of the latter allows to modulate the relative speed of convergence between the two series. In this regard, we follow the prescription in \cite{Saunders1992} for each pair of Hermite Gaussians by which in terms of \eqref{tildegamma},
\begin{equation}\label{casesgamma}
\tilde{\gamma}=\min(\gamma_{0},\mu)
\end{equation}
with the following optimized value
\begin{equation}\label{gamma03D}
\gamma_{0}=\frac{7.84}{(V^{(N_{i})})^{2/3}}
\end{equation}
For a dense enough BSE $\bm{k}-$grid the case $\tilde{\gamma}=\gamma_{0}$ will be ubiquitous, although we note that it is often desirable to include some diffuse exponents in the auxiliary GTF (for which $\mu$ is also small). In the complementary case, only the reciprocal lattice series remains. Regardless, with this choice the screening parameter will always be small for the supercell, prioritizing the convergence speed of the much denser reciprocal lattice series. 

Finally, the Ewald potential integrals in the (auxiliary) GTF basis are obtained by inserting \eqref{Ewald3DInt} in \eqref{GTF2Hermite}. 

\subsection{Ewald Potential Function in 2D}\label{Ewald2Dsec}
The Ewald formulation in the case of two-dimensional periodicity was apparently first studied by Parry \cite{Parry1975}, with a subsequent correction soon after \cite{Heyes1977}. The Ewald potential function \eqref{Ewald3D} cannot be employed in this case, which holds great practical importance, because three-dimensional periodicity was assumed in the derivation of the reciprocal lattice term. Instead, one finds the following expression for the Ewald potential function in 2D, sometimes referred to as the Parry potential function \cite{Doll2006,Harris1998,Kawata2001}
\begin{equation}\label{Ewald2D}
\begin{aligned}
&A(\bm{s})=%-2\sqrt{\frac{\gamma}{\pi}}+
\sum_{\bm{R}}^{'}\frac{\erfc(\sqrt{\gamma}\abs{\bm{s}-\bm{R}})}{\abs{\bm{s}-\bm{R}}}+\frac{\pi}{V}\sum_{\bm{G}\neq\bm{0}}\frac{e^{i\bm{Gs}}}{\abs{\bm{G}}}\cdot\\[0.2cm]
&\left[e^{\abs{\bm{G}}s_{z}}\erfc\left(\frac{\abs{\bm{G}}}{2\sqrt{\gamma}}+\sqrt{\gamma}s_{z}\right) + (s_{z}\leftrightarrow-s_{z})\right]\\[0.2cm]
&-\frac{2\pi}{\mathcal{A}}\left[s_{z}\erf\left(\sqrt{\gamma}s_{z}\right) + \frac{1}{\sqrt{\pi\gamma}}e^{-\gamma s_{z}^{2}}\right]
\end{aligned}
\end{equation}
where $\mathcal{A}$ is the unit cell area, the prime in the direct lattice sum has the same meaning as in \eqref{Ewald3D}, $s_{z}$ is the $z-$component of $\bm{s}$ (perpendicular to the lattice) and $(s_{z}\leftrightarrow-s_{z})$ represents the previous term within the brackets but with reversed $s_{z}$. $\bm{R}$ and $\bm{G}$ are here treated as 3-component vectors with null $z-$component. 

We find that the integral of \eqref{Ewald2D} with two Hermite Gaussians \eqref{HermiteGaussian} is, considering the supercell periodicity as in the 3D case above,
\begin{equation}\label{Ewald2DInt}
\begin{aligned}
&\iint\Lambda_{t,u,v}(\bm{r},\alpha,\bm{r}_{1})A(\bm{r}-\bm{r}')\Lambda_{t',u',v'}(\bm{r}',\beta,\bm{r}_{2})d\bm{r}d\bm{r}'=\\[0.2cm]
&\frac{2\pi^{5/2}(-1)^{t'+u'+v'}}{(\alpha\beta)^{3/2}}\left\{\sum_{\bm{R}^{N_{i}}}\left[\sqrt{\mu}R_{t+t',u+u',v+v'}(\mu,\bm{r}_{12}-\bm{R}^{N_{i}})  \right.  \right. \\[0.2cm]
&\left.\left.-\sqrt{\tilde{\gamma}}R_{t+t',u+u',v+v'}(\tilde{\gamma},\bm{r}_{12}-\bm{R}^{N_{i}}) \right] + \frac{\pi^{3/2}}{\mathcal{A}^{(N_{i})}}\cdot
\right. \\[0.2cm]
&\sum_{\bm{G}^{N_{i}}\neq0}^{''}\frac{(G_{x}^{N_{i}})^{t+t'}(G_{y}^{N_{i}})^{u+u'}}{\abs{\bm{G}^{N_{i}}}}\cos^{(t+t'+u+u')}(\bm{G}^{N_{i}}\bm{r}_{12}) \frac{\partial^{v+v'}}{\partial r_{12_{z}}^{v+v'}} \\[0.2cm]
&\left[e^{\:\abs{\bm{G}^{N_{i}}}r_{12_{z}}}\erfc\left(\frac{\abs{\bm{G}^{N_{i}}}}{2\sqrt{\tilde{\gamma}}}+\sqrt{\tilde{\gamma}}r_{12_{z}} \right) + (r_{12_{z}}\leftrightarrow-r_{12_{z}}) \right]\\
&\left.-\frac{\partial^{v+v'}}{\partial r_{12_{z}}^{v+v'}}\left[r_{12_{z}}\erf\left(\sqrt{\tilde{\gamma}}r_{12_{z}}\right)+\frac{1}{\sqrt{\pi\tilde{\gamma}}}e^{-\tilde{\gamma}r_{12_{z}}^{2}}\right] \right\}
\end{aligned}
\end{equation}
where $\mathcal{A}^{(N_{i})}=N_{\bm{k}}\mathcal{A}$ is the supercell area and the same considerations as for the 3D case below \eqref{Ewald3DInt} apply, including the definitions \eqref{mu}-\eqref{dercos}, only noting that here $\bm{R}^{N_{i}}$ and $\bm{G}^{N_{i}}$ are 3-component vectors with null $z-$component. 

We employ again the criterion \eqref{casesgamma} for the screening parameter, with the difference that the optimized reference value \eqref{gamma03D} is now taken as \cite{Doll2006}
\begin{equation}\label{gamma02D}
\gamma_{0}=\frac{5.76}{\mathcal{A}^{(N_{i})}}
\end{equation}

The Ewald potential integrals in the (auxiliary) GTF basis are obtained by inserting \eqref{Ewald2DInt} in \eqref{GTF2Hermite}. 

\subsection{Metrics for the RI}\label{MetricsSec}
Regarding now the metric integrals \eqref{MmatrixG}, \eqref{LmatrixG}, they can be solved analytically for different choices of metric functions, the most common thereof being the overlap $m(\bm{s})=\delta(\bm{s})$ and Coulomb $m(\bm{s})=1/\abs{\bm{s}}$ metrics. The former is the least demanding computationally, while the latter is in principle the most accurate fitting option but it presents two crucial problems in the 3-center \eqref{LmatrixG} for periodic systems: an extremely large number of integrals have non-negligible values due to the slow spatial decay, and it reintroduces the convergence issues described in \ref{EwaldTheorySec}. Attempting to employ the Ewald method results in 
\begin{equation*}\begin{aligned}
&L^{\mu\bm{k},\mu'\bm{k}'}_{P}=\sum_{\bm{R}}e^{i\bm{k}'\bm{R}}\sum_{\bm{R}'}^{\text{SUP}(\set{N_{i}})}e^{i(\bm{k}'-\bm{k})\bm{R}'}\cdot\\
&\iint\phi_{P}^{\bm{0}}(\bm{r})A(\bm{r}-\bm{r}')\varphi_{\mu}^{\bm{R}'}(\bm{r}')\varphi_{\mu'}^{\bm{R}+\bm{R}'}(\bm{r}')d\bm{r}d\bm{r}'
\end{aligned}\end{equation*}
which is impractical due to the combination of the supercell and the (quasi-)cubic scaling in the basis.

Alternative metrics have been employed such as the attenuated Coulomb\cite{AttCoulomb,RI2005} $m(\bm{s})=\erfc(\omega\abs{\bm{s}})/\abs{\bm{s}}$ with some range-tuning parameter $\omega>0$, or the truncated Coulomb\cite{Wilhelm2021,CP2KGW} $m(\bm{s})=\theta(r_{c}-\abs{\bm{s}})/\abs{\bm{s}}$ where $r_{c}$ is some cutoff radius and $\theta$ is the Heaviside function. Both of these functions serve as an intermediate parametrization between the overlap ($\omega\to\infty$, $r_{c}\to0$) and the Coulomb ($\omega\to0$, $r_{c}\to\infty$) metrics, not presenting the convergence problems of the latter and including more long-range integrals than the former. In any case, all metrics are expected to provide similar results in the limit of sufficiently complete auxiliary basis sets, although of course it is not possible to know in advance how the precision of each metric will evolve with respect to the basis in a particular system. 

In the present work we employ both the overlap and the attenuated Coulomb metrics. The integrals in the former case are computed as explained in the discussion around \eqref{HermiteInt0}, and in the latter case employing \eqref{GTF2Hermite} and \eqref{GTF3Hermite} and noting that the attenuated Coulomb operator essentially corresponds to the terms in the direct lattice sums of the Ewald potential function \eqref{Ewald3D} or \eqref{Ewald2D}, with $\omega=\sqrt{\gamma}$.

\subsection{Singularities in the BSE}\label{SecDivergence}
As explained in Section \ref{EwaldTheorySec}, for $\bm{k}=\bm{0}$ the series $\sum_{\bm{R}}J_{P,P'\bm{R}}$ resulting from \eqref{JmatrixG} corresponds to the interaction energy between $\phi_{P}^{\bm{0}}$ and the periodic arrangement $\sum_{\bm{R}}\phi_{P'}^{\bm{R}}$, which is divergent for charged distributions, i.e. if both $\phi_{P}$ and $\phi_{P'}$ are $s-$type GTF. Contrary to the $\bm{k}\neq0$ case, a supercell cannot be found such that the enclosed functions yield a chargeless density. 

In the RI-based formulation of the BSE, see Section \ref{BSEgauss}, such divergent terms arise precisely when $\bm{k}_{1}=\bm{k}_{2}$ in \eqref{vmatrix}, which in turn occurs under the TDA in 
\begin{enumerate}
\item The exchange terms \eqref{ExchangeV} for $\bm{Q}=\bm{0}$, that is $X^{c\bm{k},c'\bm{k}'}_{v\bm{k},v'\bm{k}'}$
\item The direct terms \eqref{DirectV} for $\bm{k}=\bm{k}'$, that is $D^{v'\bm{k},c'\bm{k}+\bm{Q}}_{v\bm{k},c\bm{k}+\bm{Q}}$
\end{enumerate}
This divergent behaviour is not an artifact of the present formulation, but rather an inherent feature of the BSE itself (in plane-waves or Fourier-based methods for instance, the source of the problem in the direct terms are the $\bm{G}=\bm{G}'=\bm{0}$ components \cite{strinati1988,rohlfing_louie_2000}). Therefore, we resort to solutions stemming from the general BSE derivation\footnote{A potential alternative could be formulated in terms of an orthogonal charge projection \cite{Burow2009,Burow2015,Burow2017} by incorporating the $c^{\bm{k}}_{\mu n}$ coefficients into \eqref{RI} an invoking the orthonormality of the $\ket{\psi_{n\bm{k}}}$ to fit a chargeless density. However, the orthogonality would in principle break down for the pure-diagonal $D^{v\bm{k},c\bm{k}+\bm{Q}}_{v\bm{k},c\bm{k}+\bm{Q}}$ terms.}.

It can be shown \cite{martin2016} that the $\bm{Q}\to\bm{0}$ limit in the exchange terms exists albeit it is non-analytical. In order to compute $X^{c\bm{k},c'\bm{k}'}_{v\bm{k},v'\bm{k}'}$, which is the relevant limit for optical responses (disregarding phonons, which couple to finite $\bm{Q}$), we therefore average over a set of symmetry-related, small $\bm{Q}$ values denoted by $\Delta\bm{k}$, namely
\begin{equation}\begin{aligned}\label{Xdiverg}
&X^{c\bm{k},c'\bm{k}'}_{v\bm{k},v'\bm{k}'}=\frac{1}{N_{\bm{k}}}\frac{1}{N_{\Delta\bm{k}}}\sum_{\Delta\bm{k}}\sum_{R}\sum_{\mu,\mu',\nu,\nu'}\\[0.2cm]
&c^{\bm{k}*}_{\mu v}c^{\bm{k}}_{\mu'c}v_{R}^{\mu\bm{k},\mu'\bm{k}+\Delta\bm{k}}\left(v_{R}^{\nu\bm{k}',\nu'\bm{k}'+\Delta\bm{k}}\right)^{*}c^{\bm{k}'}_{\nu v'}c^{\bm{k}'*}_{\nu'c'}
\end{aligned}\end{equation}

On the other hand, the divergence in the diagonal blocks of the direct matrix is integrable \cite{rohlfing_louie_2000,BGW2012}, being a spurious consequence of the discretization of reciprocal space. To compute these terms we average over a set of symmetry-related, small displacements $\Delta\bm{k}$ (which may in general be different from the ones in \eqref{Xdiverg}) in $\bm{k}$ and $\bm{k}'$ separately, namely
\begin{equation}\begin{aligned}\label{Ddiverg}
&D^{v'\bm{k},c'\bm{k}+\bm{Q}}_{v\bm{k},c\bm{k}+\bm{Q}}=\frac{1}{N_{\bm{k}}}\frac{1}{N_{\Delta\bm{k}}}\sum_{\Delta\bm{k}}\sum_{R,R'}\sum_{\mu,\mu',\nu,\nu'}\left\{
c^{\bm{k}*}_{\mu v}c^{\bm{k}}_{\mu' v'}\cdot\right.\\[0.2cm]
&v_{R}^{\mu\bm{k},\mu'\bm{k}+\Delta\bm{k}}\left(\epsilon^{-1}_{R,R'}(\Delta\bm{k})v_{R'}^{\nu\bm{k},\nu'\bm{k}+\Delta\bm{k}}\right)^{*}c^{\bm{k}}_{\nu c}c^{\bm{k}*}_{\nu' c'} + \\[0.2cm]
&c^{\bm{k}}_{\mu v'}c^{\bm{k}*}_{\mu' v}\left(v_{R}^{\mu\bm{k},\mu'\bm{k}+\Delta{\bm{k}}}\right)^{*} \epsilon^{-1}_{R,R'}(\Delta\bm{k})v_{R'}^{\nu\bm{k},\nu'\bm{k}+\Delta\bm{k}}\cdot\left.c^{\bm{k}*}_{\nu c'}c^{\bm{k}}_{\nu'c}
\right\}
\end{aligned}\end{equation}
This scheme preserves the hermiticity, gauge invariance with respect to the single-particle eigenstates (up to a unitary transformation, the same as the whole BSE Hamiltonian) and point-group symmetry (provided that complete stars of $\Delta\bm{k}$ are included) of the exchange and direct terms. 

In order to compute the arising $J_{P,P'\Delta\bm{k}}$ terms in both \eqref{Xdiverg} and \eqref{Ddiverg} we again employ the Ewald substitution \eqref{JmatrixGEwald}. In this context the supercell may in principle need to be larger due to the small $\abs{\Delta\bm{k}}$, however, since in practice there will be few linearly independent $\Delta\bm{k}$ one may assign a reciprocal lattice vector $\bm{G}_{a}$ to each $\Delta\bm{k}_{a}$ such that $\Delta\bm{k}_{a}=(\alpha_{a}/N_{a})\bm{G}_{a}$ for some $\alpha_{a},N_{a}\in\mathbb{N}$, in which case it would be sufficient to consider supercells just in the direction of $\bm{G}_{a}$, i.e. along $\bm{R}_{a}$ such that $\bm{R}_{a}\bm{G}_{a}=2\pi$ (albeit a different one for each linearly independent $\Delta\bm{k}$). For the record, in the calculations presented in Section \ref{ApplicationsSec} we have employed full supercells rather than this latter approach, although we have not found significant variations in the excitonic energies upon reduction of $\abs{\Delta\bm{k}}$ starting from the corresponding $(1/N_{i})\abs{\bm{G}_{i}}$ in \eqref{MPgrid}, hence the magnitudes of the displacements have not been small enough to need to resort to the alternative.

%This term is present irrespective of the metric, while the (full) Coulomb metric introduces more of these terms in \eqref{MmatrixG} and \eqref{LmatrixG}. 
%This divergence is, however, integrable \cite{rohlfing_louie_2000} and in principle one must resort to the thermodynamic limit ($N_{\bm{k}}\to\infty$) and retain the integrals in reciprocal space around the singularity ($\bm{k}=\bm{k}'$). The resonant block \eqref{HQres} then reads
%\begin{equation}\begin{aligned}\label{HresSingular}
%&H^{\text{res}}_{v\bm{k},c'\bm{k}+\bm{Q};c\bm{k}+\bm{Q},v'\bm{k}}=(\varepsilon_{c\bm{k}+\bm{Q}}-\varepsilon_{v\bm{k}})\delta_{c,c'}\delta_{v,v'}+\\
%&X_{v\bm{k},v'\bm{k}}^{c\bm{k}+\bm{Q},c'\bm{k}+\bm{Q}}
%-\left(\frac{N_{\bm{k}}}{V_{BZ}}\right)^{2}\int_{V(\bm{k})}\int_{V(\bm{k})}D_{v\bm{k}_{1},c\bm{k}_{1}+\bm{Q}}^{v'\bm{k}'_{1},c'\bm{k}'_{1}+\bm{Q}}d\bm{k}'_{1}d\bm{k}_{1} 
%\end{aligned}\end{equation}
%The integration domain $V(\bm{k})$ denotes the fraction of the BZ that is closest to $\bm{k}$ than to any other of the $N_{\bm{k}}$ grid points in the BSE grid, i.e. a Wigner-Seitz type domain. In practice we perform the integrals numerically by sampling $V(\bm{k})$ with a set of discrete points, avoiding the singularity at $\bm{k}$. Note that \eqref{HresSingular} reduces to \eqref{HQres} for $\bm{k}\neq\bm{k}'$, assuming that $N_{\bm{k}}$ is large enough that the direct term is approximately constant in $V(\bm{k})\times V(\bm{k}')$.

\section{Applications}\label{ApplicationsSec}
\subsection{Implementation}\label{ImplementationSec}
The method here laid out has been implemented in a developer version of the open-source code \texttt{XATU} \cite{Xatu}, previously dedicated to tight-binding descriptions. It is publicly available on github \cite{github}, written purely in C++ with MPI parallelization, and will be documented in the near future. The code is interfaced with the latest version of \texttt{CRYSTAL} \cite{crystal23}, from which it extracts the single-particle Hamiltonian matrix elements in the selected GTF basis, $H_{\mu,\mu'}(\bm{R})=\bra{\varphi_{\mu}^{\bm{0}}}\hat{H}\ket{\varphi_{\mu'}^{\bm{R}}}$, resulting from a self-consistent HF/DFT calculation; in addition to the basis lattice vectors $\bm{R}_{i}$, the atomic positions in the unit cell $\bm{t}_{a}$ and the total number of electrons. The auxiliary basis for the RI and the exciton parameters are given separately, in particular $N_{i}$ in \eqref{MPgrid} for the exciton basis and a different set of values for the polarizability $\bm{k}-$integration in \eqref{Pimatrix}, the set of valence and conduction bands for the exciton basis, the exciton momentum $\bm{Q}$ (which in this work is set to $\bm{0}$), and optionally a regularization parameter $\alpha$ (see Section \ref{BSEgauss}) and a scissor correction for the conduction bands. All the required matrix elements, aside from those of $\hat{H}$, have been implemented from scratch up to $g-$type GTF ($l=4$) following the guidelines in Section \ref{GaussianRIsec} and Appendix \ref{GaussianApp}.

Up to further refinements in the implementation, the calculations here presented consumed an average of $\sim30\cdot10^{3}$ core-hours each (starting right after the HF/DFT calculation and finishing after computing the optical absorption) mainly with AMD EPYC 7H12  CPUs, the main bottleneck being the calculation of the off-diagonal ($\bm{k}\neq\bm{k}'$) screening matrix elements in \eqref{DirectV}, including the irreducible polarizability matrix and the $N_{\mu}^{2}N_{P}N_{\bm{k}}^{2}$ matrix elements $v_{R}^{\mu\bm{k},\mu'\bm{k}'}$. Ultimately, the central scaling factor is the size of the auxiliary basis set, aside of course from the size of the exciton basis.

\subsection{Optical Absorption}\label{ConductivitySec}
For a more complete comparison with existing results, we compute the (real part of the) diagonal of the first-order optical conductivity tensor \cite{Pedersen2015,Ridolfi2018}
\begin{equation}\label{Kubo}
\sigma^{aa}(\omega)=\frac{e^{2}}{\hbar}\cdot\frac{\pi}{N_{\bm{k}}V\hbar\omega}\sum_{\chi}\abs{V^{a}_{\chi}}^{2}\delta(\hbar\omega-E_{\chi})
\end{equation}
Here $\omega$ is the frequency (component) of an incident electric field, $a$ labels spatial components, $V$ corresponds to unit cell volume, area or length, and we have introduced the oscillator strengths
\begin{equation}\label{OscStrength}
V^{a}_{\chi}=\sum_{\bm{k}}\sum_{v,c}\left(A_{\chi}^{v\bm{k},c\bm{k}}\right)^{*}\bra{\psi_{v\bm{k}}}\hbar\hat{v}^{a}\ket{\psi_{c\bm{k}}} 
\end{equation}
in terms of the velocity operator $\hat{\bm{v}}=(i/\hbar)[\hat{H},\hat{\bm{r}}]$. The velocity matrix elements in the GTF basis are computed as explained in \cite{garcia2023shift}, with the exception that the dipole (or position) $\hat{\bm{r}}$ matrix elements are now computed more efficiently in terms of the expansion coefficients in Hermite Gaussians, see \eqref{HermiteG2}. In particular, the calculation does not require any limit or approximation by the momentum operator. From \eqref{OscStrength} it follows that in the spinless case only the singlet states are optically active, %both spin projections contributing equally to the conductivity and thus yielding an extra factor 2 in \eqref{Kubo} upon restriction to a single spin projection in the $v,c$ sums. 
hence the figures presented in this section correspond exclusively to the singlet solutions.
The single-particle counterpart of \eqref{Kubo}, which neglects electron-hole correlations, is the standard Kubo-Greenwood expression
\begin{equation}\label{KuboSP}\begin{aligned}
&\sigma^{aa}_{\text{sp}}(\omega)=\frac{e^{2}}{\hbar}\cdot\\
&\frac{\pi}{N_{\bm{k}}V\hbar\omega}\sum_{\bm{k}}\sum_{v,c}\abs{\bra{\psi_{v\bm{k}}}\hbar\hat{v}^{a}\ket{\psi_{c\bm{k}}}}^{2}\delta(\hbar\omega-(\varepsilon_{c\bm{k}}-\varepsilon_{v\bm{k}}))
\end{aligned}\end{equation}
In both \eqref{Kubo} and \eqref{KuboSP} we consider a Gaussian broadening (standard deviation) for the Dirac delta, specified in each case. On the other hand, the optical conductivity can be directly related to the absorption spectrum, given by the imaginary part of the macroscopic dielectric function ($\bm{q}=\bm{G}=\bm{G}'=\bm{0}$) \cite{rohlfing_louie_2000,Onida1998}
\begin{equation}\label{epsilon2}
\epsilon_{2}(\omega)=\frac{1}{\epsilon_{0}}\frac{\sigma^{aa}(\omega)}{\omega}
\end{equation}
where linear light polarization along $a$ has been assumed. However, the dielectric constant obtained in this manner is dimensionless only in 3D, requiring a sample-dependent thickness factor in lower dimensions. Instead, in the 2D scenario we consider the absorption cross-section or absorbance under the same assumptions on the light polarization, namely
\begin{equation}\label{Absorbance}
\text{Abs}(\omega)=\frac{1}{\epsilon_{0}c}\sigma^{aa}(\omega)
\end{equation}
which is indeed a dimensionless quantity. In atomic units, $1/\epsilon_{0}=4\pi$, $1/c=\alpha$ (the fine structure constant) and we recover the usual expressions.

\subsection{Black Phosphorus (2D)}\label{PhoSec}
\begin{figure*}
\centering 
\begin{minipage}{0.47\textwidth}
\vspace{-0.65cm}
\begin{subfigure}{\textwidth}
    \includegraphics[width=\textwidth]{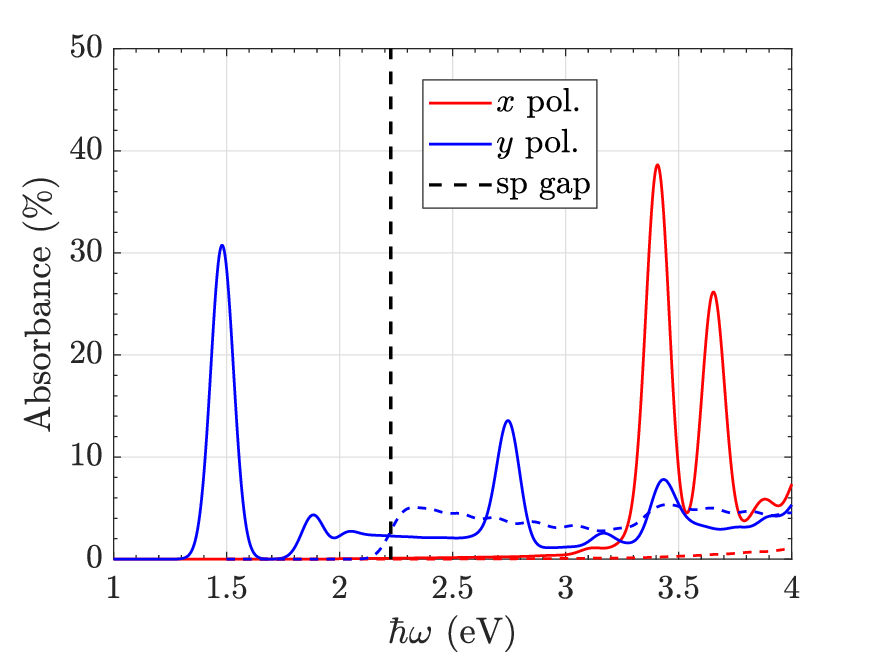}
    \caption{}
    \label{FigPho1A}
\end{subfigure}
\end{minipage} 
\begin{minipage}{0.47\textwidth}
\begin{subfigure}{0.42\textwidth}
    \includegraphics[width=\textwidth]{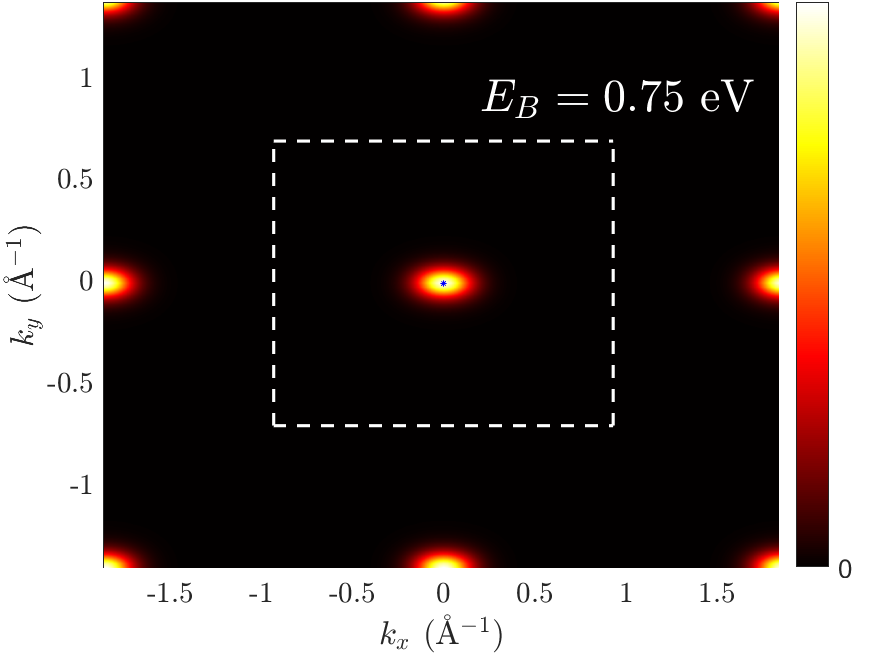}
    \caption{}
    \label{FigPho1B}
\end{subfigure}
\begin{subfigure}{0.42\textwidth}
    \includegraphics[width=\textwidth]{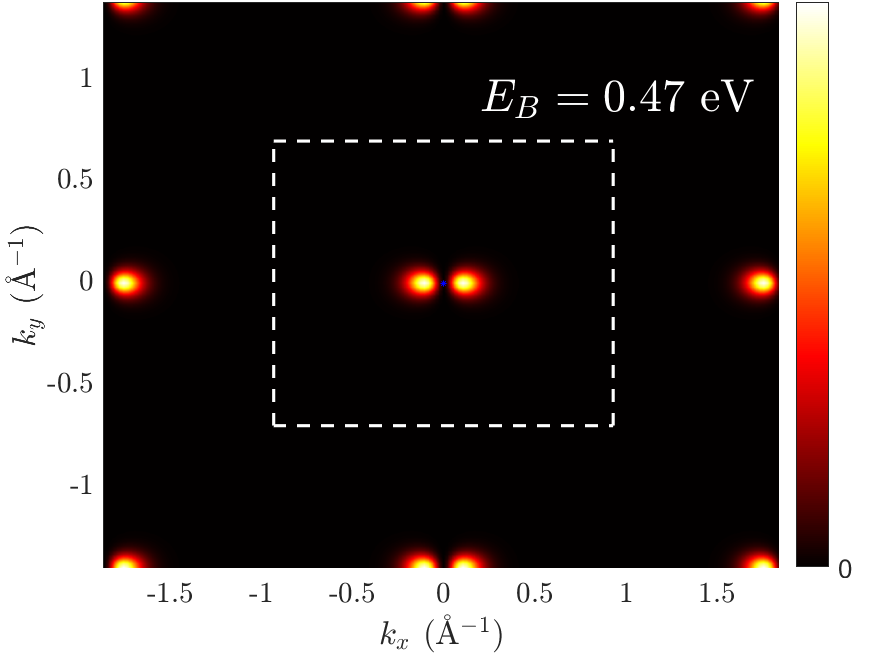}
    \caption{}
    \label{FigPho1C}
\end{subfigure}
\begin{subfigure}{0.42\textwidth}
    \includegraphics[width=\textwidth]{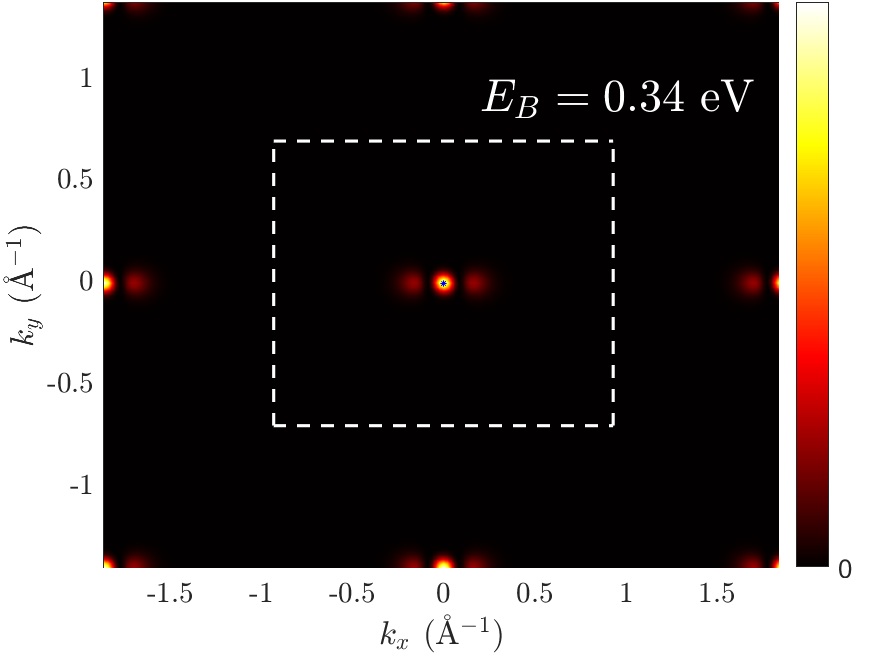}
    \caption{}
    \label{FigPho1D}
\end{subfigure}
\begin{subfigure}{0.42\textwidth}
    \includegraphics[width=\textwidth]{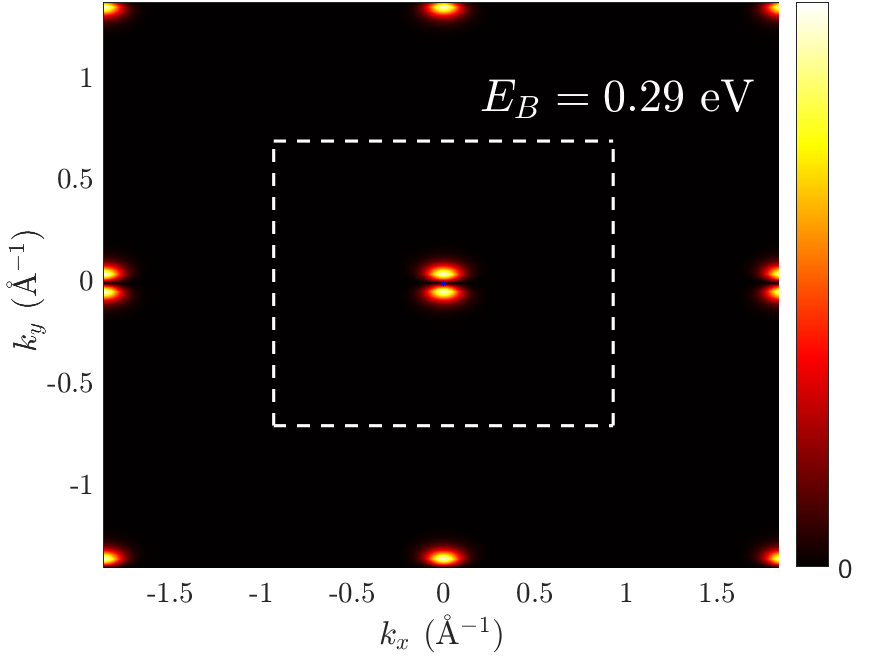}
    \caption{}
    \label{FigPho1E}
\end{subfigure}
\end{minipage}
\begin{subfigure}{0.3\textwidth}
    \includegraphics[width=\textwidth]{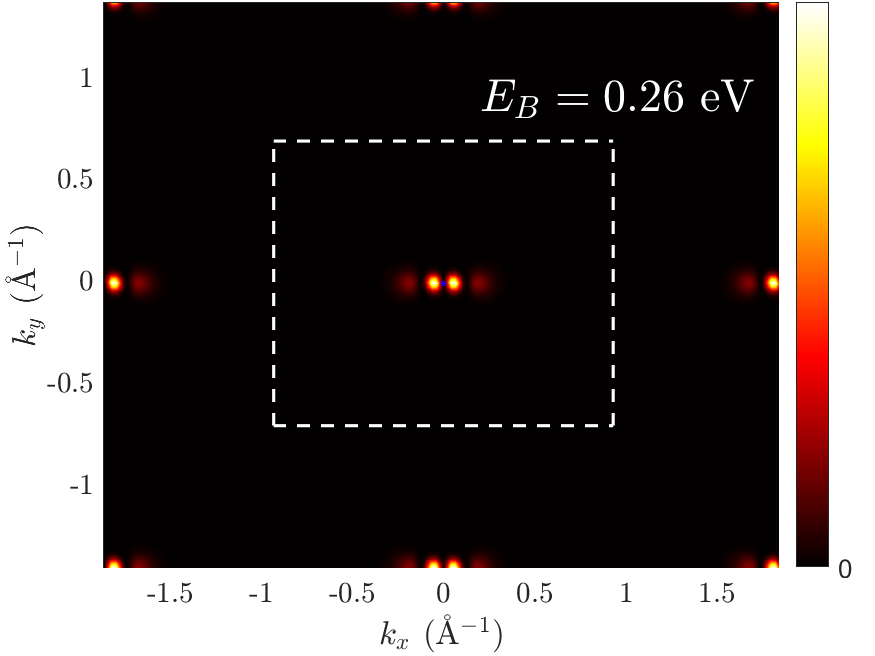}
    \caption{}
    \label{FigPho1F}
\end{subfigure}
\begin{subfigure}{0.3\textwidth}
    \includegraphics[width=\textwidth]{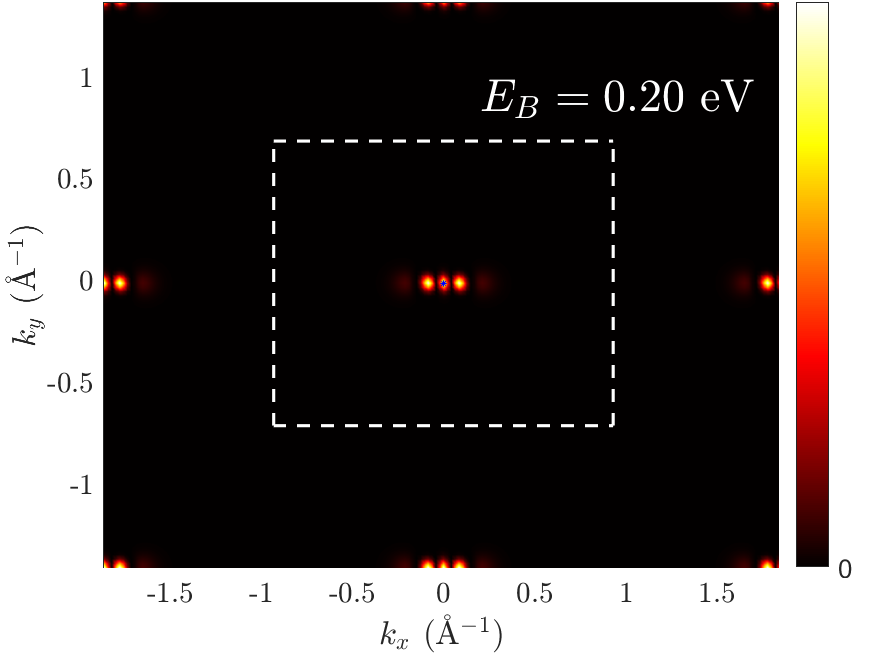}
    \caption{}
    \label{FigPho1G}
\end{subfigure}
\begin{subfigure}{0.3\textwidth}
    \includegraphics[width=\textwidth]{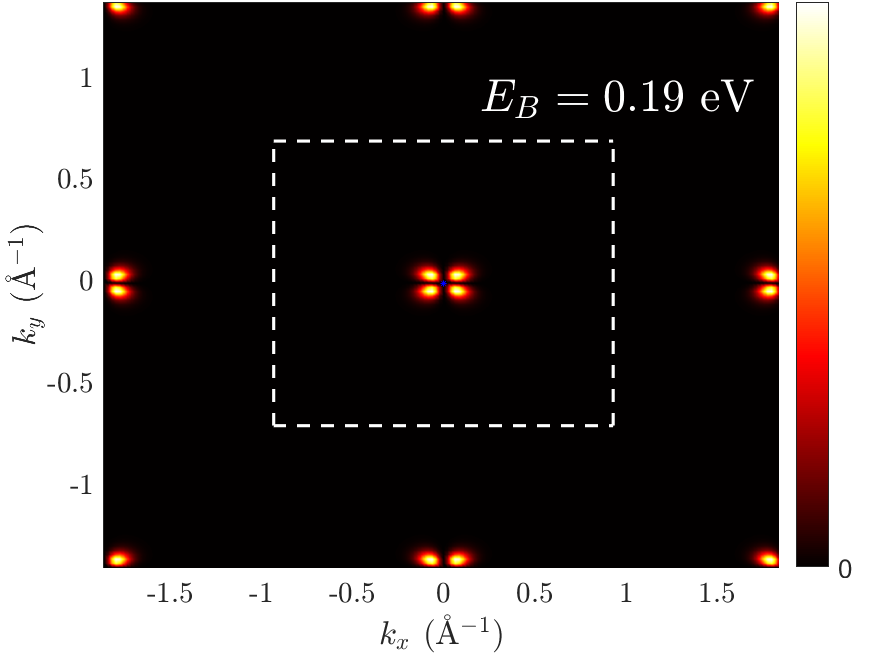}
    \caption{}
    \label{FigPho1H}
\end{subfigure}
\caption{\label{FigPho1} (a) Absorption spectrum of a single-layer of black phosphorus, with (solid lines) and without (dashed colour lines) electron-hole correlations, for different in-plane light polarizations and with a Gaussian broadening of $50$ meV. The zigzag and armchair directions correspond to $x$ and $y$, respectively. The single-particle gap is indicated by the black dashed line. (b-h) Distribution $\sum_{v,c}\abs{A_{\chi}^{v\bm{k},c\bm{k}}}^{2}$ of the exciton coefficients for the first seven singlet states, whose binding energies are indicated. Only (b), (d) and (g) are bright. The BZ is delimited by the white dashed lines, and the blue star indicates the $\Gamma$ point. $\bm{k}-$grid: $88\times 66$ for both the BSE and the absorption calculations, attenuated Coulomb metric ($\omega=0.25$ a.u.) for Gaussian density fitting.}
\end{figure*}

As a first example we consider a single-layer of black phosphorus or phosphorene \cite{PhoOriginal}, a two-dimensional non-planar lattice with four $\text{P}$ atoms per unit cell. In our chosen orientation the corresponding space group is $Pm2_{1}n$ (removing the lattice translations along $z$), with the non-symmorphic twofold rotation symmetry along the $y-$axis. The projective representation of the little group $C_{2v}$ along the $\text{XS}$ path forces a twofold band degeneracy in the spinless case, and likewise time-reversal symmetry induces a twofold degeneracy along the $\text{YS}$ path \cite{Pikus}, i.e. all the BZ edges are degenerated while the interior is not (see the Supplemental Material). 

A hybrid HF/DFT calculation was performed employing a slightly reduced def2-TZVP basis \cite{BasisExchange,def2_TZVP} (removing one $d-$ and one $f-$type shells with negligible population according to a Mulliken analysis), $169$ points in the irreducible BZ and the PBE0 functional \cite{PBE0}, with a resultant direct gap of $2.23$ eV at $\Gamma$ which is already within the experimental margin of $2.2\pm0.1$ eV measured in Ref. \cite{PhoXia2015}. For the auxiliary basis we take, conservatively, a custom set with $5$ times the dimension of the DFT basis which was created starting from a def2-TZVP-RIFIT \cite{def2_TZVP_RIFIT} and including, for each pair of angular momentum and diffuse exponent $(l_{i},\alpha_{i})$, $(l_{j},\alpha_{j})$ in the original basis an auxiliary GTF with $(l_{i}+l_{j},\alpha_{i}+\alpha_{j})$ (in order to reproduce the on-site products better), in addition to some more diffuse functions. Both basis sets can be found in the Supplemental Material with the format of the \texttt{CRYSTAL} code, see Section 2.2.1 in Ref. \cite{dovesicrystal23}. Due to the aforementioned twofold degeneracy in the BZ edges, we take $2$ valence and $2$ conduction bands for the exciton basis, in addition to a $88\times66$ grid of $\bm{k}-$points which approximately preserves the proportion of the reciprocal basis vectors. The polarizability grid in \eqref{Pimatrix} was reduced to $8\times6$. For the RI we employ the attenuated Coulomb metric with $\omega=0.25$ Bohr$^{-1}$.

Solving the BSE we find a first exciton with binding energy $E_{B}=-E_{\chi}+[\min_{\bm{k
,v,c}}(\varepsilon_{c\bm{k}}-\varepsilon_{v\bm{k}})]$ of $0.75$ eV, in decent agreement with the experimentally estimated value of $0.9\pm0.12$ eV in Ref. \cite{PhoXia2015} (actually closer to the computed value of $\sim0.8$ eV therein) and overall with other calculations based on plane-waves \cite{PhoYang2014,PhoYang2015}, albeit with slightly different single-particle gaps. It should be noted that computed values down to $\sim0.5$ eV have also been reported in the literature \cite{DianaThesis,PhoThygesen2018}. The optical absorption in Figure \ref{FigPho1A} is greatly altered by the inclusion of electron-hole correlations, and it reflects the high degree of anisotropy of the material. The first, infra-red peak for light polarization along the armchair ($y$) direction arises purely from the first exciton, whose distorted $s-$like or elliptical distribution in reciprocal space is shown in Figure \ref{FigPho1B}. Analogous figures are provided for the next six excitons, all of which exhibit the symmetries of the little group at $\bm{Q}=\bm{0}$, namely with respect to $k_{x}\to-k_{x}$ and $\bm{k}\to-\bm{k}$ (the latter due to time-reversal symmetry). More particularly, the exciton coefficients transform in reciprocal space according to irreducible representations of such group (up to a phase factor), which in this case does not impose any degeneracy in the exciton energies. The present series of seven excitons perfectly matches the one in Figure 6.4 of Ref. \cite{DianaThesis} except for the fact that the last two states are permuted, nevertheless the energy difference between them is very small in both works. From this set of states, only the $s-$like are bright and contribute to the optical absorption, namely those in Figures \ref{FigPho1B}, \ref{FigPho1D} and \ref{FigPho1G}. They are all strongly localized around the band edge at $\Gamma$, hence a high degree of spatial delocalization could be attributed to them.

The corresponding exciton energies are displayed in Table \ref{TablePhoExc}
for different RI metric choices, namely the overlap and attenuated Coulomb with different values of $\omega$ (see Section \ref{MetricsSec}). It can be observed that the choice of metric has a small impact in this case, with energy differences of a few meV. Furthermore, the impact on the absorption and reciprocal space textures is negligible. We warn that this may not be the case in general for other auxiliary bases, specially smaller ones, and even for smaller values of the attenuation parameter $\omega$ (closer to the full Coulomb metric), nevertheless it serves as an example of the potential validity of the overlap metric in the RI-based BSE context. On the other hand, a singlet-triplet splitting of $90$ meV is found for the first excited state, larger than the $50$ meV computed in Ref. \cite{PhoYang2015} but still small compared to the excitonic energy differences.

\begin{table}[h!]\caption{\label{TablePhoExc} First seven exciton energies $E_{\chi}$ (eV) for different metrics and spin blocks. The first row indicates the value of the parameter $\omega$ (a.u. or Bohr$^{-1}$) in the attenuated Coulomb metric, the $\infty$ case actually corresponding to the overlap metric.}
\centering\begin{tabular}{| c || c | c | c | c | c |} 
\hline
$\omega$ & $\infty$ (overlap) & $\infty$ & $0.5$ & $0.25$ & $0.25$  \\[0.05cm] \hline
& Triplet & Singlet & Singlet & Singlet & Triplet  \\[0.05cm] \hhline{|=|=|=|=|=|=|}
$E_{1}$ & 1.395 & 1.486 & 1.484 & 1.480 & 1.390 \\[0.08cm] \hline
$E_{2}$ & 1.763 & 1.764 & 1.762 & 1.759 & 1.758 \\[0.08cm] \hline
$E_{3}$ & 1.872 & 1.888 & 1.886 & 1.883 & 1.867 \\[0.08cm] \hline
$E_{4}$ & 1.933 & 1.936 & 1.935 & 1.932 & 1.929 \\[0.08cm] \hline
$E_{5}$ & 1.974 & 1.975 & 1.973 & 1.970 & 1.969 \\[0.08cm] \hline
$E_{6}$ & 2.021 & 2.030 & 2.029 & 2.026 & 2.017 \\[0.08cm] \hline
$E_{7}$ & 2.038 & 2.038 & 2.037 & 2.034 & 2.034 \\ \hline
\end{tabular}\end{table}

\subsection{Boron Nitride (2D)}\label{hBNSec}
\begin{figure*}
\centering 
\begin{minipage}{0.47\textwidth}
\vspace{-0.65cm}
\begin{subfigure}{\textwidth}
    \includegraphics[width=\textwidth]{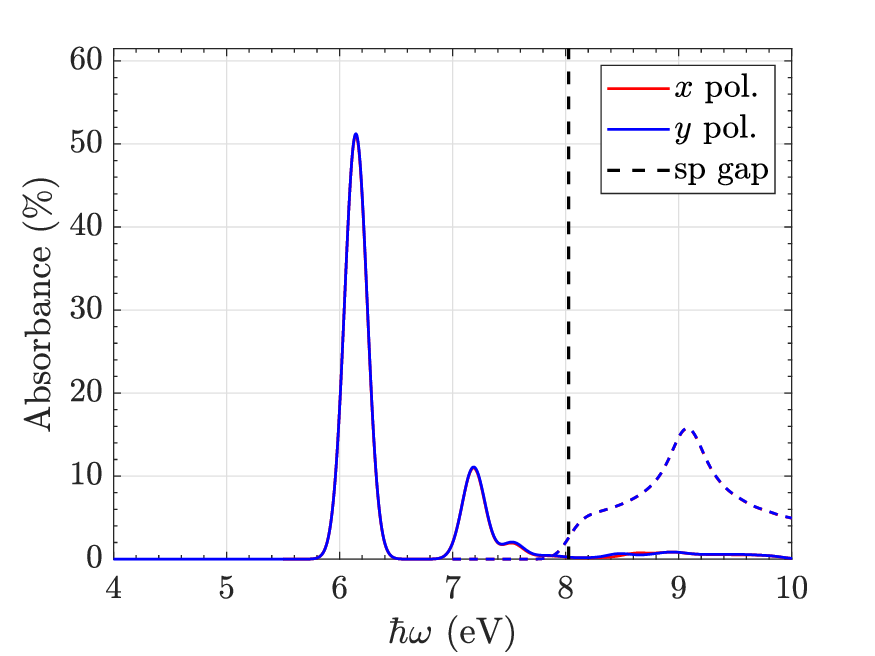}
    \caption{}
    \label{FigHBN1A}
\end{subfigure}
\end{minipage} 
\begin{minipage}{0.47\textwidth}
\begin{subfigure}{0.42\textwidth}
    \includegraphics[width=\textwidth]{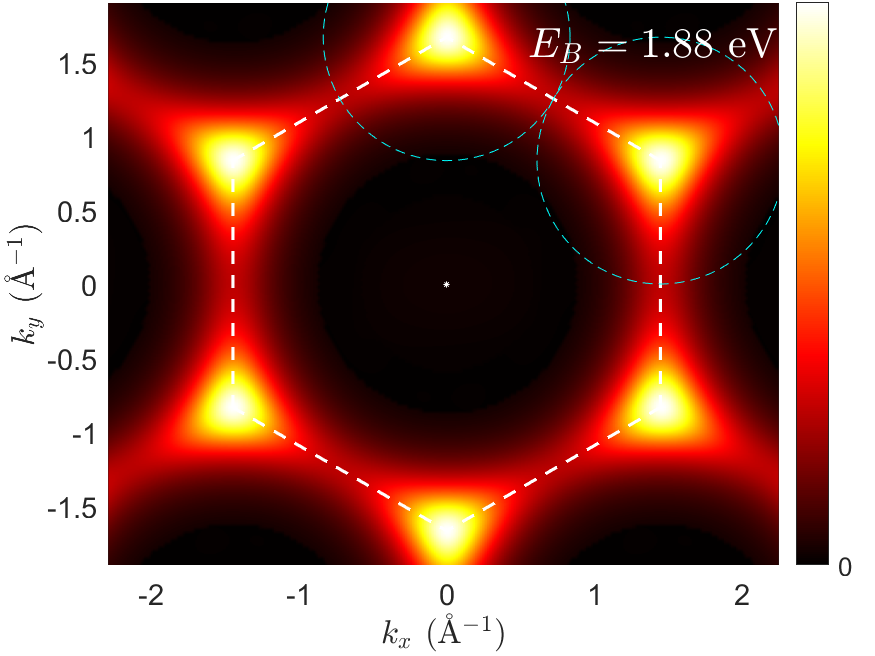}
    \caption{}
    \label{FigHBN1B}
\end{subfigure}
\begin{subfigure}{0.42\textwidth}
    \includegraphics[width=\textwidth]{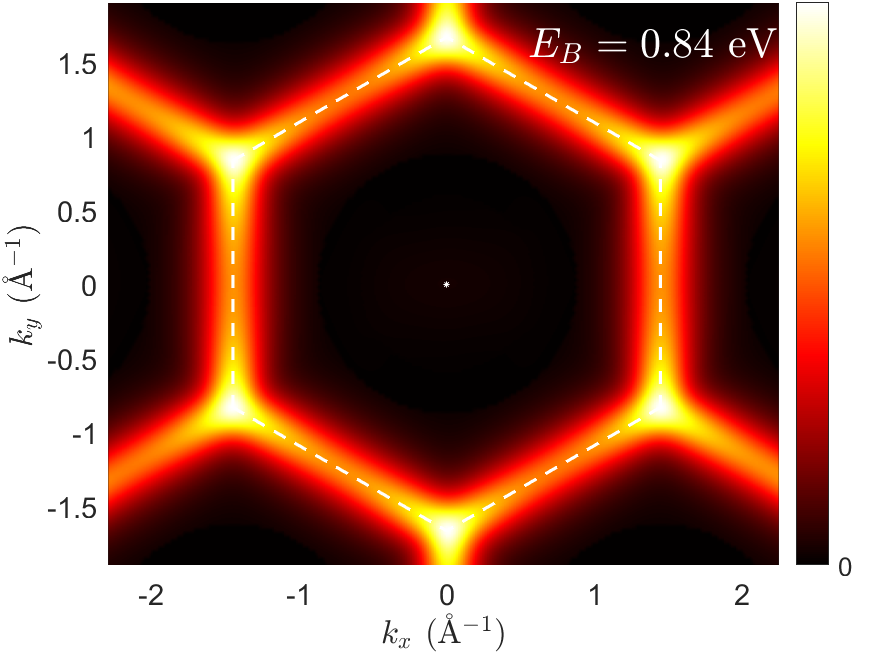}
    \caption{}
    \label{FigHBN1C}
\end{subfigure}
\begin{subfigure}{0.42\textwidth}
    \includegraphics[width=\textwidth]{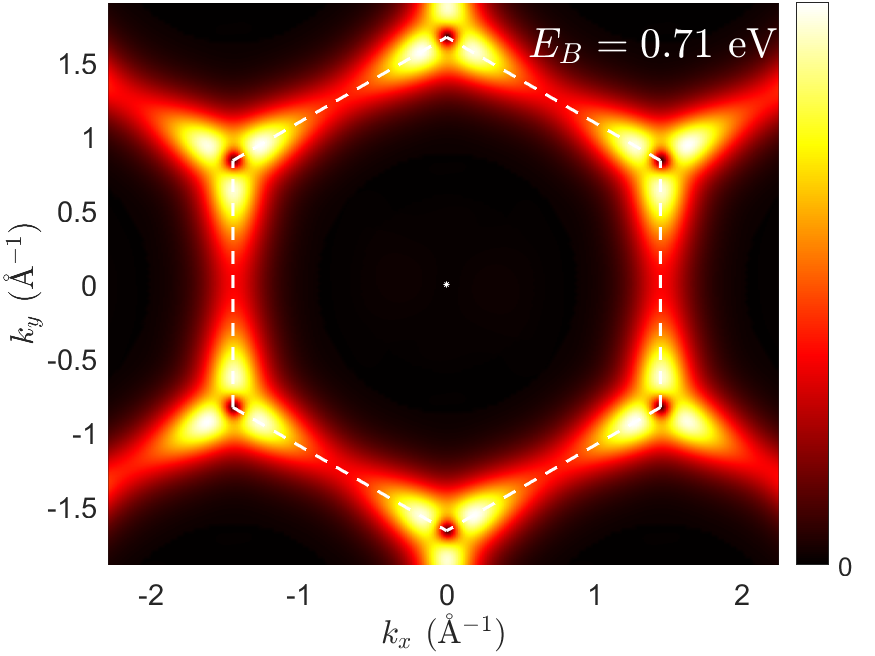}
    \caption{}
    \label{FigHBN1D}
\end{subfigure}
\begin{subfigure}{0.42\textwidth}
    \includegraphics[width=\textwidth]{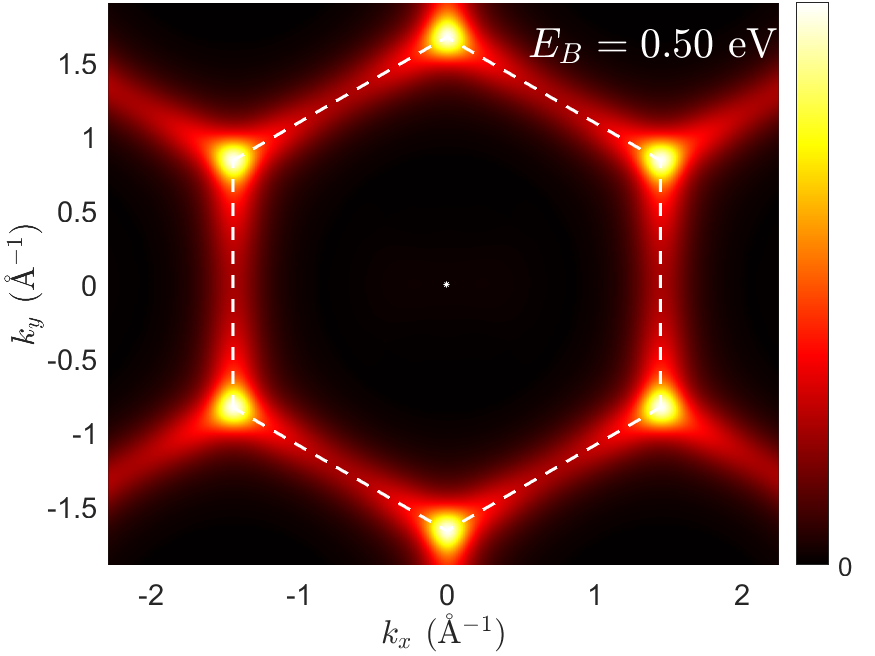}
    \caption{}
    \label{FigHBN1E}
\end{subfigure}
\end{minipage}
\begin{subfigure}{0.22\textwidth}
    \includegraphics[width=\textwidth]{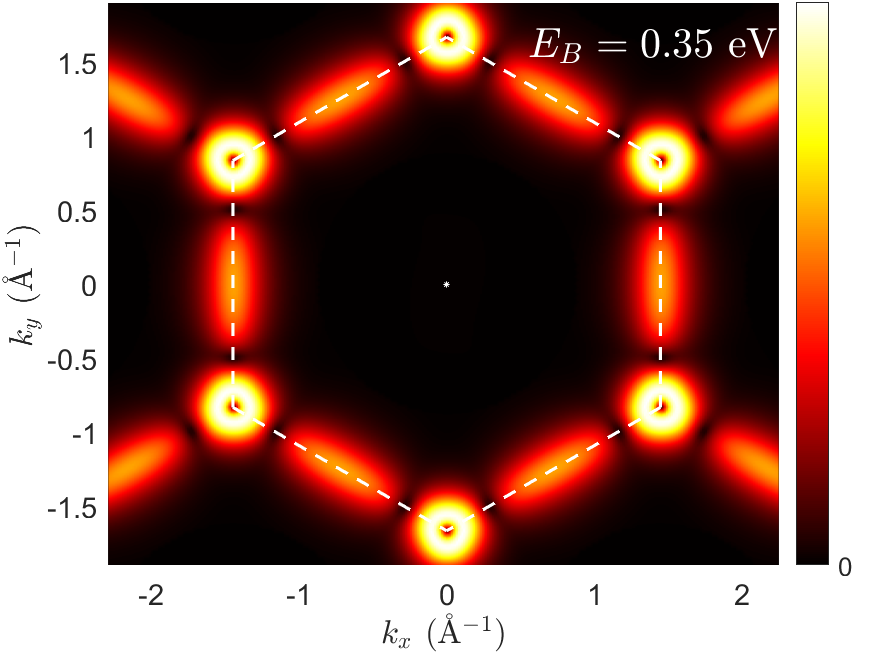}
    \caption{}
    \label{FigHBN1F}
\end{subfigure}
\begin{subfigure}{0.22\textwidth}
    \includegraphics[width=\textwidth]{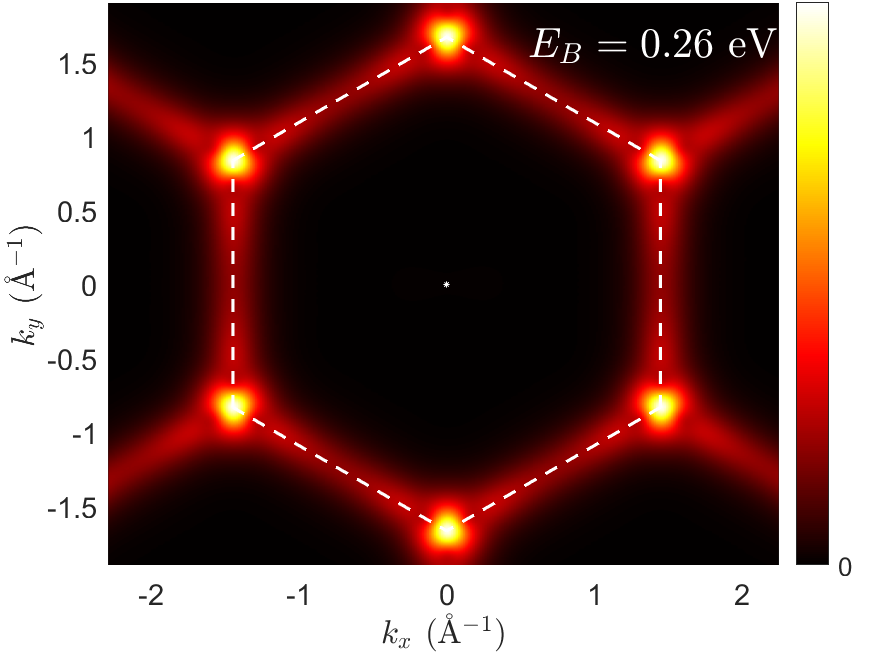}
    \caption{}
    \label{FigHBN1G}
\end{subfigure}
\begin{subfigure}{0.22\textwidth}
    \includegraphics[width=\textwidth]{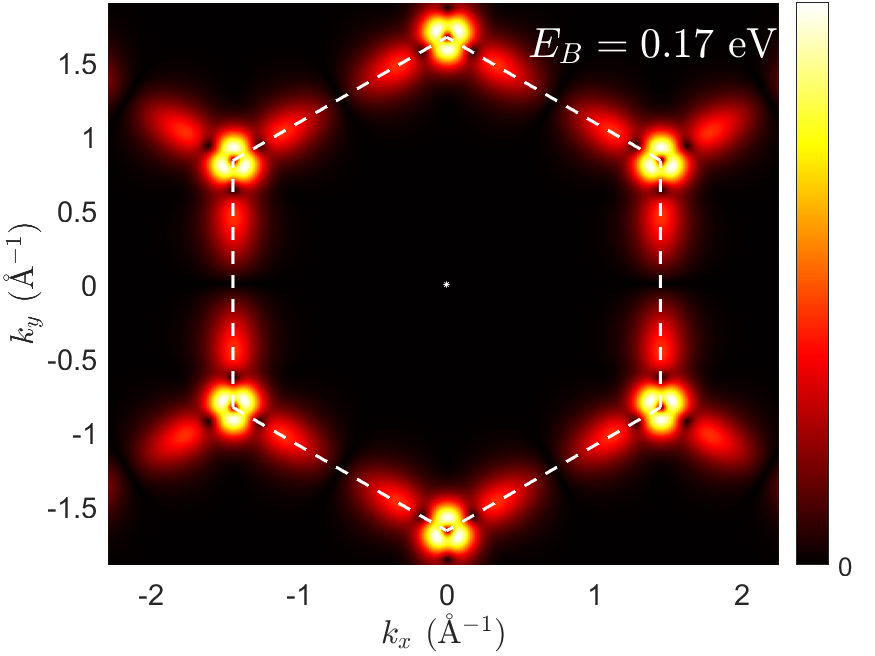}
    \caption{}
    \label{FigHBN1H}
\end{subfigure}
\begin{subfigure}{0.22\textwidth}
    \includegraphics[width=\textwidth]{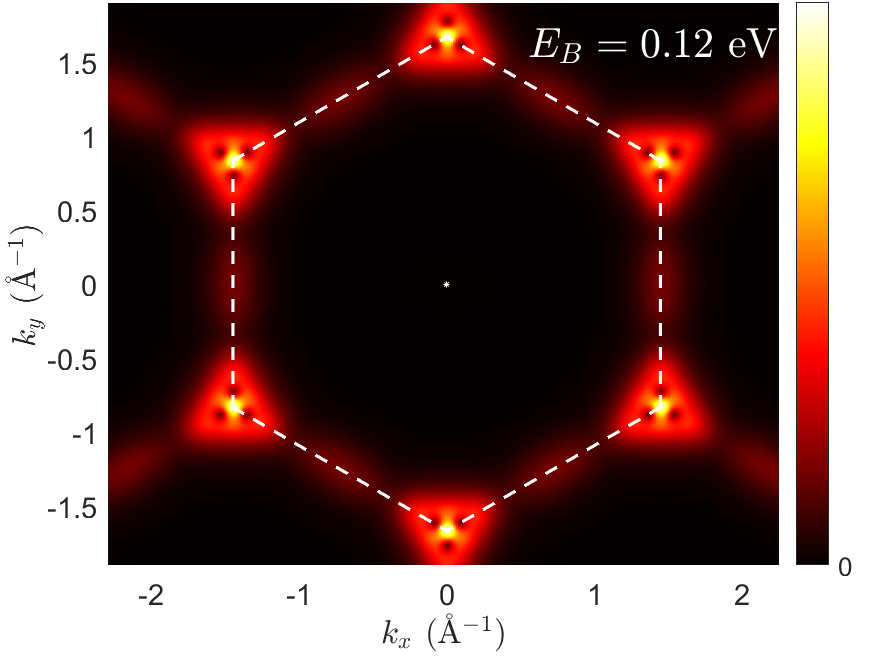}
    \caption{}
    \label{FigHBN1I}
\end{subfigure}
\caption{\label{FigHBN1} (a) Absorption spectrum of a single-layer of hBN, with (solid lines) and without (dashed colour lines) electron-hole correlations, for different in-plane light polarizations (all equal by symmetry) and with a Gaussian broadening of $100$ meV. The single-particle gap is indicated by the black dashed line. (b-i) Amplitude $(\sum_{\chi}\sum_{v,c}\vert A_{\chi}^{v\bm{k},c\bm{k}}\vert^{2})^{1/2}$ of the exciton coefficients for the first eight singlet states, where the $\chi-$sum spans the corresponding (possibly degenerated) subspace and whose binding energies are indicated. In (b) the dashed rings around $K,K'$ of radii $2\pi/3a$ are plotted for comparison with Ref. \cite{LouieHBN}. The three peaks in the excitonic absorption emerge from states (b), (c), (e), with a dark exciton (d) in between. All the states are twofold degenerated, except for (f) and (h). The BZ is delimited by the white dashed lines, and the white star indicates the $\Gamma$ point. $\bm{k}-$grid: $102\times 102$ for both the BSE and the absorption calculations, overlap metric for Gaussian density fitting.}
\end{figure*}

As a second example of a 2D material (relying also on the Parry potential function) we consider a single-layer of hBN, a planar hexagonal lattice with one $\text{B}$ and one $\text{N}$ per unit cell. The corresponding space group is $P\overline{6}m2$ with a twofold rotation symmetry along the $x$ direction, hence all exciton states will exhibit one or twofold degeneracy as imposed by $D_{3h}$, the little group at $\bm{Q}=\bm{0}$. A hybrid HF/DFT calculation was performed employing the unmodified def2-TZVP basis \cite{BasisExchange,def2_TZVP}, $217$ points in the irreducible BZ and the HSE06 functional \cite{HSE06}, with a resultant gap direct gap of $6.08$ eV at $K$ which is $\sim25\%$ lower than the calculated GW result at this $\bm{k}-$point \cite{GuoHBN,LouieHBN}. Since the binding energies are dependent on the quasi-particle gap, for consistency we apply a scissor correction of $1.95$ eV to the conduction bands energies. Regarding the auxiliary basis set, we employ a modified def2-QZVP-RIFIT \cite{def2_QZVP_RIFIT} with an additional diffuse $s$, $p$ and $d-$type GTF per chemical species and removing the $h-$type functions, resulting in a dimension $\sim4.5$ times larger than the original one (see the Suplemental Material). The exciton basis in this case is built from 2 valence bands, in order to account for the twofold degeneracy at $\Gamma$ of the highest occupied state, 1 conduction band and a $102\times102$ grid of $\bm{k}-$points, while the polarizability grid was reduced to $9\times9$. For the RI we only consider the overlap metric in this case.

From the BSE we find a binding energy of $E_{B}=1.88$ eV for the first exciton, in between the  $1.81$ and $2.08$ eV values computed in Refs. \cite{GuoHBN} and \cite{LouieHBN} respectively. This doubly-degenerated state is depicted in Figure \ref{FigHBN1B}, exhibiting a pronounced trigonal warping (deviation from circular symmetry to a triangular shape) and a remarkable degree of delocalization from the band edge at $K,K'$ into the $MK,M'K'$ and symmetry-related segments, which hints to a strong localization in real space. The main, ultraviolet absorption peak in Figure \ref{FigHBN1A} emerges exclusively from this first exciton. Analogously, the second and fourth excitons depicted respectively in Figures \ref{FigHBN1C} and \ref{FigHBN1E} are the dominant source of the second and third absorption peaks, with a dark state (Figure \ref{FigHBN1D}) intercalated. All these features are reproduced in Ref. \cite{LouieHBN}, notably the exciton amplitudes in reciprocal space (see Figure 2a, 2c and 3 therein for comparison with the present first four excitons, where we note than the degree of delocalization is also very similar).

\subsection{Solid Argon (3D)}\label{ArSec}
\begin{figure*}
\centering 
\begin{minipage}{0.47\textwidth}
\vspace{-0.65cm}
\begin{subfigure}{\textwidth}
    \includegraphics[width=\textwidth]{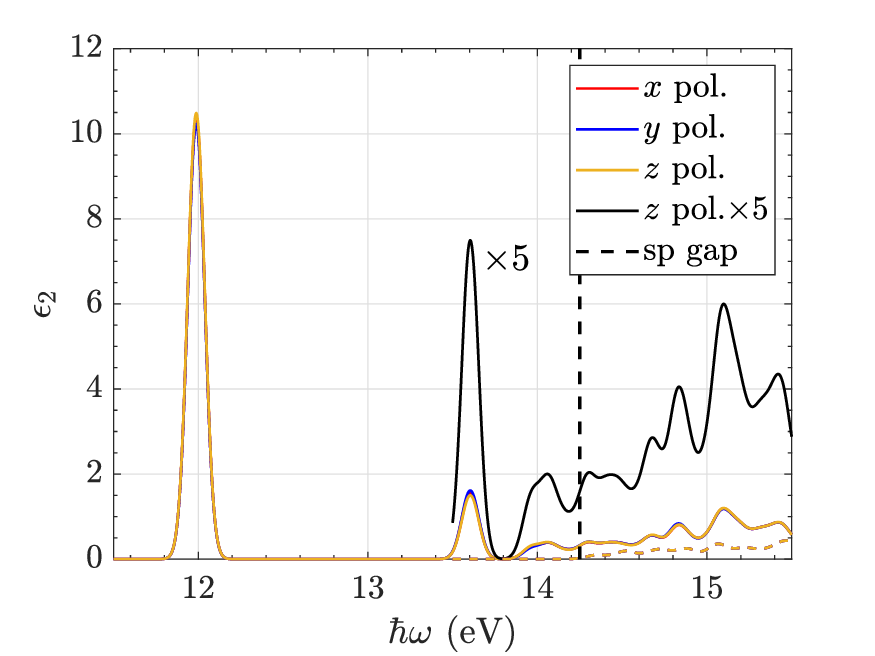}
    \caption{}
    \label{FigAr1A}
\end{subfigure}
\end{minipage} 
\begin{minipage}{0.47\textwidth}
\begin{subfigure}{0.42\textwidth}
    \includegraphics[width=\textwidth]{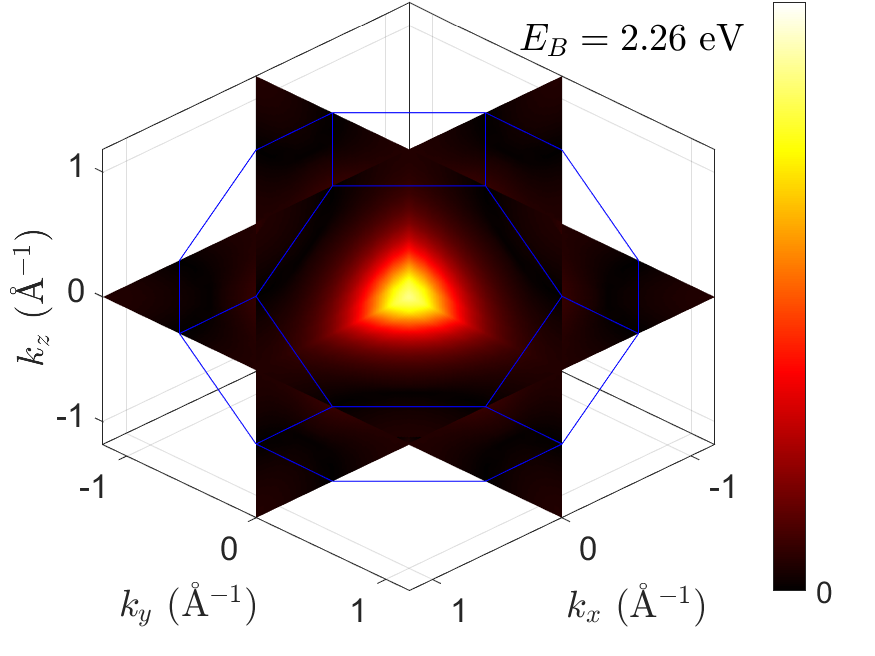}
    \caption{}
    \label{FigAr1B}
\end{subfigure}
\begin{subfigure}{0.42\textwidth}
    \includegraphics[width=\textwidth]{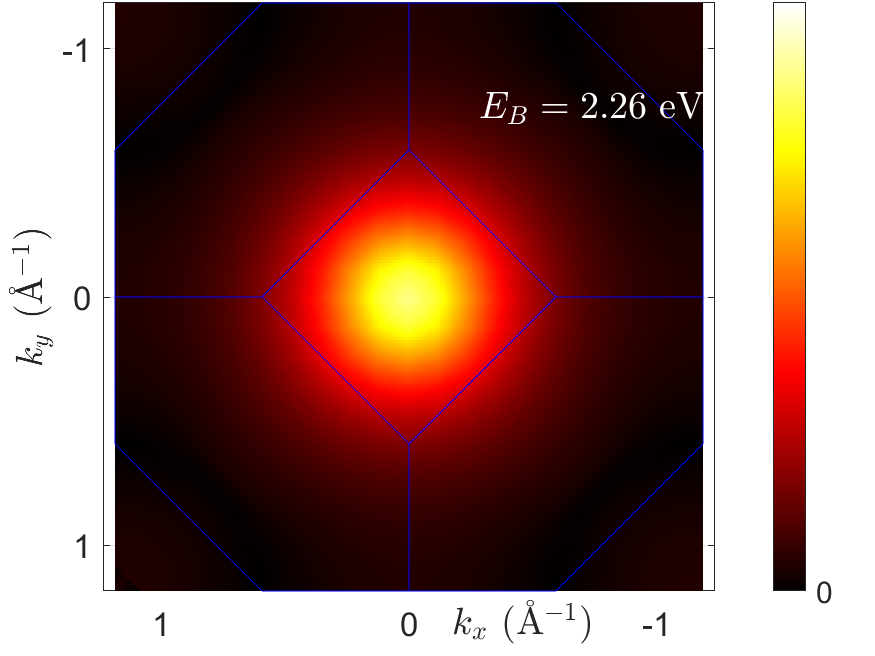}
    \caption{}
    \label{FigAr1C}
\end{subfigure}
\begin{subfigure}{0.42\textwidth}
    \includegraphics[width=\textwidth]{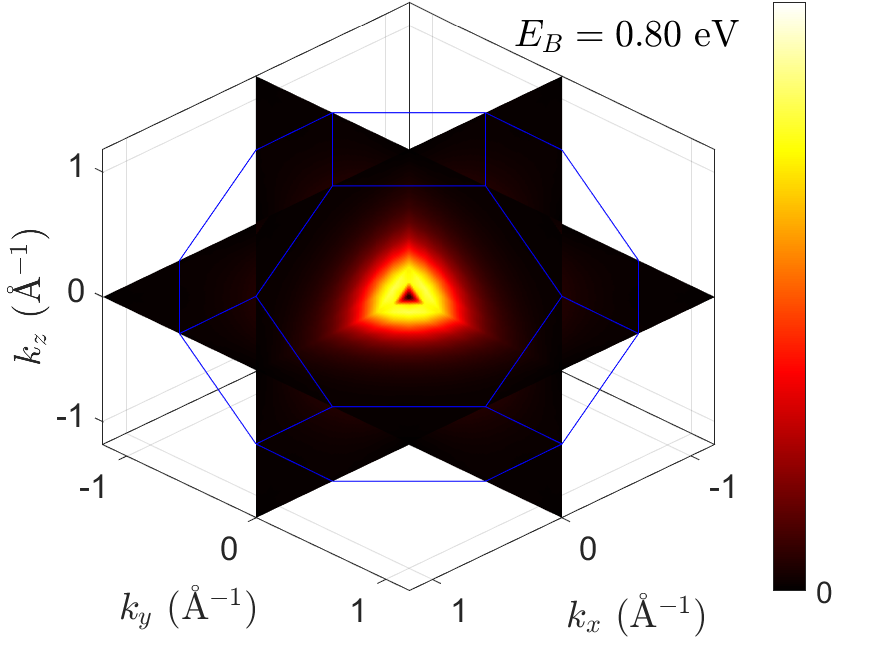}
    \caption{}
    \label{FigAr1D}
\end{subfigure}
\begin{subfigure}{0.42\textwidth}
    \includegraphics[width=\textwidth]{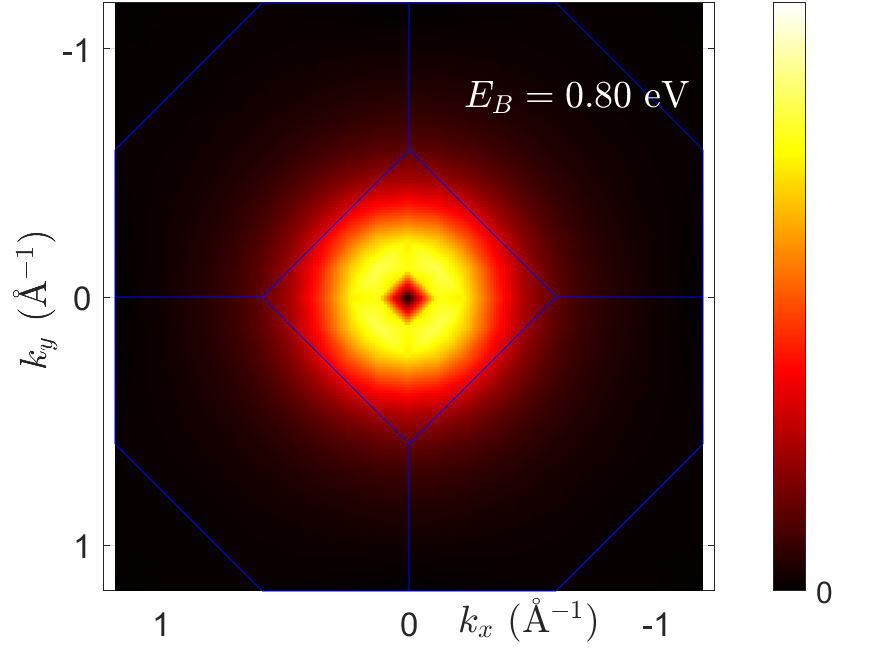}
    \caption{}
    \label{FigAr1E}
\end{subfigure}
\end{minipage}
\begin{subfigure}{0.22\textwidth}
    \includegraphics[width=\textwidth]{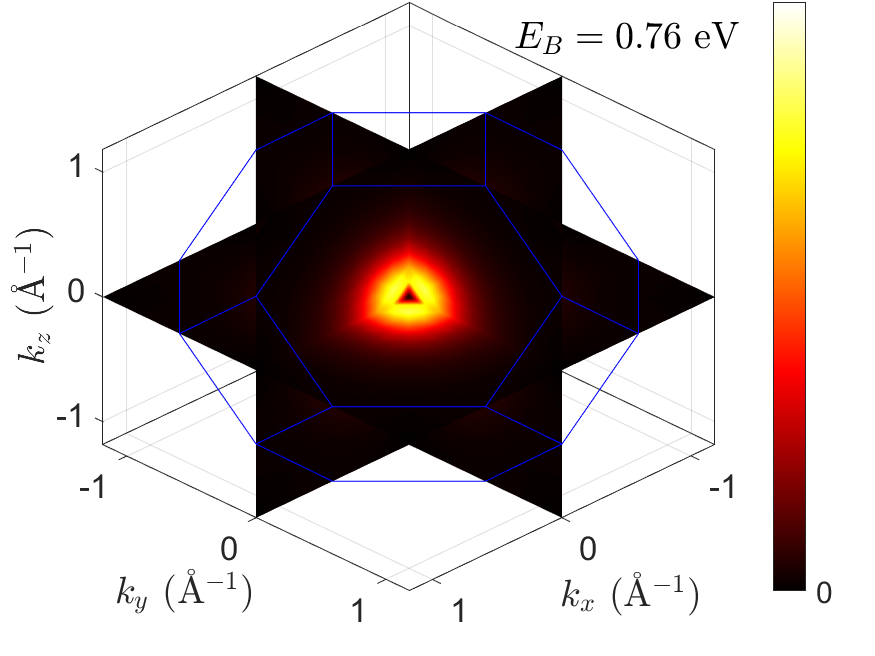}
    \caption{}
    \label{FigAr1F}
\end{subfigure}
\begin{subfigure}{0.22\textwidth}
    \includegraphics[width=\textwidth]{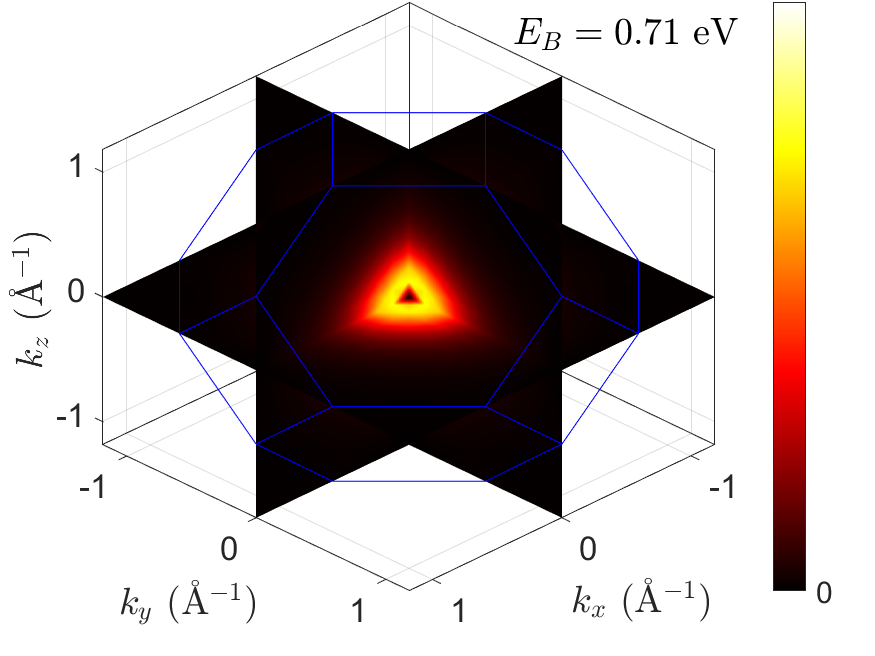}
    \caption{}
    \label{FigAr1G}
\end{subfigure}
\begin{subfigure}{0.22\textwidth}
    \includegraphics[width=\textwidth]{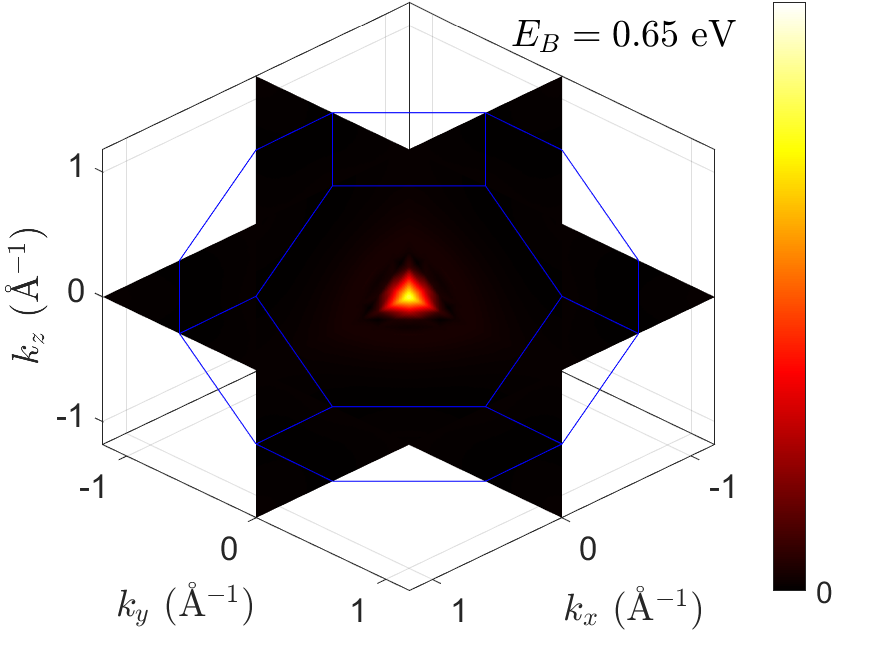}
    \caption{}
    \label{FigAr1H}
\end{subfigure}
\begin{subfigure}{0.22\textwidth}
    \includegraphics[width=\textwidth]{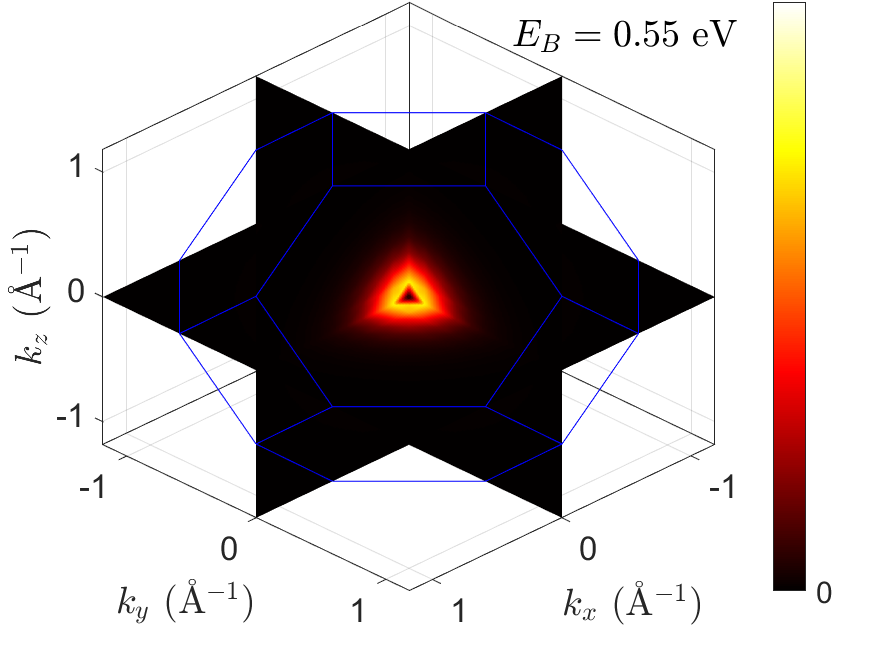}
    \caption{}
    \label{FigAr1I}
\end{subfigure}
\caption{\label{FigAr1} (a) Absorption spectrum of FCC argon, with (solid lines) and without (dashed colour lines) electron-hole correlations, for different light polarizations (all equal by symmetry) and with a Gaussian broadening of $50$ meV. The single-particle gap is indicated by the black dashed line. The amplified $\times5$ curve is included for a direct comparison with previous experimental and computational results, see \cite{PattersonAR}. (b-i) Amplitude $(\sum_{\chi}\sum_{v,c}\vert A_{\chi}^{v\bm{k},c\bm{k}}\vert^{2})^{1/2}$ of the exciton coefficients for the first six singlet states, where the $\chi-$sum spans the corresponding (possibly degenerated) subspace and whose binding energies are indicated, projected in the three orthogonal cartesian planes. (b,c) and (d,e) correspond to different perspectives of the same exciton, the first and second respectively.
The two main peaks in the excitonic absorption emerge from states (b) and (h), the other displayed excitons being dark. The BZ is indicated by solid blue lines. $\bm{k}-$grid: $20\times20\times20$ for both the BSE and the absorption calculations, overlap metric for Gaussian density fitting.}
\end{figure*}

In order to illustrate the 3D periodicity we consider solid argon, a stable FCC structure (symmorphic) with one $\text{Ar}$ atom per unit cell that is energetically favorable at relatively low temperatures \cite{RareGasSolids}. In this case we carried out a hybrid HF/DFT calculation employing the def2-TZVPPD basis \cite{def2_TZVPPD} with a slightly less diffuse lowest exponent in the $p-$type (see the Supplemental Material), $2769$ points in the irreducible BZ and the HSE06 functional, with a resultant direct gap of of $10.39$ eV at $\Gamma$ which is $\sim27\%$ lower than the estimated experimental value \cite{ARexp1962,ARexp1969} in the extreme ultraviolet. A scissor correction of $3.86$ eV is thus applied to the single-particle conduction energies. As an auxiliary basis set we employ an enlarged def2-TZVPPD-RIFIT \cite{def2_TZVPPD_RIFIT} including product functions (as done for black phosphorus) in addition to some more diffuse exponents, with a total dimension of $\sim4.9$ times that of the HF/DFT basis. The exciton basis is built from 3 valence bands, in order to properly include the threefold degenerated (highest occupied) state at $\Gamma$, 1 conduction band and a $20\times20\times20$ grid of $\bm{k}-$points, while the polarizability grid was reduced to $6\times6\times6$. For the RI we employ the overlap metric. 

Solving the BSE we obtain a binding energy of $E_{B}=2.26$ eV for the first exciton, in decent agreement with the experimentally estimated value of $2.1$ eV \cite{ARexp1969}. This many-body state is the only source of the main absorption peak, see Figure \ref{FigAr1A}, and as can be seen from Figures \ref{FigAr1B} and \ref{FigAr1C} it exhibits an spherical $s-$like symmetry in reciprocal space. The second peak in the absorption is found $\sim1.6$ eV higher with $\sim1/7$ of intensity, and it is associated with another $s-$like exciton depicted in Figure \ref{FigAr1H}. Both excitons are triply degenerated (as allowed by $O_{h}$, the little group at $\bm{Q}=\bm{0}$), and between them we find three dark excitons with cubic symmetry and a vanishing amplitude at $\Gamma$ displayed in Figures \ref{FigAr1D}-\ref{FigAr1G}. In order to compare with experimental absorption measurements, it should be noted that SOC is not entirely negligible and plays a qualitative role in this system, namely splitting the main absorption peak due to the single-particle splitting of the threefold (sixfold distinguishing spins) degenerated highest occupied state into one fourfold and one twofold states. Performing a relativistic calculation with \texttt{CRYSTAL} \cite{DesmaraisSOC1,DesmaraisSOC2} we found a valence splitting of $\sim0.35$ eV at $\Gamma$ (see the Supplemental Material), which is larger than the singlet-triplet splitting of $\sim0.2$ eV for the first exciton (see Table \ref{TableArExc}). Following Ref. \cite{BSE_SOC} this suggests that a perturbative correction could be applied to the exciton energies resulting from a configuration with half the exchange than the singlet, nonetheless we leave spin treatments for a future study. In any case, taking this into account it can be checked that the absorption spectrum is in good agreement with previous experimental \cite{ARexp1962,ARexp1969} and computational \cite{ARcalc2007} results, in particular we underline Ref. \cite{PattersonAR} (see Figure 2 therein) which includes an explicit comparison with the measurements in Ref. \cite{RareGasSolids} and can be directly compared with our Figure \ref{FigAr1A}. 

To complete the 3D case we additionally considered the attenuated Coulomb metric, although with a reduced auxiliary basis consisting of the regular def2-TZVPPD-RIFIT with just one added $p-$type GTF with a diffuse $\alpha=0.05$ a.u. exponent. The dimension is increased by $\sim2.8$ with respect to the HF/DFT basis. As in the case of black phosphorus we found almost identical results between both metrics, with energy differences of the order of meV (see Table \ref{TableArExc}) and virtually unaltered absorption and reciprocal space exciton textures. 

\begin{table}[h!]\caption{\label{TableArExc} First six exciton energies $E_{\chi}$ (eV) for two different metrics and spin blocks. $d$ indicates the degeneracy of each state. The first row indicates the value of the parameter $\omega$ (a.u. or Bohr$^{-1}$) in the attenuated Coulomb metric, the $\infty$ case actually corresponding to the overlap metric. The auxiliary basis in the overlap case is larger, see main text.}
\centering\begin{tabular}{| c || c | c | c | c |} 
\hline
$\omega$ & $\infty$ (overlap) & $\infty$ & $0.5$ & $0.5$  \\[0.05cm] \hline
& Triplet & Singlet & Singlet & Triplet  \\[0.05cm] \hhline{|=|=|=|=|=|}
$E_{1}$ ($d=3$) & 11.785 & 11.988 & 11.975 & 11.772 \\[0.08cm] \hline
$E_{2}$ ($d=3$) & 13.445 & 13.445 & 13.438 & 13.438 \\[0.08cm] \hline
$E_{3}$ ($d=2$) & 13.471 & 13.488 & 13.480 & 13.464 \\[0.08cm] \hline
$E_{4}$ ($d=3$) & 13.523 & 13.539 & 13.532 & 13.516 \\[0.08cm] \hline
$E_{5}$ ($d=3$) & 13.580 & 13.604 & 13.598 & 13.574 \\[0.08cm] \hline
$E_{6}$ ($d=1$) & 13.594 & 13.696 & 13.690 & 13.587 \\ \hline
\end{tabular}\end{table}

\section{Conclusion}\label{ConclusionSec}
We have presented a theoretical and computational method to solve the BSE from first principles employing a pure Gaussian basis set, based on the density fitting technique (or RI) to reduce the scaling on the single-particle dimension and in combination with Ewald potential functions to automatically turn the arising, conditionally convergent Coulomb series into exponentially convergent for any lattice dimensionality. This is at the cost of computing the Ewald-type integrals in a direct space supercell given by the BSE $\bm{k}-$grid, which does not represent a serious obstacle since these are two-center integrals independent from the more computationally demanding three-center counterparts, treated as regular lattice series. An important advantage of this method is the avoidance of the lattice replication for periodicity in less than 3 dimensions and its associated range-truncation of the Coulomb interaction. 

Two metric options for the RI have been considered, the overlap and the attenuated Coulomb with different attenuation parameters, the latter being in principle more precise at the cost of increased memory requirements. For the considered configurations in both 2D and 3D, we found negligible differences between the results provided by both metrics, with changes in exciton energies of the order of meV. This, however, may not be a global tendency and it is most likely dependent on the quality of the auxiliary basis for the fitting. In this work we have employed relatively large such bases, possibly more than actually needed, to ensure stability in the first set of calculations, hence further studies on the optimization of the basis size for both metrics should be carried out, in particular to treat larger unit cells.   

It is important to note that the single-particle spectra employed in the calculations of this work have been obtained directly from hybrid HF/DFT. In this regard, further studies on the impact of different hybrid functionals in materials with different bonding characters would be convenient. On the other hand, the neglect of GW may seem troublesome since it is assumed to formally derive the interaction kernel \eqref{kernel}, and it could indeed cause some numerical deviations in our results. Nevertheless, the standard use of GW is typically restricted to the correction of the single-particle energies only, something that can be emulated to some extent with the combination of a hybrid functional (introducing some effective mass renormalization with respect to pure DFT) and, if necessary, a rigid shift in the unoccupied states to match a desired band gap. It should be noted that since the pioneering work of Rohlfing, Krüger and Pollmann \cite{rohlfing1993}, in recent years there has been remarkable progress in the GTF-based GW formulations for periodic systems (also employing the RI approximation) \cite{RIG0W0,Wilhelm2021}, and they are now accessible in some codes \cite{PySCFGW2022,CP2KGW}. Regardless, as shown in Section \ref{ApplicationsSec} our results are consistent and in overall good agreement with those in the computational and experimental literature for the present selection of materials. 

%In this work we have not considered SOC explicitly, hence the selection of materials with relatively light elements where this effect is not strong.  

\appendix 
\section{Derivation of the screening matrix elements \eqref{ScreenedInt}}\label{ScreeningApp}
The static RPA polarizability \eqref{P0} in a periodic system reads
\begin{equation}\label{P0solid}\begin{aligned}
&P_{0}(\bm{x}_{1},\bm{x}_{2};0)=\frac{N_{\bm{k}}}{N_{\tilde{\bm{k}}}}\sum_{\tilde{\bm{k}},\tilde{\bm{k}}'}\sum_{\tilde{v},\tilde{c}}\frac{1}{\varepsilon_{\tilde{v}\tilde{\bm{k}}}-\varepsilon_{\tilde{c}\tilde{\bm{k}}'}}\cdot \\
&\left[\psi^{*}_{\tilde{v}\tilde{\bm{k}}}(\bm{x}_{1})\psi_{\tilde{c}\tilde{\bm{k}}'}(\bm{x}_{1})\psi^{*}_{\tilde{c}\tilde{\bm{k}}'}(\bm{x}_{2})\psi_{\tilde{v}\tilde{\bm{k}}}(\bm{x}_{2})+ \right. \\
&\left.
\psi^{*}_{\tilde{c}\tilde{\bm{k}}'}(\bm{x}_{1})\psi_{\tilde{v}\tilde{\bm{k}}}(\bm{x}_{1})\psi^{*}_{\tilde{v}\tilde{\bm{k}}}(\bm{x}_{2})\psi_{\tilde{c}\tilde{\bm{k}}'}(\bm{x}_{2})\right]
\end{aligned}\end{equation}
where in practice $\tilde{\bm{k}},\tilde{\bm{k}}'$ span a different grid than $\bm{k},\bm{k}'$ in the BSE Hamiltonian. It is widely found that values $N_{\tilde{\bm{k}}}\ll N_{\bm{k}}$ are enough to yield good convergence in the irreducible polarizability \eqref{Pimatrix}. On the other hand, $\tilde{v},\tilde{c}$ span the set of occupied and unoccupied single-particle states, respectively. In a Gaussian basis the dimension is relatively small, hence we include all bands given by the original $\set{\varphi_{\mu}}$ basis in the $\tilde{v},\tilde{c}$ sums, as advocated in \cite{RIG0W0}, which should account for a completeness relation. 

Employing \eqref{P0solid} in the series expansion of \eqref{WDyson}, recalling the notation of \eqref{CoulombInt} and the selection rule therein for the crystal momenta, and noting that $J^{1/2}_{R,Q-\bm{k}}=J^{1/2}_{Q,R\bm{k}}$, it follows that
\begin{equation}\begin{aligned}\label{AppW}
&\iint\psi^{*}_{m\bm{k}}(\bm{r})\psi_{n'\bm{k}'}(\bm{r})W(\bm{r},\bm{r}')\psi_{n\bm{k}+\bm{Q}}(\bm{r}')\psi^{*}_{m'\bm{k}'+\bm{Q}}(\bm{r}')d\bm{r}d\bm{r}'\\
&=(m\bm{k},n'\bm{k}'\vert n\bm{k}+\bm{Q},m'\bm{k}'+\bm{Q})+\frac{N_{\bm{k}}}{N_{\tilde{\bm{k}}}}\sum_{\tilde{\bm{k}},\tilde{\bm{k}}'}\sum_{\tilde{v},\tilde{c}} \frac{1}{\varepsilon_{\tilde{v}\tilde{\bm{k}}}-\varepsilon_{\tilde{c}\tilde{\bm{k}}'}}\cdot \\
&\left[ (m\bm{k},n'\bm{k}'\vert \tilde{c}\tilde{\bm{k}}',\tilde{v}\tilde{\bm{k}})(\tilde{c}\tilde{\bm{k}}',\tilde{v}\tilde{\bm{k}}\vert n\bm{k}+\bm{Q},m'\bm{k}'+\bm{Q})+ \right. \\
&\left. (m\bm{k},n'\bm{k}'\vert \tilde{v}\tilde{\bm{k}},\tilde{c}\tilde{\bm{k}}')(\tilde{v}\tilde{\bm{k}},\tilde{c}\tilde{\bm{k}}'\vert n\bm{k}+\bm{Q},m'\bm{k}'+\bm{Q}) \right] +\dots \\
&=\sum_{R}v_{R}^{m\bm{k},n'\bm{k}'}v_{R}^{m'\bm{k}'+\bm{Q},n\bm{k}+\bm{Q}}+\sum_{R,R'}v_{R}^{m\bm{k},n'\bm{k}'}\left[\frac{N_{\bm{k}}}{N_{\tilde{\bm{k}}}}\sum_{\tilde{\bm{k}}}\sum_{\tilde{v},\tilde{c}} \right. \\
&\left. \frac{v_{R}^{\tilde{v}\tilde{\bm{k}},\tilde{c}\tilde{\bm{k}}+\bm{k}-\bm{k}'}v_{R'}^{\tilde{c}\tilde{\bm{k}}+\bm{k}-\bm{k}',\tilde{v}\tilde{\bm{k}}}}{\varepsilon_{\tilde{v}\tilde{\bm{k}}}-\varepsilon_{\tilde{c}\tilde{\bm{k}}+\bm{k}-\bm{k}'}} + 
\frac{v_{R}^{\tilde{c}\tilde{\bm{k}}-(\bm{k}-\bm{k}'),\tilde{v}\tilde{\bm{k}}}v_{R'}^{\tilde{v}\tilde{\bm{k}},\tilde{c}\tilde{\bm{k}}-(\bm{k}-\bm{k}')}}{\varepsilon_{\tilde{v}\tilde{\bm{k}}}-\varepsilon_{\tilde{c}\tilde{\bm{k}}-(\bm{k}-\bm{k}')}}
\right]\cdot \\[0.1cm]
&v_{R'}^{m'\bm{k}'+\bm{Q},n\bm{k}+\bm{Q}}+\dots= 
\sum_{R}v_{R}^{m\bm{k},n'\bm{k}'}v_{R}^{m'\bm{k}'+\bm{Q},n\bm{k}+\bm{Q}}+ \\ 
&\sum_{R,R'}v_{R}^{m\bm{k},n'\bm{k}'}\left[
\sum_{p=1}^{\infty}\Pi^{p}_{\bm{k}-\bm{k}'}
\right]_{R,R'}v_{R'}^{m'\bm{k}'+\bm{Q},n\bm{k}+\bm{Q}}= \\
&\sum_{R,R'}v_{R}^{m\bm{k},n'\bm{k}'}\left[\sum_{p=0}^{\infty}\Pi^{p}_{\bm{k}-\bm{k}'}\right]_{R,R'}v_{R'}^{m'\bm{k}'+\bm{Q},n\bm{k}+\bm{Q}}
\end{aligned}\end{equation}
where by employing \eqref{vsymm} we have introduced the following matrix in the auxiliary basis, representing the irreducible polarizability
\begin{equation}\begin{aligned}\label{AppPi}
&\Pi_{R,R'\bm{k}}\equiv\\
&\frac{N_{\bm{k}}}{N_{\tilde{\bm{k}}}}\sum_{\tilde{\bm{k}}}\sum_{\tilde{v},\tilde{c}} \left[\frac{v_{R}^{\tilde{v}\tilde{\bm{k}},\tilde{c}\tilde{\bm{k}}+\bm{k}}\left(v_{R'}^{\tilde{v}\tilde{\bm{k}},\tilde{c}\tilde{\bm{k}}+\bm{k}}\right)^{*}}{\varepsilon_{\tilde{v}\tilde{\bm{k}}}-\varepsilon_{\tilde{c}\tilde{\bm{k}}+\bm{k}}} + (\bm{k}\leftrightarrow-\bm{k})^{*} \right] 
\end{aligned}\end{equation}
The second term inside the brackets represents the conjugate of the first term evaluated at $-\bm{k}$ instead of $\bm{k}$. Note that $\Pi_{R,R'\bm{k}}$ is hermitian and $\Pi_{R,R'-\bm{k}}=\Pi_{R,R'\bm{k}}^{*}$. In the presence of inversion or time-reversal symmetry (irrespective of whether their product is a symmetry) or a negligible SOC and other hypothetical complex terms in the Hamiltonian, the energy spectrum is centrosymmetric and $c^{-\bm{k}}_{\mu n}=e^{i\theta_{n\bm{k}}}c^{\bm{k}*}_{\mu n}$ with an arbitrary phase factor and with a spin reversal in the presence of both SOC and time-reversal (both of which factors are unimportant here). Then, assuming a centrosymmetric $\tilde{\bm{k}}-$grid e.g. as in \eqref{MPgrid}, the second term is equal to the first one and \eqref{AppPi} reduces to \eqref{Pimatrix}. Under this assumption, an analogous result was found in Refs. \cite{Scheffler2012,Wilhelm2016,RIG0W0} in the context of GW.

In practice, we sum the geometric series in \eqref{AppW} as indicated in \eqref{ScreenedInt}. Note that this is valid if every eigenvalue of $\Pi_{\bm{k}-\bm{k}'}$ is smaller than $1$ in absolute value\footnote{Actually, larger than $-1$ since the irreducible polarizability matrix is negative definite}, otherwise the series does not converge. However, the origin of this divergence would ultimately lie (assuming proper convergence in the other parameters) in the series defined by the Hedin's equation \eqref{WDyson}, which is implicitly assumed to be convergent in the GW-BSE perturbative framework.

\section{From integrals over GTF to integrals over Hermite Gaussians (2 and 3$-$center)}\label{GaussianApp}
The GTF $\varphi_{\mu}$ (and analogously $\phi_{P}$) employed in Section \ref{GaussianRIsec} are parametrized in terms of the multi-index $\mu=\set{a,\lambda,l,m,s}$ labelling the atoms in the unit cell ($a$), the shell of basis functions ($\lambda=\lambda(a)$) defined by a set of $\text{NG}(\lambda)$ pairs of exponents $\alpha_{h(\lambda)}>0$ and contraction coefficients $d_{h(\lambda)}\in\mathbb{R}$, the orbital angular momentum quantum numbers ($l=l(\lambda)\in\mathbb{N}\cup\set{0}$, $m=m(l)\in\set{-l,\dots,l}$) and the spin quantum number with respect to a given axis ($s\in\set{\uparrow,\downarrow}$). The GTF can be expressed in terms of cartesian Gaussians, whence
\begin{equation}\begin{aligned}\label{GTFdef}
&\varphi_{\mu}^{\bm{R}}(\bm{r})=\mathcal{N}(l,m,\lambda)\sum_{h(\lambda)=1}^{\text{NG}(\lambda)}d_{h(\lambda)}\alpha_{h(\lambda)}^{\frac{1}{2}(l+\frac{3}{2})}\cdot\\
&\sum_{i,j,k=0}^{l}\mathfrak{g}^{l,m}_{i,j,k}G_{i,j,k}(\alpha_{h(\lambda)}, \bm{r}-(\bm{t}_{a}+\bm{R}))
\end{aligned}\end{equation}
where $\mathcal{N}(l,m,\lambda)>0$ is a normalization factor, $\bm{t}_{a}$ is the coordinates vector of atom $a$ in the unit cell, and we have introduced the cartesian GTF
\begin{equation}\label{CartesianGTF}
G_{i,j,k}(\alpha,\bm{r})=x^{i}y^{j}z^{k}e^{-\alpha\:\abs{\bm{r}}^{2}} \:,\;\; \bm{r}=(x,y,z)
\end{equation}
and the tensor $\mathfrak{g}^{l,m}_{i,j,k}\in\mathbb{Z}$ relating the cartesian $G_{i,j,k}$ and the real-valued solid-spherical-harmonic $G_{l,m}$ (see page 27 in Ref. \cite{dovesicrystal23})
\begin{equation*}
G_{l,m}(\alpha,\bm{r})=\sum_{i,j,k=0}^{l}\mathfrak{g}^{l,m}_{i,j,k}G_{i,j,k}(\alpha,\bm{r})
\end{equation*}

The integrals in Section \ref{GaussianIntSec} are computed by expanding the cartesian GTF \eqref{CartesianGTF} in terms of Hermite Gaussians \eqref{HermiteGaussian}, namely 
\begin{equation}\begin{aligned}\label{HermiteG2}
&G_{i,j,k}(\alpha,\bm{r}-\bm{A})G_{i',j',k'}(\beta,\bm{r}-\bm{B})=\\[0.2cm]
&\sum_{t=0}^{i+i'}E^{i,i'}_{t}(p,\text{P}A_{x},\text{P}B_{x})\sum_{u=0}^{j+j'}E^{j,j'}_{u}(p,\text{P}A_{y},\text{P}B_{y})\cdot\\[0.2cm]
&\sum_{v=0}^{k+k'}E^{k,k'}_{v}(p,\text{P}A_{z},\text{P}B_{z})\Lambda_{t,u,v}(\bm{r},p,\mathbf{P})
\end{aligned}\end{equation}
where $p=\alpha+\beta$, $\mathbf{P}=\frac{\alpha\bm{A}+\beta\bm{B}}{\alpha+\beta}$, $\mathbf{P}-\bm{A}=(\text{P}A_{x},\text{P}A_{y},\text{P}A_{z})$ (likewise for $\text{P}B$) and the $E^{a,b}_{s}$ coefficients can be determined by the McMurchie-Davidson recurrence relations \cite{McMurchie1978}. In practice, we symbolically obtain an explicit expression for all coefficients required by the maximum angular momentum included in the auxiliary basis. In particular, for a single cartesian GTF
\begin{equation}\begin{aligned}\label{HermiteG1}
&G_{i,j,k}(\alpha,\bm{r}-\bm{A})=G_{i,j,k}(\alpha,\bm{r}-\bm{A})G_{0,0,0}(0,\bm{r})=\\[0.2cm]
&\sum_{t=0}^{i}E^{i,0}_{t}(\alpha,0,A_{x})\sum_{u=0}^{j}E^{j,0}_{u}(\alpha,0,A_{y})
\sum_{v=0}^{k}E^{k,0}_{v}(\alpha,0,A_{z})\cdot\\[0.2cm]
&\Lambda_{t,u,v}(\bm{r},\alpha,\bm{A})
\end{aligned}
\end{equation}
where it can be shown that $E^{i,0}_{t}(\alpha,0,A_{x})$ is independent of $A_{x}$ (likewise for the other components). Inserting \eqref{GTFdef} and \eqref{HermiteG1} in a general 2-center matrix representing \eqref{MmatrixG}, \eqref{JmatrixG} or \eqref{JmatrixGEwald} and assigning $\alpha_{h},d_{h}$ ($\alpha_{h'},d_{h'}$) to $\phi_{P}$ ($\phi_{P'}$, respectively) one therefore gets
\begin{equation}\begin{aligned}\label{GTF2Hermite}
&\iint\phi_{P}^{\bm{0}}(\bm{r})\mathcal{O}(\bm{r}-\bm{r}')\phi_{P'}^{\bm{R}}(\bm{r}')d\bm{r}d\bm{r}'=\mathcal{N}(l,m)\mathcal{N}(l',m')\cdot \\[0.2cm]
&\sum_{h=1}^{\text{NG}}d_{h}\alpha_{h}^{\frac{1}{2}(l+\frac{3}{2})} 
\sum_{h'=1}^{\text{NG}'}d_{h'}\alpha_{h'}^{\frac{1}{2}(l'+\frac{3}{2})} 
\sum_{i,j,k=0}^{l}\mathfrak{g}^{l,m}_{i,j,k}\cdot\\[0.2cm]
&\sum_{i',j',k'=0}^{l'}\mathfrak{g}^{l',m'}_{i',j',k'}\sum_{t=0}^{i}E_{t}^{i,0}(\alpha_{h})
\sum_{u=0}^{j}E_{u}^{j,0}(\alpha_{h})
\sum_{v=0}^{k}E_{v}^{k,0}(\alpha_{h})\cdot \\[0.2cm]
&\sum_{t'=0}^{i'}E_{t'}^{i',0}(\alpha_{h'})
\sum_{u'=0}^{j'}E_{u'}^{j',0}(\alpha_{h'})
\sum_{v'=0}^{k'}E_{v'}^{k',0}(\alpha_{h'})\cdot\\[0.2cm]
&\iint\Lambda_{t,u,v}(\bm{r},\alpha_{h},\bm{t}_{a})\mathcal{O}(\bm{r}-\bm{r}')\Lambda_{t',u',v'}(\bm{r}',\beta_{h'},\bm{t}_{a'}+\bm{R})d\bm{r}d\bm{r}'
\end{aligned}\end{equation}
where the shell indices $\lambda,\lambda'$ have been omitted to alleviate the notation. 

Analogously, in the general 3-center case representing \eqref{LmatrixG} and assigning $\alpha_{h},d_{h}$ to $\phi_{P}$ and $\beta_{\overline{h}},e_{\overline{h}}$ ($\beta_{\overline{h}'},e_{\overline{h}'}$) to $\varphi_{\mu}$ ($\varphi_{\mu'}$, respectively),
\begin{equation}\begin{aligned}\label{GTF3Hermite}
&\iint\phi_{P}^{\bm{0}}(\bm{r})\mathcal{O}(\bm{r}-\bm{r}')\varphi_{\mu}^{\bm{R}}(\bm{r}')\varphi_{\mu'}^{\bm{R}'}(\bm{r}')d\bm{r}d\bm{r}'=\\[0.2cm]
&\mathcal{N}(l,m)\mathcal{N}(\overline{l},\overline{m})\mathcal{N}(\overline{l}',\overline{m}')
\sum_{h=1}^{\text{NG}}d_{h}\alpha_{h}^{\frac{1}{2}(l+\frac{3}{2})}
\sum_{\overline{h}=1}^{\text{NG}}e_{\overline{h}}\beta_{\overline{h}}^{\frac{1}{2}(\overline{l}+\frac{3}{2})}\cdot \\[0.2cm]
&\sum_{\overline{h}'=1}^{\text{NG}}e_{\overline{h}'}\beta_{\overline{h}'}^{\frac{1}{2}(\overline{l}'+\frac{3}{2})}
\sum_{i,j,k=0}^{l}\mathfrak{g}^{l,m}_{i,j,k}
\sum_{\overline{i},\overline{j},\overline{k}=0}^{\overline{l}}\mathfrak{g}^{\overline{l},\overline{m}}_{\overline{i},\overline{j},\overline{k}}
\sum_{\overline{i}',\overline{j}',\overline{k}'=0}^{\overline{l}'}\mathfrak{g}^{\overline{l}',\overline{m}'}_{\overline{i}',\overline{j}',\overline{k}'}\cdot\\[0.2cm]
&\sum_{t=0}^{i}E_{t}^{i,0}(\alpha_{h})
\sum_{u=0}^{j}E_{u}^{j,0}(\alpha_{h})
\sum_{v=0}^{k}E_{v}^{k,0}(\alpha_{h})\cdot\\[0.2cm]
&\sum_{\overline{t}=0}^{\overline{i}+\overline{i}'}E_{\overline{t}}^{\overline{i},\overline{i}'}(\beta_{\overline{h}}+\beta_{\overline{h}'},\text{P}^{(1)}_{x},\text{P}^{(2)}_{x})\sum_{\overline{u}=0}^{\overline{j}+\overline{j}'}E_{\overline{u}}^{\overline{j},\overline{j}'}(\beta_{\overline{h}}+\beta_{\overline{h}'},\text{P}^{(1)}_{y},\text{P}^{(2)}_{y}) \\[0.2cm]
&\cdot\sum_{\overline{v}=0}^{\overline{k}+\overline{k}'}E_{\overline{v}}^{\overline{k},\overline{k}'}(\beta_{\overline{h}}+\beta_{\overline{h}'},\text{P}^{(1)}_{z},\text{P}^{(2)}_{z})\cdot\\[0.2cm]
&\iint\Lambda_{t,u,v}(\bm{r},\alpha_{h},\bm{t}_{a})\mathcal{O}(\bm{r}-\bm{r}')\Lambda_{\overline{t},\overline{u},\overline{v}}(\bm{r}',\beta_{\overline{h}}+\beta_{\overline{h}'},\mathbf{P})d\bm{r}d\bm{r}'
\end{aligned}\end{equation}
where in this case
\begin{equation*}
\mathbf{P}=\frac{\beta_{\overline{h}}(\bm{t}_{\overline{a}}+\bm{R}) + \beta_{\overline{h}'}(\bm{t}_{\overline{a}'}+\bm{R}')}{\beta_{\overline{h}} + \beta_{\overline{h}'}}
\end{equation*}
\begin{equation*}
\mathbf{P}^{(1)}=\frac{\beta_{\overline{h}'}(\bm{t}_{\overline{a}'}-\bm{t}_{\overline{a}}+\bm{R}'-\bm{R})}{\beta_{\overline{h}} + \beta_{\overline{h}'}}
\end{equation*}
\begin{equation*}
\mathbf{P}^{(2)}=-\frac{\beta_{\overline{h}}(\bm{t}_{\overline{a}'}-\bm{t}_{\overline{a}}+\bm{R}'-\bm{R})}{\beta_{\overline{h}} + \beta_{\overline{h}'}}
\end{equation*}

In the limit case $\mathcal{O}(\bm{r}-\bm{r}')=\delta(\bm{r}-\bm{r}')$, the corresponding overlap integrals in \eqref{GTF2Hermite} can instead be computed by performing the expansion \eqref{HermiteG2} directly for the product $\phi^{\bm{0}}_{P}(\bm{r})\phi^{\bm{R}}_{P'}(\bm{r})$, and in \eqref{GTF3Hermite} by expanding $\Lambda_{\overline{t},\overline{u},\overline{v}}(\bm{r},\beta_{\overline{h}}+\beta_{\overline{h}'},\mathbf{P})$ back into cartesian GTF, and expanding again the resulting products with $\phi^{\bm{0}}_{P}(\bm{r})$ in Hermite Gaussians. In both cases, we employ the identity
\begin{equation}\label{HermiteInt0}
\int_{\mathbb{R}^{3}}\Lambda_{t,u,v}(\bm{r},p,\mathbf{P})d\bm{r}=\left(\frac{\pi}{p}\right)^{3/2}\delta_{t,0}\delta_{u,0}\delta_{v,0}
\end{equation}
which follows directly from the definition \eqref{HermiteGaussian}.

For Coulomb-related operators $\mathcal{O}$, such as the Ewald potential functions, the Hermite Gaussian integrals involve Hermite Coulomb integrals \eqref{HermiteCoulomb}. We compute them with explicit expressions obtained symbolically from the McMurchie-Davidson recursive scheme \cite{McMurchie1978}, which ultimately yield linear combinations of higher order Boys functions \cite{Helgaker}
\begin{equation}
F_{n}(x)=\left\{\begin{aligned}
&\frac{\gamma\left(n+\frac{1}{2},x\right)}{2x^{n+\frac{1}{2}}}\;, &\text{ if }x>0 \\[0.2cm]
&\frac{1}{2n+1}\;, &\text{ if }x=0
\end{aligned}\right.
\end{equation}
with $\gamma(n,x)=\int_{0}^{x}t^{n-1}e^{-t}dt$ the lower incomplete gamma function.

\section{Beyond the Tamm-Dancoff aproximation}\label{TDAApp}
For completeness, we explicitly include here the coupling and anti-resonant Hamiltonian blocks that are disregarded in the TDA. Recalling \eqref{Hcoupl}, we can write the remaining sections of \eqref{Qblock} as
\begin{equation}\begin{aligned}
&H^{\text{coupl}}(\bm{Q})_{v\bm{k},c'\bm{k}';c\bm{k},v'\bm{k}'}\equiv
H^{\text{coupl}}_{v\bm{k},c'\bm{k}';c\bm{k}+\bm{Q},v'\bm{k}'+\bm{Q}}=\\[0.2cm]
&\tilde{D}_{v\bm{k},c\bm{k}+\bm{Q}}^{c'\bm{k}',v'\bm{k}'+\bm{Q}}+\tilde{X}_{v\bm{k},c'\bm{k}'}^{c\bm{k}+\bm{Q},v'\bm{k}'+\bm{Q}},
\end{aligned}\end{equation}
\begin{equation}\begin{aligned}
&H^{\text{ares}}(\bm{Q})_{v\bm{k},c'\bm{k}';c\bm{k},v'\bm{k}'}\equiv
H^{\text{res}}_{c\bm{k},v'\bm{k}'+\bm{Q};v\bm{k}+\bm{Q},c'\bm{k}'}=\\[0.2cm]
&(\varepsilon_{c,\bm{k}+\bm{Q}}-\varepsilon_{v,\bm{k}})\delta_{c,c'}\delta_{v,v'}\delta_{\bm{k},\bm{k}'}\\[0.2cm]
&-D_{v\bm{k},c\bm{k}+\bm{Q}}^{v'\bm{k}',c'\bm{k}'+\bm{Q}}+X_{v\bm{k},v'\bm{k}'}^{c\bm{k}+\bm{Q},c'\bm{k}'+\bm{Q}}, 
\end{aligned}\end{equation}
with the off-diagonal kernel terms (of exchange type)
\begin{equation}\begin{aligned}\label{CouplingD}
&\tilde{D}_{v\bm{k},c\bm{k}+\bm{Q}}^{c'\bm{k}',v'\bm{k}'+\bm{Q}}=\\
&\iint\psi^{*}_{v\bm{k}}(\bm{r})\psi_{c'\bm{k}'}(\bm{r})W(\bm{r},\bm{r}')\psi_{c\bm{k}+\bm{Q}}(\bm{r}')\psi^{*}_{v'\bm{k}'+\bm{Q}}(\bm{r}')d\bm{r}d\bm{r}', 
\end{aligned}\end{equation}
\begin{equation}\begin{aligned}\label{CouplingX}
&\tilde{X}_{v\bm{k},c'\bm{k}'}^{c\bm{k}+\bm{Q},v'\bm{k}'+\bm{Q}}=\\
&\iint\psi^{*}_{v\bm{k}}(\bm{r})\psi_{c\bm{k}+\bm{Q}}(\bm{r})v(\bm{r},\bm{r}')\psi_{c'\bm{k}'}(\bm{r}')\psi^{*}_{v'\bm{k}'+\bm{Q}}(\bm{r}')d\bm{r}d\bm{r}'
\end{aligned}\end{equation}
It can be seen that the blocks in the full BSE Hamiltonian \eqref{fullHBSE} are pseudo-hermitian with $\boxed{Q_{i}}^{\dagger}=\begin{pmatrix}I&0\\0&-I\end{pmatrix}\boxed{Q_{i}}\begin{pmatrix}I&0\\0&-I\end{pmatrix}$. Furthermore, upon some additional basis reordering within the $\boxed{-Q_{i}}$ block (in particular permuting the resonant and anti-resonant sections), one can write $\boxed{-Q_{i}}=-\boxed{Q_{i}}^{*}$. Hence, the solutions for $-\bm{Q}_{i}$ are automatically constructed from those of $\bm{Q}_{i}$, provided that the TDA is not employed.

In the RI-based formulation of the BSE given in Section \ref{BSEgauss}, the off-diagonal kernel (screening) terms in \eqref{CouplingD} can be expressed as
\begin{equation}\begin{aligned}\label{DirectOffV}
\tilde{D}_{v\bm{k},c\bm{k}+\bm{Q}}^{c'\bm{k}',v'\bm{k}'+\bm{Q}}=\sum_{R,R'}v_{R}^{v\bm{k},c'\bm{k}'}\epsilon^{-1}_{R,R'}(\bm{k}-\bm{k}')v_{R'}^{v'\bm{k}'+\bm{Q},c'\bm{k}'}
\end{aligned}\end{equation}
Similarly, the off-diagonal kernel (Coulomb) terms in \eqref{CouplingX} read
\begin{equation}\label{ExchangeOffV}
\tilde{X}_{v\bm{k},c'\bm{k}'}^{c\bm{k}+\bm{Q},v'\bm{k}'+\bm{Q}}=\sum_{R}v_{R}^{v\bm{k},c\bm{k}+\bm{Q}}v_{R}^{v'\bm{k}'+\bm{Q},c'\bm{k}'}
\end{equation}

\begin{acknowledgments}
M. A. García-Blázquez is deeply grateful to A. J. Uría-Álvarez, J. K. Desmarais and J. J. Esteve-Paredes for their truly helpful discussions. The authors acknowledge financial support from the Spanish MICINN (grants nos. PID2019-109539GB-C43, TED2021-131323B-I00, and PID2022-141712NB-C21), the Mar\'ia de Maeztu Program for Units of Excellence in R\&D (grant no. CEX2018-000805-M), Comunidad Aut\'onoma de Madrid through the Nanomag COST-CM Program (grant no. S2018/NMT-4321) and the Recovery, Transformation and Resilience Plan from Spain, and by NextGeneration EU from the European Union  (MAD2D-CM-UAM7), the Generalitat Valenciana through the Program Prometeo (2021/017). The authors thankfully acknowledge Red Española de Supercomputación for the computational resources provided by Universidad de Málaga through the project FI-2024-2-0016.
\end{acknowledgments}

%%%%%%%%%%%%%%%%%%%%%%%%%%%%%%%%%%%%%%%%%%%%%%%%%%%%%%%%%%%%%%%%%%%%%%%%%%%%%%%%%
\newpage
\bibliographystyle{apsrev4-2}
\bibliography{main}
\end{document}

\title{\large First Principles Excitons in Periodic Systems with Gaussian Density Fitting and Ewald Potential Functions - SM}

\author{M. A. García-Blázquez}
\email{manuelantonio.garcia@estudiante.uam.es}
\affiliation{Department of Condensed Matter Physics,
Universidad Aut\'onoma de Madrid, Madrid 28049, Spain}

\author{J. J. Palacios}
\email{juanjose.palacios@uam.es}
\affiliation{Department of Condensed Matter Physics,
Universidad Aut\'onoma de Madrid, Madrid 28049, Spain}
\affiliation{Instituto Nicol\'as Cabrera (INC), Universidad Aut\'onoma de Madrid, Madrid 28049, Spain}
\affiliation{Condensed Matter Physics Center (IFIMAC), Universidad Aut\'onoma de Madrid, Madrid 28049, Spain}
\maketitle

%%%%%%%%%%%%%%%%%%%%%%%%%%%%%%%%%%%%%%%%%%%%%%%%%%%%%%%%%%%%%%%%%%%%

\onecolumngrid
Note: the bands here displayed do not contain any scissor correction, regardless of whether one is employed in the BSE calculation (see main text).
\section{Band structures, \texttt{CRYSTAL} input files \& auxiliary basis sets}
\subsection{Black Phosphorus (single layer)}
\begin{figure}[h!]
\centering 
\includegraphics[width=0.7\textwidth]{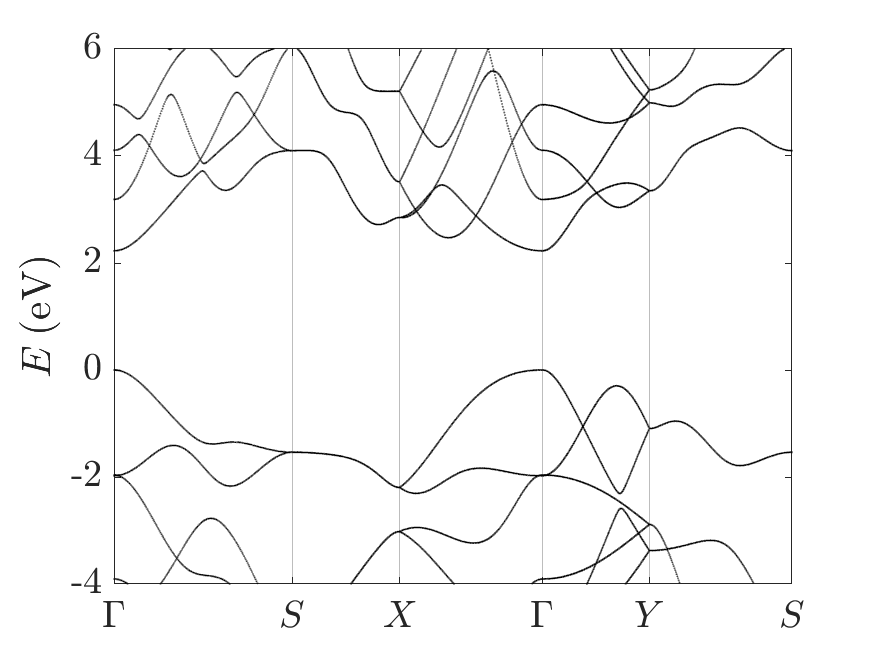}
\caption{Fractional coordinates: $\Gamma=(0,0,0)$, $S=(1/2,1/2,0)$, $X=(1/2,0,0)$, $Y=(0,1/2,0)$. Lattice parameters: $a=3.380\:$\AA, $b=4.496\:$\AA. Functional: PBE0.}
\end{figure}
\noindent
\texttt{Black Phosphorus \newline 
CRYSTAL \newline 
0 0 0 \newline 
19 \newline 
4.496  3.380   10.577 \newline 
2 \newline 
15          0.417902891    0.000000000      0.602252216 \newline 
15          0.082097110    0.500000000      0.602252216 \newline 
SLABCUT \newline 
0 0 1 \newline 
2 2 \newline 
COORPRT \newline 
END \newline 
15 11 \newline 
0 0 7 2.0 1.0 \newline 
\indent  52426.9992330              0.55207164100D-03 \newline 
\indent   7863.2660552              0.42678595308D-02 \newline 
\indent   1789.5227333              0.21931529186D-01 \newline 
\indent    506.27300165             0.85667168373D-01 \newline 
\indent    164.60698546             0.24840686605 \newline 
\indent     58.391918722            0.46336753971 \newline 
\indent     21.643663201            0.35350558156 \newline 
0 0 3 2.0 1.0 \newline 
\indent     99.013837620            0.21895679958D-01 \newline 
\indent     30.550439817            0.95650470295D-01 \newline 
\indent      5.4537087661          -0.29454270186 \newline 
0 0 2 2.0 1.0 \newline 
\indent      2.6503362563           1.3294381200 \newline 
\indent      1.2726688867           0.66109396473 \newline 
0 0 1 0 1.0 \newline 
\indent      0.31645005203          1.0000000 \newline 
0 0 1 0 1.0 \newline 
\indent      0.11417466938          1.0000000 \newline 
0 2 5 6.0 1.0 \newline 
\indent    472.27219248             0.25710623052D-02 \newline 
\indent    111.58882756             0.20250297999D-01 \newline 
\indent     35.445936418            0.91580716787D-01 \newline 
\indent     12.990776875            0.25749454014 \newline 
\indent      5.0486221658           0.42862899758 \newline 
0 2 1 3.0 1.0 \newline 
\indent      1.9934049566            1.0 \newline  
0 2 1 0 1.0 \newline 
\indent      0.66527284430          1.0000000 \newline 
0 2 1 0 1.0 \newline 
\indent      0.25516832128          1.0000000 \newline 
0 2 1 0 1.0 \newline 
\indent      0.90357762251D-01      1.0000000 \newline 
0 3 1 0 1.0 \newline 
\indent      0.218000000            1.0000000 \newline 
99  0 \newline 
END \newline 
DFT \newline 
PBE0 \newline 
END \newline 
TOLINTEG \newline 
10 10 10 10 30 \newline 
SHRINK \newline 
24 24 24 \newline 
MAXCYCLE \newline 
250 \newline 
FMIXING \newline 
80 \newline 
PPAN \newline 
TOLDEE \newline 
9 \newline 
END }
\newline 
\\
\newline 
\vspace{0.2cm} \noindent
\underline{Auxiliary basis:} \newline
\texttt{
15 37   \newline 
0 0 2 0 1.0  \newline 
\indent    299.02239127             1.18280040  \newline 
\indent    105.53104017              .06130569  \newline 
0 0 1 0 1.0  \newline 
\indent     38.45666140            1.0 \newline 
0 0 1 0 1.0 \newline 
\indent     13.18151043             1.0000000 \newline 
0 0 1 0 1.0 \newline 
\indent      7.67135153             1.0000000 \newline 
0 0 1 0 1.0 \newline 
\indent      4.03409415             1.0000000 \newline 
0 0 1 0 1.0 \newline 
\indent      1.69455023             1.0000000 \newline 
0 0 1 0 1.0 \newline 
\indent       .72764573             1.0000000 \newline 
0 0 1 0 1.0 \newline 
\indent       0.63290010406             1.0 \newline 
0 0 1 0 1.0 \newline 
\indent       0.43062472141             1.0 \newline   
0 0 1 0 1.0 \newline 
\indent       .31718400             1.0000000 \newline 
0 0 1 0 1.0 \newline 
\indent       0.22834933876            1.00 \newline 
0 0 1 0 1.0 \newline 
\indent       .144             1.0 \newline 
0 0 1 0 1.0 \newline 
\indent       .06             1.0 \newline 
0 2 1 0 1.0 \newline 
\indent     89.31440733             1.0 \newline 
0 2 1 0 1.0 \newline 
\indent     12.45122715             1.0 \newline 
0 2 1 0 1.0 \newline 
\indent      6.18569317             1.0 \newline 
0 2 1 0 1.0 \newline 
\indent      2.20221854             1.0 \newline 
0 2 1 0 1.0 \newline 
\indent      1.36302664895             1.0000000 \newline 
0 2 1 0 1.0 \newline 
\indent       .87430756             1.0000000 \newline 
0 2 1 0 1.0 \newline 
\indent       0.57161837331             1.0 \newline 
0 2 1 0 1.0 \newline  
\indent       0.38807540247           1.0 \newline 
0 2 1 0 1.0 \newline 
\indent       0.204532431631            1.0 \newline 
0 2 1 0 1.0 \newline 
\indent       0.13            1.0 \newline 
0 2 1 0 1.0 \newline 
\indent       0.07           1.0 \newline 
0 3 2 0 1.0 \newline 
\indent     54.52241961              .63626601 \newline 
\indent     17.58552162             1.54574911 \newline 
0 3 1 0 1.0 \newline 
\indent      5.05626997              1.0 \newline 
0 3 1 0 1.0 \newline 
\indent      2.78509427              1.0 \newline 
0 3 1 0 1.0 \newline 
\indent       .86436165             1.0000000 \newline 
0 3 1 0 1.0 \newline 
\indent       .48264476             1.0000000 \newline 
0 3 1 0 1.0 \newline 
\indent       0.33217466938             1.00 \newline 
0 3 1 0 1.0 \newline 
\indent       0.180715524502             1.00 \newline 
0 4 1 0 1.0 \newline 
\indent      9.0738102843           1.0000000 \newline 
0 4 1 0 1.0 \newline 
\indent      3.5046468322           1.0000000 \newline 
0 4 1 0 1.0 \newline 
\indent      1.3807901291           1.0000000 \newline 
0 4 1 0 1.0 \newline 
\indent      0.75615139217          1.0000000 \newline 
0 4 1 0 1.0 \newline 
\indent      0.38481386692          1.0000000 \newline 
0 5 1 0 1.0 \newline 
\indent      0.55          1.00 
}

\newpage
\subsection{Boron Nitride (single layer)}
\begin{figure}[h!]
\centering 
\includegraphics[width=0.7\textwidth]{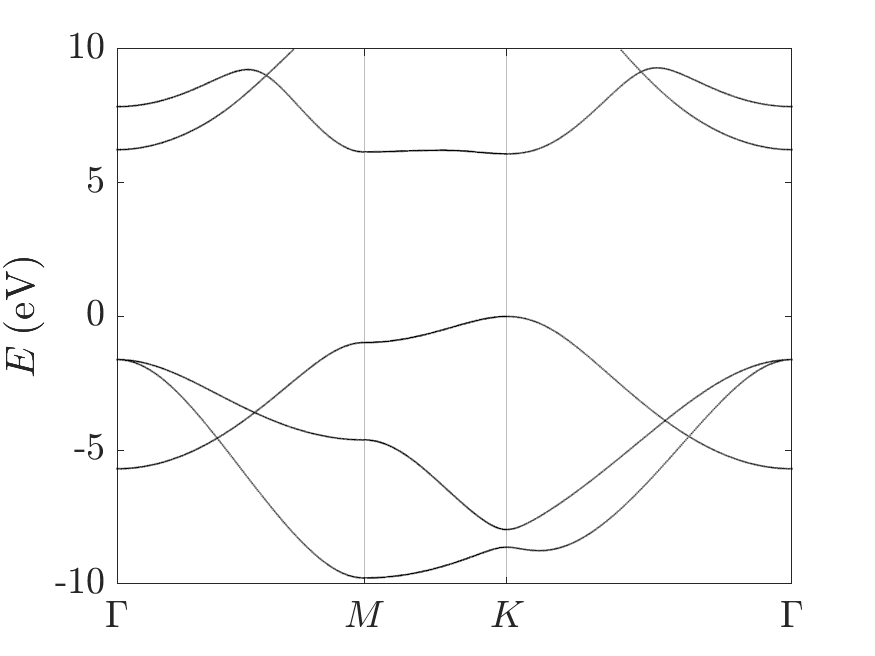}
\caption{Fractional coordinates: $\Gamma=(0,0,0)$, $M=(1/2,0,0)$, $K=(1/3,1/3,0)$. Lattice parameter: $a=2.510\:$\AA. Functional: HSE06.}
\end{figure}
\noindent
\texttt{hBN       \newline                      
SLAB             \newline                        
69               \newline                       
2.51             \newline                      
2                \newline                        
5  0.33333333333 0.66666666667 0.   \newline   
7  0.66666666667 0.33333333333 0. \newline 
END    \newline 
5 11 \newline 
0 0 6 2.0 1.0 \newline 
\indent   8564.8660687              0.22837198155D-03 \newline 
\indent   1284.1516263              0.17682576447D-02 \newline 
\indent    292.27871604             0.91407080516D-02 \newline 
\indent     82.775469176            0.36342638989D-01 \newline 
\indent     27.017939269            0.11063458441 \newline 
\indent      9.8149619660           0.23367344321 \newline 
0 0 2 2.0 1.0 \newline 
\indent      3.9318559059           0.41818777978 \newline 
\indent      1.6595599712           0.22325473798 \newline 
0 0 1 0 1.0 \newline 
\indent      0.35762965239          1.0000000 \newline 
0 0 1 0 1.0 \newline 
\indent      0.14246277496          1.0000000 \newline 
0 0 1 0 1.0 \newline 
\indent      0.60560594768D-01      1.0000000 \newline 
0 2 4 1.0 1.0 \newline 
\indent     22.453875803            0.50265575179D-02 \newline 
\indent      5.1045058330           0.32801738965D-01 \newline 
\indent      1.4986081344           0.13151230768 \newline 
\indent      0.50927831315          0.33197167769 \newline 
0 2 1 0 1.0 \newline 
\indent      0.18147077798           1.0 \newline 
0 2 1 0 1.0 \newline 
\indent      0.64621893904D-01       1.0 \newline  
0 3 1 0 1.0 \newline 
\indent      0.66100000             1.0000000 \newline 
0 3 1 0 1.0 \newline  
\indent      0.19900000             1.0000000 \newline 
0 4 1 0 1.0 \newline 
\indent      0.49000000             1.0000000 \newline 
7 11 \newline 
0 0 6 2.0 1.0 \newline 
\indent  19730.8006470              0.21887984991D-03 \newline 
\indent   2957.8958745              0.16960708803D-02 \newline 
\indent    673.22133595             0.87954603538D-02 \newline 
\indent    190.68249494             0.35359382605D-01 \newline 
\indent     62.295441898            0.11095789217 \newline 
\indent     22.654161182            0.24982972552 \newline 
0 0 2 2.0 1.0 \newline 
\indent      8.9791477428           0.40623896148 \newline 
\indent      3.6863002370           0.24338217176 \newline 
0 0 1 0 1.0 \newline 
\indent      0.84660076805          1.0000000 \newline 
0 0 1 0 1.0 \newline 
\indent      0.33647133771          1.0000000 \newline 
0 0 1 0 1.0 \newline 
\indent      0.13647653675          1.0000000 \newline 
0 2 4 3.0 1.0 \newline 
\indent     49.200380510            0.55552416751D-02 \newline 
\indent     11.346790537            0.38052379723D-01 \newline 
\indent      3.4273972411           0.14953671029 \newline  
\indent      1.1785525134           0.34949305230 \newline 
0 2 1 0 1.0 \newline 
\indent      0.41642204972          1.0 \newline 
0 2 1 0 1.0 \newline 
\indent      0.14260826011          1.0 \newline 
0 3 1 0 1.0 \newline 
\indent      1.65400000             1.0000000 \newline 
0 3 1 0 1.0 \newline 
\indent      0.46900000             1.0000000 \newline 
0 4 1 0 1.0 \newline 
\indent      1.09300000             1.0000000 \newline 
99  0     \newline 
PRINT \newline 
END \newline 
DFT  \newline 
XXLGRID                      \newline              
HSE06                                 \newline 
END                                   \newline   
TOLINTEG                           \newline       
12 12 12 12 36 \newline 
SHRINK                      \newline             
48 48 48                      \newline      
MAXCYCLE                    \newline              
500  \newline 
FMIXING                        \newline         
95     \newline 
TOLDEE                             \newline       
10         \newline 
NODIIS           \newline 
END}
\newline 
\\
\newline 
\vspace{0.2cm} \noindent
\underline{Auxiliary basis:} \newline
\texttt{
5 35  \newline
0 0 1 0 1.0  \newline
\indent     252.54                   1.0000000  \newline
0 0 1 0 1.0  \newline
\indent     57.1851                 1.0000000  \newline
0 0 1 0 1.0  \newline
\indent     18.2486                 1.0000000  \newline
0 0 1 0 1.0  \newline
\indent      7.67715                1.0000000  \newline
0 0 1 0 1.0  \newline
\indent      3.91872                1.0000000  \newline
0 0 1 0 1.0  \newline
\indent      1.74456                1.0000000  \newline
0 0 1 0 1.0  \newline
\indent      0.704034               1.0000000  \newline
0 0 1 0 1.0  \newline
\indent      0.323573               1.0000000  \newline
0 0 1 0 1.0  \newline 
\indent      0.163085               1.0000000  \newline
0 0 1 0 1.0  \newline
\indent      0.06               1.0000000  \newline
0 2 1 0 1.0  \newline
\indent     37.0411                 1.0000000  \newline
0 2 1 0 1.0  \newline
\indent     10.4481                 1.0000000  \newline
0 2 1 0 1.0  \newline
\indent      5.04018                1.0000000  \newline
0 2 1 0 1.0  \newline
\indent      2.24762                1.0000000  \newline
0 2 1 0 1.0  \newline
\indent      1.06158                1.0000000  \newline
0 2 1 0 1.0  \newline
\indent      0.509525               1.0000000  \newline
0 2 1 0 1.0  \newline
\indent      0.261888               1.0000000  \newline
0 2 1 0 1.0  \newline
\indent      0.135744               1.0000000  \newline
0 2 1 0 1.0  \newline 
\indent      0.05               1.0000000  \newline
0 3 1 0 1.0  \newline
\indent     14.9067                 1.0000000  \newline
0 3 1 0 1.0  \newline
\indent      5.86893                1.0000000  \newline
0 3 1 0 1.0  \newline
\indent      2.63244                1.0000000  \newline
0 3 1 0 1.0  \newline
\indent      1.386                  1.0000000  \newline
0 3 1 0 1.0  \newline
\indent      0.742048               1.0000000  \newline
0 3 1 0 1.0  \newline
\indent      0.506888               1.0000000  \newline
0 3 1 0 1.0  \newline
\indent      0.245784               1.0000000  \newline 
0 3 1 0 1.0  \newline
\indent      0.11              1.0000000  \newline
0 4 1 0 1.0  \newline
\indent      4.29775693             1.0000000  \newline
0 4 1 0 1.0  \newline
\indent      1.75710885             1.0000000  \newline
0 4 1 0 1.0  \newline
\indent      1.09073089             1.0000000  \newline
0 4 1 0 1.0  \newline
\indent      0.47390324             1.0000000  \newline
0 4 1 0 1.0  \newline
\indent      0.21274953             1.0000000  \newline
0 5 1 0 1.0  \newline
\indent      1.70494610             1.0000000  \newline
0 5 1 0 1.0  \newline
\indent      0.87447217             1.0000000  \newline
0 5 1 0 1.0  \newline
\indent      0.47441651             1.0000000  \newline
7 35  \newline
0 0 1 0 1.0  \newline
\indent    270.16                   1.0000000  \newline
0 0 1 0 1.0  \newline
\indent     65.5655                 1.0000000  \newline  
0 0 1 0 1.0  \newline
\indent     26.1581                 1.0000000  \newline
0 0 1 0 1.0   \newline
\indent     12.1539                 1.0000000  \newline
0 0 1 0 1.0  \newline
\indent      2.93899                1.0000000  \newline
0 0 1 0 1.0  \newline
\indent      1.74277                1.0000000  \newline
0 0 1 0 1.0  \newline
\indent      0.970756               1.0000000  \newline
0 0 1 0 1.0  \newline
\indent      0.431318               1.0000000  \newline
0 0 1 0 1.0  \newline
\indent      0.265722               1.0000000  \newline
0 0 1 0 1.0  \newline
\indent      0.12               1.0000000  \newline
0 2 1 0 1.0  \newline  
\indent     75.0137                 1.0000000  \newline
0 2 1 0 1.0  \newline
\indent     20.646                  1.0000000  \newline
0 2 1 0 1.0  \newline  
\indent      8.50289                1.0000000  \newline
0 2 1 0 1.0  \newline
\indent      3.51896                1.0000000  \newline
0 2 1 0 1.0  \newline
\indent      2.17404                1.0000000  \newline
0 2 1 0 1.0  \newline
\indent      0.852831               1.0000000  \newline
0 2 1 0 1.0  \newline
\indent      0.453657               1.0000000  \newline
0 2 1 0 1.0  \newline
\indent      0.323403               1.0000000  \newline
0 2 1 0 1.0  \newline
\indent      0.15               1.0000000  \newline
0 3 1 0 1.0  \newline
\indent     21.1146                 1.0000000  \newline
0 3 1 0 1.0  \newline
\indent      7.10001                1.0000000  \newline  
0 3 1 0 1.0  \newline
\indent      3.57592                1.0000000  \newline
0 3 1 0 1.0  \newline
\indent      2.04643                1.0000000  \newline
0 3 1 0 1.0  \newline
\indent      1.25245                1.0000000  \newline
0 3 1 0 1.0  \newline
\indent      0.631629               1.0000000  \newline
0 3 1 0 1.0  \newline
\indent      0.311297               1.0000000  \newline
0 3 1 0 1.0  \newline 
\indent      0.16               1.0000000  \newline
0 4 1 0 1.0  \newline 
\indent      8.47976509             1.0000000  \newline
0 4 1 0 1.0  \newline
\indent      3.61735700             1.0000000  \newline
0 4 1 0 1.0  \newline
\indent      2.49535994             1.0000000  \newline
0 4 1 0 1.0  \newline
\indent      1.03313399             1.0000000  \newline
0 4 1 0 1.0  \newline
\indent      0.48153351             1.0000000  \newline
0 5 1 0 1.0  \newline
\indent      4.18508470             1.0000000  \newline
0 5 1 0 1.0  \newline
\indent      1.92553488             1.0000000  \newline
0 5 1 0 1.0  \newline
\indent      0.97077363             1.0000000  \newline
}

\newpage
\subsection{Solid Argon}
\begin{figure}[h!]
\centering 
\includegraphics[width=0.7\textwidth]{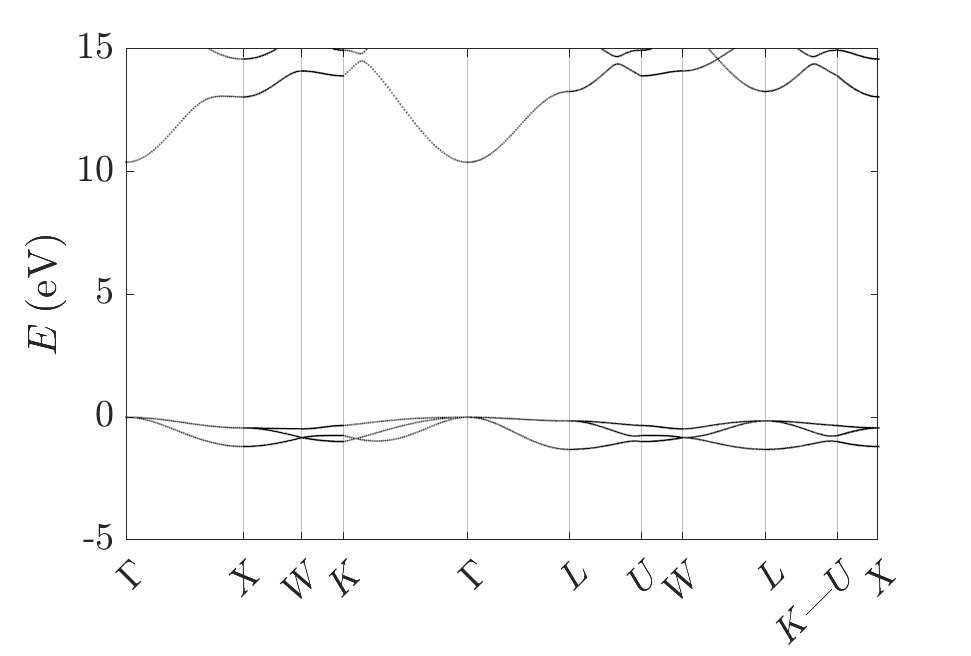}
\caption{Fractional coordinates: $\Gamma=(0,0,0)$, $X=(1/2,0,1/2)$, $W=(1/2,1/4,3/4)$, $K=(3/8,3/8,3/4)$, $L=(1/2,1/2,1/2)$, $U=(5/8,1/4,5/8)$. Lattice parameters: $a=5.31\:$\AA  (crystallographic, i.e. non-primitive cell). Functional: HSE06.}
\end{figure}
\noindent
\texttt{Ar FCC      \newline        
CRYSTAL \newline 
0 0 0               \newline                   
225               \newline                       
5.31                \newline               
1                           \newline            
18    0.0  0.0  0.0 \newline 
COORPRT              \newline              
END   \newline 
18 17 \newline 
0 0 7 2.0 1.0 \newline 
\indent  79111.4229980              0.51029325002D-03 \newline 
\indent  11864.7460090              0.39463036981D-02 \newline 
\indent   2700.1642973              0.20307073910D-01 \newline 
\indent    763.95943485             0.79691825214D-01 \newline 
\indent    248.45150561             0.23420623836 \newline 
\indent     88.283581000            0.44833849481 \newline 
\indent     32.948607069            0.36408167400 \newline 
0 0 3 2.0 1.0 \newline 
\indent    142.55358000             0.26387407001D-01 \newline 
\indent     44.163688009            0.11226433999 \newline 
\indent      8.9524995000          -0.26178922001 \newline 
0 0 2 2.0 1.0 \newline 
\indent      4.5546920941           1.3002484998 \newline 
\indent      2.1444079001           0.67197237009 \newline 
0 0 1 0 1.0 \newline 
\indent      0.60708777004          1.0000000 \newline 
0 0 1 0 1.0 \newline 
\indent      0.21651431999          1.0000000 \newline 
0 0 1 0 1.0 \newline 
\indent      0.73732186754D-01      1.0000000 \newline 
0 2 5 6.0 1.0 \newline 
\indent    776.77541998             0.22028005003D-02 \newline 
\indent    183.80107018             0.17694180008D-01 \newline 
\indent     58.694003175            0.82431293717D-01 \newline  
\indent     21.701591695            0.24207278863 \newline 
\indent      8.5821489635           0.42263558251 \newline 
0 2 1 6.0 1.0 \newline 
\indent      3.4922679161            1.0 \newline 
0 2 1 0 1.0 \newline 
\indent      1.2637426998           1.0000000 \newline 
0 2 1 0 1.0 \newline 
\indent      0.46607870005          1.0000000 \newline 
0 2 1 0 1.0 \newline 
\indent      0.15766003000          1.0000000 \newline 
0 2 1 0 1.0 \newline 
\indent      0.61D-01      1.0000000 \newline 
0 3 1 0 1.0 \newline 
\indent      5.55100000             1.0000000 \newline 
0 3 1 0 1.0 \newline 
\indent      1.23500000             1.0000000 \newline 
0 3 1 0 1.0 \newline 
\indent      0.412000000            1.0000000 \newline 
0 3 1 0 1.0 \newline 
\indent      0.12039311859          1.0000000 \newline 
0 4 1 0 1.0 \newline 
\indent      0.890000000            1.0000000 \newline 
99  0     \newline 
PRINT \newline 
END \newline 
DFT     \newline 
HUGEGRID                              \newline 
HSE06                        \newline 
END                                     \newline 
TOLINTEG                                 \newline 
12 12 12 12 36  \newline 
SHRINK                           \newline         
48 48 48             \newline                    
MAXCYCLE                \newline                  
300 \newline 
FMIXING              \newline                   
80 \newline 
TOLDEE             \newline                      
10               \newline 
PPAN         \newline            
END       }
\newline 
\\
\newline 
\vspace{0.2cm} \noindent
\underline{Auxiliary basis:} \newline
\texttt{
18 59 \newline 
0 0 1 0 1.0 \newline 
\indent    581.184                  1.0000000 \newline 
0 0 1 0 1.0 \newline 
\indent    165.101                  1.0000000 \newline 
0 0 1 0 1.0 \newline 
\indent     60.0502                 1.0000000 \newline 
0 0 1 0 1.0 \newline 
\indent     15.2676                 1.0000000 \newline 
0 0 1 0 1.0 \newline 
\indent      7.87286                1.0000000 \newline 
0 0 1 0 1.0 \newline 
\indent      4.81212                1.0000000 \newline 
0 0 1 0 1.0 \newline 
\indent      2.51481                1.0000000 \newline 
0 0 1 0 1.0 \newline 
\indent      1.43151                1.0000000 \newline 
0 0 1 0 1.0 \newline 
\indent      1.21417554008          1.00 \newline 
0 0 1 0 1.0 \newline 
\indent      0.82360209003          1.00 \newline  
0 0 1 0 1.0 \newline 
\indent      0.680819956794         1.00 \newline 
0 0 1 0 1.0 \newline 
\indent      0.43302863998          1.00 \newline 
0 0 1 0 1.0 \newline 
\indent      0.290246506744         1.00 \newline 
0 0 1 0 1.0 \newline 
\indent      0.147464373508         1.00 \newline 
0 0 1 0 1.0 \newline 
\indent      0.0675396              1.0000000 \newline 
0 2 1 0 1.0 \newline 
\indent     95.7281                 1.0000000 \newline 
0 2 1 0 1.0 \newline  
\indent     23.7126                 1.0000000 \newline 
0 2 1 0 1.0 \newline 
\indent      9.50489                1.0000000 \newline 
0 2 1 0 1.0 \newline 
\indent      3.87991                1.0000000 \newline 
0 2 1 0 1.0 \newline 
\indent      2.08975                1.0000000 \newline 
0 2 1 0 1.0 \newline 
\indent      1.07316647009        1.00 \newline 
0 2 1 0 1.0 \newline 
\indent      0.76474780004        1.00 \newline 
0 2 1 0 1.0 \newline 
\indent      0.68259302004        1.00 \newline 
0 2 1 0 1.0 \newline 
\indent      0.539810886804        1.00 \newline 
0 2 1 0 1.0 \newline 
\indent      0.37417435          1.00 \newline 
0 2 1 0 1.0 \newline 
\indent      0.27751432          1.00 \newline 
0 2 1 0 1.0 \newline 
\indent      0.231392216754         1.00 \newline 
0 2 1 0 1.0 \newline 
\indent      0.134732186754         1.00 \newline 
0 2 1 0 1.0 \newline 
\indent      0.05               1.0000000 \newline 
0 3 1 0 1.0 \newline 
\indent     82.5392                 1.0000000 \newline 
0 3 1 0 1.0 \newline 
\indent     26.276                  1.0000000 \newline 
0 3 1 0 1.0 \newline 
\indent      9.40744                1.0000000 \newline 
0 3 1 0 1.0 \newline 
\indent      3.91259                1.0000000 \newline 
0 3 1 0 1.0 \newline 
\indent      1.62726                1.0000000 \newline 
0 3 1 0 1.0 \newline 
\indent      1.30873218675        1.00 \newline 
0 3 1 0 1.0 \newline 
\indent      0.9321574001         1.00 \newline 
0 3 1 0 1.0 \newline 
\indent      0.72748088863        1.00 \newline 
0 3 1 0 1.0 \newline 
\indent      0.62373873005        1.00 \newline 
0 3 1 0 1.0 \newline 
\indent      0.506405443402       1.00 \newline 
0 3 1 0 1.0 \newline 
\indent      0.32611374929        1.00 \newline 
0 3 1 0 1.0 \newline 
\indent      0.206392667672        1.00 \newline 
0 3 1 0 1.0 \newline 
\indent      0.122                  1.00 \newline 
0 4 1 0 1.0 \newline 
\indent     15.586304207            1.0000000 \newline 
0 4 1 0 1.0 \newline 
\indent      6.0271769796           1.0000000 \newline 
0 4 1 0 1.0 \newline 
\indent      2.3410113748           1.0000000 \newline 
0 4 1 0 1.0 \newline 
\indent      1.70107870005          1.00 \newline 
0 4 1 0 1.0 \newline 
\indent      1.344330015            1.00 \newline 
0 4 1 0 1.0 \newline 
\indent      0.963732186754         1.00 \newline 
0 4 1 0 1.0 \newline 
\indent      0.87807870005          1.00 \newline 
0 4 1 0 1.0 \newline 
\indent      0.70370885563          1.0000000 \newline 
0 4 1 0 1.0 \newline 
\indent      0.57806592432          1.00 \newline 
0 4 1 0 1.0 \newline  
\indent      0.473                  1.00 \newline 
0 4 1 0 1.0 \newline 
\indent      0.27805314859          1.00 \newline 
0 4 1 0 1.0 \newline 
\indent      0.18139311859          1.00 \newline 
0 5 1 0 1.0 \newline 
\indent      4.6932752519           1.0000000 \newline 
0 5 1 0 1.0 \newline 
\indent      1.35539311859           1.00 \newline 
0 5 1 0 1.0 \newline 
\indent      0.8875         1.00 \newline 
0 5 1 0 1.0 \newline 
\indent      0.53239311859         1.00 \newline 
0 5 1 0 1.0 \newline 
\indent      0.24078623718          1.00 \newline 
}

\newpage
\section{Spin-orbit coupling splitting in Solid Argon}
\begin{figure}[h!]
\centering 
\includegraphics[width=0.7\textwidth]{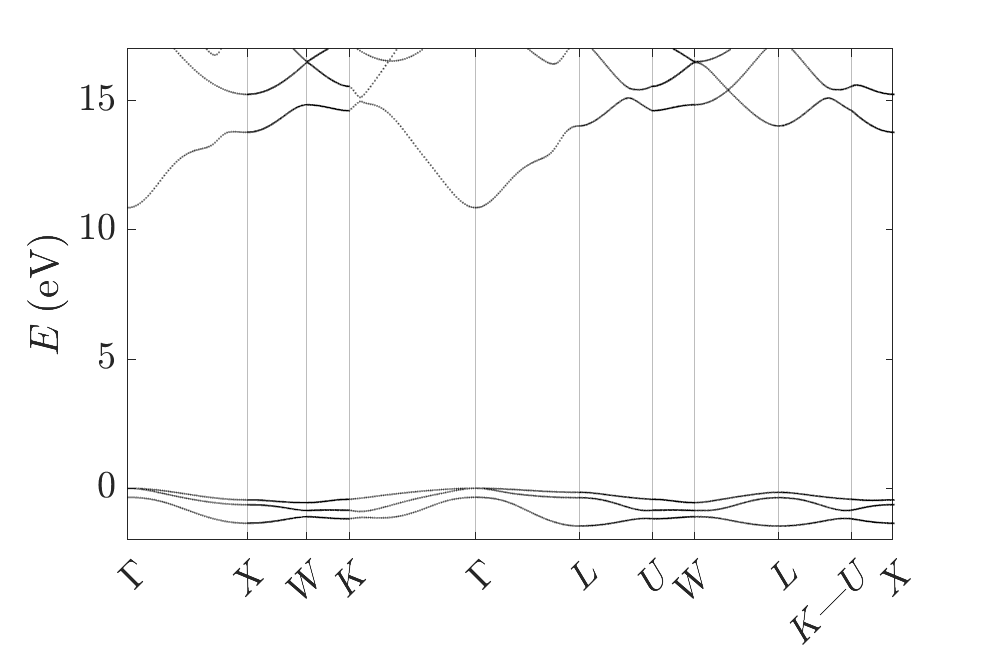}
\caption{Functional: PBE0, with relativistic pseudo-potential including SOC.}
\end{figure}

\noindent
In this case we modify the functional to comply with \texttt{CRYSTAL}'s current availability within the two-component framework, in addition to other parameters to ease the convergence. The input file we employed is the following:
\newline 
\\
\newline 
\vspace{0.2cm} 
\noindent
\texttt{  
Ar FCC SOC     \newline 
CRYSTAL \newline 
0 0 0                   \newline               
225                    \newline                  
5.31                    \newline            
1                         \newline               
218    0.0  0.0  0.0 \newline 
COORPRT \newline 
SYMMREMO                          \newline 
END   \newline 
218 14 \newline 
COLUSC \newline 
0 0 2 2.0 1.0 \newline 
\indent      4.5546920941           1.3002484998 \newline 
\indent      2.1444079001           0.67197237009 \newline 
0 0 1 0 1.0 \newline 
\indent      0.60708777004          1.0000000 \newline 
0 0 1 0 1.0 \newline 
\indent      0.21651431999          1.0000000 \newline 
0 0 1 0 1.0 \newline 
\indent     0.73732186754D-01      1.0000000 \newline 
0 2 1 6.0 1.0 \newline 
\indent      3.4922679161            1.0 \newline 
0 2 1 0 1.0 \newline 
\indent      1.2637426998           1.0000000 \newline 
0 2 1 0 1.0 \newline 
\indent      0.46607870005          1.0000000 \newline 
0 2 1 0 1.0 \newline 
\indent      0.15766003000          1.0000000 \newline 
0 2 1 0 1.0 \newline 
\indent      0.61     1.0000000 \newline 
0 3 1 0 1.0 \newline 
\indent      5.55100000             1.0000000 \newline 
0 3 1 0 1.0 \newline 
\indent      1.23500000             1.0000000 \newline 
0 3 1 0 1.0 \newline 
\indent      0.412000000            1.0000000 \newline 
0 3 1 0 1.0 \newline 
\indent      0.12039311859          1.0000000 \newline  
0 4 1 0 1.0 \newline 
\indent      0.890000000            1.0000000 \newline 
99  0     \newline 
END \newline 
TWOCOMPON \newline 
SOC \newline 
ENDTWO \newline 
DFT    \newline 
SPIN  \newline 
XXLGRID                              \newline 
PBE0                       \newline 
ENDDFT                          \newline             
TOLINTEG                        \newline          
8 8 8 8 16 \newline 
SHRINK                        \newline            
4 4 4                        \newline  
MAXCYCLE                          \newline        
300 \newline 
FMIXING                        \newline         
95 \newline 
TOLDEE                       \newline            
9                \newline 
NODIIS                \newline    
END                                  
}

%%%REFERENCES%%%

%\bibliographystyle{apsrev4-2}
%\bibliography{BibFile}